\documentclass{statsoc}
\usepackage{geometry}

\usepackage[utf8]{inputenc}
\usepackage{lipsum}

\usepackage{amsmath,amssymb}
\usepackage{natbib}
\usepackage{soul}
\usepackage{color, xcolor}
\usepackage{enumerate}
\usepackage{appendix}
\usepackage{graphicx}
\usepackage{float}

\newtheorem{theorem}{\sc Theorem}

\newtheorem{assumption}{\sc Assumption}
\newtheorem{lemma}{\sc Lemma}

\title[Statistical Inference for High-dimensional Matrix-variate Factor Models with Missing Observations]{Statistical Inference for High-dimensional Matrix-variate Factor Models with Missing Observations}
\author[]{Yongxia \textsc{Zhang}}
\address{College of Science, North China University of Technology, Beijing, China}

\author[]{Jinwen \textsc{Liang}}
\address{School of Mathematics, Statistics and Mechanics, Beijing University of Technology, Beijing, China}

\author[]{Liwen \textsc{Xu}}
\address{College of Science, North China University of Technology, Beijing, China}

\author[]{Keming \textsc{Yu}}
\address{Mathematical Sciences, Brunel University London, London, United Kingdom}

\author[]{Maozai \textsc{Tian}*}
\address{Center for Applied Statistics, School of Statistics, Renmin University of China, Beijing, China.}

\begin{document}

\begin{abstract}
This paper develops an inferential theory for high-dimensional matrix-variate factor models with missing observations. We propose an easy-to-use all-purpose method that involves two straightforward steps. First, we perform principal component analysis on two re-weighted covariance matrices to obtain the row and column loadings. Second, we utilize these loadings along with the matrix-variate data to derive the factors. We develop an inferential theory that establishes the consistency and the rate of convergence under general conditions and missing patterns. The simulation results demonstrate the adequacy of the asymptotic results in approximating the properties of a finite sample. Finally, we illustrate the application of our method using a real numerical dataset.
\end{abstract}


\maketitle
\section{Introduction}\label{sec:1}
 Large-scale matrix-variate data have become increasingly common across various fields, such as neuroscience, healthcare, economics, and social networking. For instance, quarterly economic indicators from multiple countries at different time points naturally create a dynamic sequence of matrix variables. In this setup, the rows represent different countries, while the columns represent measurements such as GDP and CPI at specific times. Additionally, it is common to encounter matrix-variate data with missing entries, as seen in the multinational macroeconomic indices dataset collected from the OECD (https://data-explorer.oecd.org/). Dealing with matrix-variate data with missing has attracted increasing attention from researchers. This is a fundamental problem that has applications in social science, statistics, and computer science. The development of statistical methods for analyzing large-scale matrix-variate data with missing values remains in its early stages. Consequently, scientists often address the issue by replacing missing values with zero and then applying principal component analysis (PCA) to the covariance matrices estimated from this imputed data. There is a significant amount of literature on PCA applied to latent factor models \citep{https://doi.org/10.1111/1468-0262.00273, https://doi.org/10.1111/1468-0262.00392, 10.1111/rssb.12016, CHANG2015297, Pelger03072022}.

This paper develops inferential theory for latent factor models estimated from high-dimensional matrix-variate factor models with missing observations. To the best of our knowledge, we are the first to develop the method to deal with matrix-variate latent factor models with data missing. We propose a novel and user-friendly approach to estimate matrix-variate latent factor models by applying PCA to adjusted covariance matrices derived from partially observed data. Furthermore, we derive the theoretical properties of the estimated factors and loadings.

 Our method is very simple to implement and works under general assumptions. We present an ‘‘all-purpose’’ estimator that performs effectively across various empirically relevant missing data patterns. Our estimation process consists of two simple steps. First, we apply PCA to two re-weighted covariance matrices to derive the row and column loadings. In the second step, we utilize these loadings along with the matrix-variate data to extract the factors. Importantly, our estimator does not require the estimation of the observation pattern itself. Our framework is distinguished by its ability to accommodate a wide range of missing observation patterns. We address common scenarios, including missing at random and simultaneous missingness, in which a unit that cannot be observed remains unavailable for a period of time.

Research on statistical inference for high-dimensional matrix-variate factor models, there are many related researches. \cite{WANG2019231} and \cite{Chen02042020} considered the matrix-variate factor models in which the error term is assumed to be white noise \citep{10.1093/biomet/asr048,10.1214/12-AOS970}. \cite{Chen02012022} further expanded upon these results to time series of tensor observations, under the assumption that the noise tensors are not temporally correlated. This line of research discards contemporaneous covariance and utilizes only the auto-covariance of the data at time $t$ and time $t-h$ with $h\geq 1$.
 \cite{Chen03042023} generalized the assumptions by allowing the error term to be weakly correlated across rows, columns, and observations. Additionally, tensor decomposition has been applied to matrix-variate observations \citep{Kolda2009TensorDA, Kolda2006Multilinear}. Note that
 $\{\mathbf{Y}_t\}_{1\leq t\leq T}$ form an order-3 tensor with dimensions $a \times b \times T$ by stacking $\mathbf{Y}_t$ along the third mode for $1\leq t\leq T$. Statistical convergence rates in the Frobenius norm have been investigated by \cite{8368145} under the assumption that the tensor entries are homogeneous. However, none of these studies have addressed the issue of missing observations in matrix-variate data.

 There have been several attempts to address large-dimensional vector data with missing entries. \cite{JIN2021745} presented the inferential theory for an estimated factor model using the expectation–maximization (EM) algorithm, assuming that the values are missing at random. This represents a significant advancement in the literature concerning the use of the EM algorithm for imputing missing values in cross-sectional data \citep{https://doi.org/10.1002/j.2333-8504.1975.tb01053.x, https://doi.org/10.1111/j.2517-6161.1977.tb01600.x}. Additionally, \cite{Bai02102021} offered an inferential theory for factor-based imputed values by introducing the innovative approach of shuffling rows and columns to create fully observed TALL and WIDE blocks, which are used for estimating the factor model. Their TALL-WIDE algorithm involves applying PCA to the two fully observed blocks. \cite{CAHAN2023113} developed the TALL-PROJECT estimator, which extends the TALL-WIDE estimator by initially using only the fully observed TALL block for PCA to estimate the factors, and then deriving the loadings from a time-series regression that incorporates all observed entries. Each of these estimators was designed for a specific missing pattern, under which they perform particularly well, but they may not generalize to other patterns. \cite{XIONG2023271} introduced an easy-to-use, all
purpose estimator for a latent factor model. This method applies PCA to an adjusted covariance matrix that is estimated from partially observed panel data and generalizes the approach for causal inference.

  The inferential theory of matrix-variate factor models with missing observations is a rapidly evolving area of research. Our paper closely relates to the work of \cite{XIONG2023271} and \cite{Chen03042023}. This paper makes three key contributions. First, we extend the inferential theory of latent factors to matrix-variate data with general patterns of missing entries. This significantly broadens the applicability of \cite{Chen03042023} and related studies, making our theories and methods relevant for a wider range of applications. Second, our method allows for the imputation of matrix-variate data with missing entries, based on the assumption of a low-rank structure. Third, we establish new results on the asymptotic normality and the convergence rate of the estimators. They are useful in constructing the confidence intervals of the estimators.
\subsection{Notation and Organization}
We use the lowercase letter $x$, boldface letter $\mathbf{x}$, and boldface capital letter $\mathbf{X}$ to denote a scalar, vector and matrix, respectively. We use $\mathbf{X}_{i},\mathbf{X}_{\cdot j}$, and $x_{ij}$ to denote the $i$th row, $j$th column, and $\left(i,j\right)$th element of a matrix $\mathbf{X}$, respectively. For a matrix $\mathbf{X}$, we define the following matrix norms: maximum norm $\Vert \mathbf{X}\Vert_{\max}\triangleq \max_{ij}|x_{ij}|, l_1$ norm $\Vert \mathbf{X}\Vert_{1} \triangleq \max_j \sum_i |x_{ij}|, l_{\infty}$-norm $\Vert \mathbf{X}\Vert_{\infty}\triangleq \max_i \sum_j |x_{ij}|$, and $l_2$-norm $\Vert \mathbf{X}\Vert_{2}\triangleq \lambda_1$, where $\lambda_1$ is the largest singular value $\{\lambda_i\}$ of $\mathbf{X}$ with $\lambda_i$ being the $i$th largest square root of eigenvalues of $\mathbf{X}^{\top}\mathbf{X}$. We also use $\Vert \mathbf{X}\Vert$ for $l_2$ norm. When $\mathbf{X}$ is a square matrix, we denote by $Tr\left(\mathbf{X}\right), \sigma_{\max}\left(\mathbf{X}\right)$, and $\sigma_{\min}\left(\mathbf{X}\right)$ the trace, maximum and minimum singular value of $\mathbf{X}$, respectively. We let $[n]\triangleq \{1,\cdots,n\}$ denote the set of integers from $1$ to $n$.

The rest of this paper is organized as follows. In Section \ref{sec:2}, we introduce the model and the estimation method. We develop the asymptotic theoretical results for the estimated loading matrices and latent factors in Section \ref{sec:3}. In Section \ref{s:sim}, we study the finite sample performance of our estimation through simulation. Section \ref{sec:App} provides empirical studies using data collected from OECD. Finally, Section \ref{sec:Con} concludes the paper. All proofs and technique lemmas are included in Appendix \ref{sec:A}, while Appendix \ref{sec:A2} shows more simulation results, Appendix \ref{A:3} shows the details of multinational macroeconomic indexes.
\section{Model and Estimation}\label{sec:2}
\subsection{Model}
We specifically explore the following matrix-variate factor model for observations $\mathbf{Y}_t \in \mathbb{R}^{a \times b}, t \in \{1,2,\cdots,T\}$:
\begin{equation}\label{eq:2-1}
\mathbf{Y}_t=\mathbf{R}\mathbf{F}_t \mathbf{C}^{\top}+\mathbf{E}_t.
\end{equation}
We only observe $\mathbf{Y}_t$, while everything on the right side of model \eqref{eq:2-1} remains unknown. The matrix $\mathbf{Y}_t$ is driven by the latent factor matrix $\mathbf{F}_t \in \mathbb{R}^{k \times r}$, which has lower dimensions, specifically, $k<<a$ and $r<<b$. The matrices $\mathbf{R} \in \mathbb{R}^{a \times k}$ and $\mathbf{C} \in \mathbb{R}^{b \times r}$ are the row and column loading matrices, respectively. The noise matrix $\mathbf{E}_t \in \mathbb{R}^{a \times b}$ is assumed to be uncorrelated with $\mathbf{F}_t$.

We can separate the signal component $\mathbf{S}_t=\mathbf{R}\mathbf{F}_t \mathbf{C}^{\top}$ from the noise component $\mathbf{E}_t$ by relying on the pervasiveness assumption regarding the loading matrices $\mathbf{R}$ and $\mathbf{C}$, as well as the bounded eigenvalues assumption for the row and column covariances of the noise.
Large-scale matrix-variate data with missing entries are prevalent. Let $W_{t,ij}=1$ indicate that the ${i,j}$th entry of $\mathbf{Y}_t$ is observed, while $W_{t,ij}=0$ indicates that the ${i,j}$th entry of $\mathbf{Y}_t$ is missing. This paper develops the inferential theory for latent factor models estimated from large-scale matrix-variate data with missing observations.
\subsection{Missing Observations}
We allow for very general patterns in the missing observations. The first one is a randomly missing pattern, meaning that whether an entry is observed or not does not depend on other entries or observable covariates. The second pattern is simultaneous missingness, in which a unit that cannot be observed remains unavailable for a period of time, this will be modeled as missing value.

Let $Q_{R,ij}=\{t,m: W_{t,im}=1, W_{t,jm}=1\}$ denote the set of time periods $t$ and dimension $m$ for both units $i$ and $j$ are observed. Let $Q_{C,ij}=\{t,l: W_{t,li}=1 , W_{t,lj}=1\}$ denote the set of time periods $t$ and the unit $l$ for both dimensions $i$ and $j$ are observed. The cardinality of $Q_{R,ij}$ and $Q_{C,ij}$  is represented by$\left|Q_{R,ij}\right|$ and $\left|Q_{C,ij}\right|$, respectively.
\subsection{Model Identification}
We only observe $\mathbf{Y}_t$, and everything on the right hand side of model \eqref{eq:2-1} is unknown. The latent factor matrix $\mathbf{F}_t$ and the loading matrices $\mathbf{R}$ and $\mathbf{C}$ are not individually identifiable. However, they can be estimated up to an invertible matrix transformation. Therefore, rather than aiming to determine the true values of $\mathbf{R},\mathbf{F}_t$ and $\mathbf{C}$, our objective is to estimate their transformations. Without loss of generality, we restrict our estimator $\widehat{\mathbf{R}}$ and $\widehat{\mathbf{C}}$ such that:
\begin{equation}\label{eq:2-2}
\frac{1}{a}\widehat{\mathbf{R}}^{\top}\widehat{\mathbf{R}}=\mathbf{I};\\
\frac{1}{b}\widehat{\mathbf{C}}^{\top}\widehat{\mathbf{C}}=\mathbf{I}.
\end{equation}
As shown in Theorems \ref{Th1}, and \ref{Th3}, for any ground truth matrices $\mathbf{R},\mathbf{F}_t$ and $\mathbf{C}$, our estimators $\widehat{\mathbf{R}}$ and $\widehat{\mathbf{C}}$ can be expressed using two invertible matrices, $\mathbf{H}_R$ and $\mathbf{H}_C$, given in \eqref{eq:3-1}, \eqref{eq:3-2}. Specifically, $\widehat{\mathbf{R}}$ is a close estimator of $\mathbf{R}\mathbf{H}_R$, $\widehat{\mathbf{C}}$ is a close estimator of $\mathbf{C}\mathbf{H}_C$. Additionally, $\widehat{\mathbf{F}}_t$ is an estimator of $\mathbf{H}_R^{-1}\mathbf{F}_t\mathbf{H}_C^{{-1}\top}$. Knowing $\mathbf{R}\mathbf{H}_R$, $\mathbf{C}\mathbf{H}_C$ and $\mathbf{H}_R^{-1}\mathbf{F}_t\mathbf{H}_C^{{-1}\top}$ is effectively equivalent to knowing the true matrices $\mathbf{R},\mathbf{C}$ and $\mathbf{F}_t$ for many practical applications.
\subsection{Estimation}
 We propose a novel and user-friendly approach to estimate a latent factor model for partially observed matrix-variate data, by applying PCA to two adjusted covariance matrices estimated from partially observed data. Define $\hat{M}_{R,ij} \triangleq \frac{1}{a}\frac{1}{\left|Q_{R,ij}\right|}\mathop{\sum}_{t,m\in Q_{R,ij}}Y_{t,im}Y_{t,jm}$, $\hat{M}_{C,ij} \triangleq \frac{1}{b}\frac{1}{\left|Q_{C,ij}\right|}\mathop{\sum}_{\scriptstyle t,l\in Q_{C,ij}}Y_{t,li}Y_{t,lj}$. The estimators $\widehat{\mathbf{R}}$ and $\widehat{\mathbf{C}}$ are then computed as $\sqrt{a}$ times the top largest $k$ eigenvectors of $\widehat{\mathbf{M}}_R$ and $\sqrt{b}$ times the top largest $r$ eigenvectors of $\widehat{\mathbf{M}}_C$  respectively.

We need to note that the observations $\mathbf{Y}_t$ are demeaned, which means that $\overline{\mathbf{Y}}_t=\frac{1}{T}\sum_{t=1}^T \mathbf{Y}_t=0$. We can impute the missing values with 0 in the observation data $\mathbf{Y}_t$, and denote the imputed data as $\widetilde{\mathbf{Y}}_t$, then, $\widetilde{\mathbf{Y}}_t=\mathbf{Y}_t \odot \mathbf{W}_t$,
\begin{align}\label{eq:2-3}
\widehat{\mathbf{M}}_R=\frac{1}{a}\sum_{t=1}^T \widetilde{\mathbf{Y}}_t \widetilde{\mathbf{Y}}_t^{\top}\odot \left[\frac{1}{\left|Q_{R,ij}\right|}\right],\\
\widehat{\mathbf{M}}_C=\frac{1}{b}\sum_{t=1}^T \widetilde{\mathbf{Y}}_t^{\top} \widetilde{\mathbf{Y}}_t\odot \left[\frac{1}{\left|Q_{C,ij}\right|}\right].
\end{align}
Where $\left[\frac{1}{\left|Q_{R,ij}\right|}\right]$ and $\left[\frac{1}{\left|Q_{C,ij}\right|}\right]$ are $a \times a$ and $b \times b$ dimensional matrices in which the $(i,j)-th$ entries are $\frac{1}{\left|Q_{R,ij}\right|}$ and $\frac{1}{\left|Q_{C,ij}\right|}$, respectively. Then, the estimators $\widehat{\mathbf{R}}$, $\widehat{\mathbf{C}}$, $\widehat{\mathbf{F}}_t$ of $\mathbf{R}$, $\mathbf{C}$, $\mathbf{F}_t$ satisfy the following formulas:
\begin{align}\label{eq:2-4}
\widehat{\mathbf{M}}_R\widehat{\mathbf{R}} =\widehat{\mathbf{R}}V_{R,abT},\\
\widehat{\mathbf{M}}_C\widehat{\mathbf{C}} =\widehat{\mathbf{C}}V_{C,abT}.
\end{align}
Where $V_{R,abT}$, $V_{C,abT}$ are the diagonal matrices of the first $k$ and $r$ largest eigenvalues of $\widehat{\mathbf{M}}_R$ and $\widehat{\mathbf{M}}_C$ in decreasing order, respectively.
After estimating $\widehat{\mathbf{R}}$ and $\widehat{\mathbf{C}}$, we obtain an estimator of $\mathbf{F}_t$ using the following formula:
\begin{equation}\label{eq:2-5}
\widehat{\mathbf{F}}_t=\frac{1}{ab}\widehat{\mathbf{R}}^{\top}\widetilde{\mathbf{Y}}_t\widehat{\mathbf{C}}.
\end{equation}
The signal part $\mathbf{S}_t=\mathbf{R}\mathbf{F}_t \mathbf{C}^{\top}$ can be estimated by
\begin{equation}\label{eq:2-6}
\widehat{\mathbf{S}}_t=\frac{1}{ab}\widehat{\mathbf{R}}\widehat{\mathbf{R}}^{\top}\widetilde{\mathbf{Y}}_t\widehat{\mathbf{C}}\widehat{\mathbf{C}}^{\top}.
\end{equation}
The estimation procedure outlined above assumes that the latent dimensions $k$ and $r$ are known. However, in practice, we also need to estimate these parameters. To determine the values of $k$ and $r$, we use the eigenvalue ratio-based estimator proposed by \cite{https://doi.org/10.3982/ECTA8968}. Let $\hat{\lambda}_1\geq \hat{\lambda}_2 \geq \cdots \geq \hat{\lambda}_k \geq 0$ represent the ordered eigenvalues of $\widehat{\mathbf{M}}_R$. The ratio-based estimator for $k$ is defined as follows:
\begin{equation}\label{eq:2-7}
\hat{k}=argmax_{1\leq j\leq k_{max}}\frac{\hat{\lambda}_j}{\hat{\lambda}_{j+1}},
\end{equation}
where $k_{max}$ is a given upper bound. In practice we may take $k_{max}=\lceil a/2 \rceil$ or $k_{max}=\lceil a/3 \rceil$ according to \cite{https://doi.org/10.3982/ECTA8968}. Ratio estimator $\hat{r}$ is defined similarly with respect to $\widehat{\mathbf{M}}_C$.

In the next section, we present theoretical results showing that in high dimensional settings, $\widehat{\mathbf{R}}, \widehat{\mathbf{C}}, \widehat{\mathbf{F}}_t$ are consistent estimators when $k$ and $r$ are known and fixed. Moreover, we obtain the asymptotic distributions for $\widehat{\mathbf{R}}$ and $\widehat{\mathbf{C}}$.
\section{Theoretical results}\label{sec:3}
We first outline all the necessary assumptions used in the following analysis. Let $\overline{\mathbf{F}}=\frac{1}{T}\sum_{t=1}^T\mathbf{F}_t$ and $\overline{\mathbf{E}}=\frac{1}{T}\sum_{t=1}^T\mathbf{E}_t$ represent the sample means of the factors and the noise, respectively. The entries in the matrices are denoted as $\bar{f}_{ij}$ and $\bar{e}_{ij}$, respectively.
\begin{assumption}\label{As:3-1}
$\alpha$-\textbf{mixing}. The vectorized factor $VEC\left(\mathbf{F}_t\right)$ and noise $VEC\left(\mathbf{E}_t\right)$ are $\alpha$-mixing. Specifically, a vector process $\{\mathbf{x}_t, t=0,\pm 1, \pm 2, \cdots\}$ is $\alpha$-mixing if, for some $\gamma >2$, the mixing coefficients satisfy the condition that
\begin{equation*}
\sum_{h=1}^{\infty}\alpha\left(h\right)^{1-2/\gamma}<\infty,
\end{equation*}
where $\alpha\left(h\right)=\mathop{sup}\limits_{\tau}\mathop{sup}\limits_{A\in \mathcal{F}_{-\infty}^{\tau}, B\in \mathcal{F}^{\infty}_{\tau+h}}|P(A\cap B)-P(A)P(B)|$ and $\mathcal{F}_{\tau}^s$ is $\sigma$-field generated by $\{\mathbf{x}_t:\tau\leq t\leq s\}$.
\end{assumption}
\begin{assumption}\label{As:C1}  \textbf{General Assumption:}\\
(1) $\forall t, \mathbf{W}_t$ is independent of $\mathbf{F}_t, \mathbf{E}_t$.\\
(2) $\forall t, \mathbf{R}, \mathbf{C}, \mathbf{F}_t$ and $\mathbf{E}_t$ are independent.\\
(3) $\frac{\left|Q_{R,ij}\right|}{bT}\geq q_R>0, \frac{\left|Q_{C,ij}\right|}{aT}\geq q_C>0$ for all $i, j$. There exist constants $q_{R,ij}, q_{R,ij,kl}$ for all $i,j,k,l$ such that $q_{R,ij}=lim_{b,T \to \infty}\frac{\left|Q_{R,ij}\right|}{bT}$ and $q_{R,ij,kl}=lim_{b,T \to \infty}\frac{\left|Q_{R,ij}\cap Q_{R,kl}\right|}{bT}$. There exist constants $q_{C,ij}, q_{C,ij,kl}$ for all $i,j,k,l$ such that $q_{C,ij}=lim_{a,T \to \infty}\frac{\left|Q_{c,ij}\right|}{aT}$ and $q_{C,ij,kl}=lim_{a,T \to \infty}\frac{\left|Q_{C,ij}\cap Q_{C,kl}\right|}{aT}$.
\end{assumption}
\begin{assumption}\label{As:1}
\textbf{Factors:} $\forall t,m, \mathbb{E}\left[\Vert \mathbf{F}_t \Vert^4\right]\leq \bar{F}<\infty, \mathbb{E}\left[\Vert \mathbf{F}_t \mathbf{C}_m\Vert^4\right]\leq \bar{F}_C<\infty$. There exists a positive definite $k\times k$ matrix $\Sigma_{F}$, such that $\frac{1}{T}\sum_{t=1}^T\mathbf{F}_t\mathbf{F}_t^{\top} \xrightarrow{p}\Sigma_{F}$, there exists a positive definite $r\times r$ matrix $\Sigma_{F^{\top}}$, such that $\frac{1}{T}\sum_{t=1}^T\mathbf{F}_t^{\top}\mathbf{F}_t \xrightarrow{p}\Sigma_{F^{\top}}$.
There exists a positive definite $k\times k$ matrix $\Sigma_{FC}$, such that $\frac{1}{bT}\sum_{t=1}^T\mathbf{F}_t\mathbf{C}^{\top} \mathbf{C}\mathbf{F}_t^{\top} \xrightarrow{p}\Sigma_{FC}$, and $\mathbb{E}\Vert \sqrt{bT}\left(\frac{1}{bT}\sum_{t=1}^T \mathbf{F}_t\mathbf{C}^{\top} \mathbf{C}\mathbf{F}_t^{\top}-\Sigma_{FC}\right) \Vert^2\leq M$.\\
There exists a positive definite $r\times r$ matrix $\Sigma_{FR}$, such that $\frac{1}{aT}\sum_{t=1}^T\mathbf{F}_t^{\top}\mathbf{R}^{\top} \mathbf{R}\mathbf{F}_t \xrightarrow{p}\Sigma_{FR}$, and $\mathbb{E}\Vert \sqrt{aT}\left(\frac{1}{aT}\sum_{t=1}^T \mathbf{F}_t^{\top}\mathbf{R}^{\top} \mathbf{R}\mathbf{F}_t-\Sigma_{FR}\right) \Vert^2\leq M$.\\
Furthermore, for any $Q_{R,ij}$, $\frac{1}{|Q_{R,ij}|}\sum_{t,m\in Q_{R,ij}} \mathbf{F}_t\mathbf{C}_m \mathbf{C}_m^{\top}\mathbf{F}_t^{\top}\xrightarrow{p}\Sigma_{FC} $, and \\
$\mathbb{E}\Vert \sqrt{|Q_{R,ij}|}\left(\frac{1}{|Q_{R,ij}|}\sum_{t,m\in Q_{R,ij}} \mathbf{F}_t\mathbf{C}_m \mathbf{C}_m^{\top}\mathbf{F}_t^{\top}-\Sigma_{FC}\right) \Vert^2\leq M$.
\end{assumption}
\begin{assumption}\label{As:2}
\textbf{Loading matrix:} $\mathbb{E}\left[\Vert \mathbf{R}_i \Vert^4 \right]\leq \bar{R}<\infty$.
For each row of $\mathbf{R}, \Vert \mathbf{R}_{i}\Vert=O(1)$, and as $a,b \rightarrow \infty$, we have $a^{-1}\mathbf{R}^{\top}\mathbf{R} \xrightarrow{p}\Sigma_{R}$ for a $k \times k$ positive definite matrix $\Sigma_{R}$. For each row of $\mathbf{C}, \Vert \mathbf{C}_{i}\Vert=O(1)$, and as $a,b \rightarrow \infty$, we have $b^{-1}\mathbf{C}^{\top}\mathbf{C}\xrightarrow{p}\Sigma_{C}$ for a $r \times r$ positive definite matrix $\Sigma_{C}$.
\end{assumption}
\begin{assumption}\label{As:3} \textbf{Error:}  $\mathbb{E}\left[\Vert \mathbf{E}_t \Vert^2 \right]\leq \bar{E}<\infty$, and there exists a positive constant $M < \infty$, such that for all $a,b$:\\
(1) $\mathbb{E}\left[e_{t,ij}\right]=0, \mathbb{E}|e_{t,ij}|^8\leq M$ .\\
(2) $\mathbb{E}\left[e_{t,ij}e_{s,ij}\right]=\gamma_{ts,ij}$, with $|\gamma_{ts,ij}|\leq \gamma_{ts}$ for some $\gamma_{ts}$ and all $i$ and $j$. For all $t$, $\sum_{s=1}^T \gamma_{ts} \leq M$.\\
(3) $\mathbb{E}\left[e_{t,im}e_{t,jm}\right]=\tau_{t,ijm}, |\tau_{t,ijm}|\leq \tau_{ij}$ for some $\tau_{ij}$ and all $t,m$. For all $j, \sum_{i=1}^a\tau_{ij}\leq M$.\\
(4) For all $i$ and $j$, $\mathbb{E}|\frac{1}{|Q_{R,ij}|^{1/2}}\sum_{t,m\in Q_{R,ij}}\left(e_{t,im}e_{t,jm}-\mathbb{E}\left[e_{t,im}e_{t,jm}\right]\right)|^4\leq M$.
\end{assumption}
\begin{assumption}\label{As:4}
\textbf{Weak Dependence between Factor and Idiosyncratic Errors:} For every $(i,j)$,
\begin{equation*}
\mathbb{E}\Vert \frac{1}{\sqrt{|Q_{R,ij}|}}\sum_{t,m\in Q_{R,ij}} \mathbf{F}_t\mathbf{C}_m e_{t,jm} \Vert^4 \leq M,\\
\mathbb{E}\Vert \frac{1}{\sqrt{|Q_{R,ij}|}}\sum_{t,m\in Q_{R,ij}} \mathbf{F}_t\mathbf{C}_m e_{t,im} \Vert^4 \leq M.
\end{equation*}
\end{assumption}
\begin{assumption}\label{As:6}
 \textbf{Moments and Central Limit Theorems:} For all $a$ and $T$:\\
(1)\begin{equation*}
\mathbb{E}\left[\Vert \sqrt{\frac{bT}{a}}\sum_{i=1}^a \frac{1}{|Q_{R,ij}|}\sum_{t,m \in Q_{R,ij}}\mathbf{R}_i \left(e_{t,im}e_{t,jm}-\mathbb{E}\left(e_{t,im}e_{t,jm}\right)\right\Vert^2\right] \leq M.
\end{equation*}
(2) \begin{equation*}
\mathbb{E}\left[\Vert \sqrt{\frac{bT}{a}}\sum_{i=1}^a \frac{1}{|Q_{R,ij}|}\sum_{t,m \in Q_{R,ij}}\mathbf{R}_i \mathbf{C}_m^{\top}\mathbf{F}_t^{\top}e_{t,im}\Vert^2\right] \leq M.
\end{equation*}
(3) \begin{equation*}
\sqrt{\frac{bT}{a}}\sum_{i=1}^a \mathbf{R}_i \mathbf{R}_i^{\top} \frac{1}{|Q_{R,ij}|}\sum_{t,m \in Q_{R,ij}}\mathbf{F}_t \mathbf{C}_m e_{t,jm} \xrightarrow {d} N\left(\mathbf{0},\mathbf{\Gamma}_{R_j}^{obs}\right).
\end{equation*}
\begin{equation*}
\sqrt{\frac{aT}{b}}\sum_{i=1}^b \mathbf{C}_i \mathbf{C}_i^{\top} \frac{1}{|Q_{C,ij}|}\sum_{t,m \in Q_{C,ij}}\mathbf{F}_t \mathbf{R}_m e_{t,jm} \xrightarrow {d} N\left(\mathbf{0},\mathbf{\Gamma}_{C_j}^{obs}\right).
\end{equation*}
(4) For every $i$, they hold:\\
\small{$\frac{\sqrt{bT}}{a}\sum_{i=1}^a\mathbf{R}_i\mathbf{R}_i^{\top}\left(\frac{1}{|Q_{R,ij}|}\sum_{t,m\in Q_{R,ij}}\mathbf{F}_t \mathbf{C}_m\mathbf{C}_m^{\top}\mathbf{F}_t^{\top}-\frac{1}{bT}\sum_{t=1}^T\sum_{m=1}^b\mathbf{F}_t \mathbf{C}_m\mathbf{C}_m^{\top}\mathbf{F}_t^{\top}\right)\mathbf{R}_j\\
\xrightarrow{\mathcal{D}}N\left(\mathbf{0},\mathbf{\Gamma}_{R_j}^{miss}\right)$;\\
$\frac{\sqrt{aT}}{b}\sum_{i=1}^b\mathbf{C}_i\mathbf{C}_i^{\top}\left(\frac{1}{|Q_{C,ij}|}\sum_{t,l\in Q_{C,ij}}\mathbf{F}_t^{\top} \mathbf{R}_l\mathbf{R}_l^{\top}\mathbf{F}_t-\frac{1}{aT}\sum_{t=1}^T\sum_{l=1}^a\mathbf{F}_t^{\top} \mathbf{R}_l\mathbf{R}_l^{\top}\mathbf{F}_t\right)\mathbf{C}_j\\
\xrightarrow{\mathcal{D}}N\left(\mathbf{0},\mathbf{\Gamma}_{C_j}^{miss}\right).$
}
\end{assumption}
\begin{theorem}\label{Th1}
Under Assumptions \ref{As:3-1}$\sim$\ref{As:4}, we have, as $k,r$ fixed and $a,b,T \rightarrow \infty$, let $\delta_{a,bT}=min\left(a,bT\right),\delta_{b,aT}=min\left(b,aT\right)$, then
\begin{align*}
&\frac{1}{a}\sum_{j=1}^a \left\Vert \widehat{\mathbf{R}}_j-\mathbf{H}_R^{\top}\mathbf{R}_j \right\Vert^2 =\mathcal{O}_p\left(\frac{1}{\delta_{a,bT}}\right),\\
&\frac{1}{b}\sum_{j=1}^b \left\Vert \widehat{\mathbf{C}}_j-\mathbf{H}_C^{\top}\mathbf{C}_j \right\Vert^2 =\mathcal{O}_p\left(\frac{1}{\delta_{b,aT}}\right),\\
&\frac{1}{a}\left\Vert \widehat{\mathbf{R}} - \mathbf{R} \mathbf{H}_R\right\Vert_{F}^2=\mathcal{O}_p\left(\frac{1}{\delta_{a,bT}}\right),\\
&\frac{1}{b}\left\Vert \widehat{\mathbf{C}} - \mathbf{C} \mathbf{H}_C\right\Vert_{F}^2=\mathcal{O}_p\left(\frac{1}{\delta_{b,aT}}\right),\\
&\frac{1}{a}\left\Vert \widehat{\mathbf{R}} - \mathbf{R} \mathbf{H}_R\right\Vert^2=\mathcal{O}_p\left(\frac{1}{\delta_{a,bT}}\right),\\
&\frac{1}{b}\left\Vert \widehat{\mathbf{C}} - \mathbf{C} \mathbf{H}_C\right\Vert^2=\mathcal{O}_p\left(\frac{1}{\delta_{b,aT}}\right).
\end{align*}
\end{theorem}
where
\begin{equation}\label{eq:3-1}
\mathbf{H}_R=\frac{1}{abT}\sum_{t=1}^T \mathbf{F}_t \mathbf{C}^{\top} \mathbf{C}\mathbf{F}_t^{\top}\mathbf{R}^{\top}\widehat{\mathbf{R}}V_{R,abT}^{-1}\in \mathbb{R}^{k \times k},
\end{equation}
\begin{equation}\label{eq:3-2}
\mathbf{H}_C=\frac{1}{abT}\sum_{t=1}^T \mathbf{F}_t^{\top} \mathbf{R}^{\top} \mathbf{R}\mathbf{F}_t\mathbf{C}^{\top}\widehat{\mathbf{C}}V_{C,abT}^{-1} \in \mathbb{R}^{r \times r}.
\end{equation}
The estimators $\widehat{\mathbf{R}}$ and $\widehat{\mathbf{C}}$ satisfy:
\begin{equation}\label{eq:3-3}
\widehat{\mathbf{R}}=\frac{1}{a} \sum_{t=1}^T \widetilde{\mathbf{Y}}_t \widetilde{\mathbf{Y}}_t^{\top} \odot \left[\frac{1}{\left|Q_{R,ij}\right|}\right] \widehat{\mathbf{R}} V_{R,abT}^{-1},
\end{equation}
\begin{equation}\label{eq:3-4}
\widehat{\mathbf{C}}=\frac{1}{b} \sum_{t=1}^T \widetilde{\mathbf{Y}}_t^{\top} \widetilde{\mathbf{Y}}_t \odot \left[\frac{1}{\left|Q_{C,ij}\right|}\right] \widehat{\mathbf{C}} V_{C,abT}^{-1}.
\end{equation}
\begin{theorem}\label{Th3}
Under Assumptions \ref{As:C1}$\sim$\ref{As:6}, as $k,r$ fixed and $a,b,T\rightarrow \infty$, we have
\begin{equation*}
\widehat{\mathbf{F}}_t-\mathbf{H}_R^{-1}\mathbf{F}_t\mathbf{H}_C^{-1{\top}}=\mathcal{O}_p\left(\frac{1}{min\left(\sqrt{a},\sqrt{b}\right)}\right).
\end{equation*}
\end{theorem}
\begin{theorem}\label{Th4}
For row loading matrix $\mathbf{R}$, if $\frac{\sqrt{bT}}{a}\rightarrow 0$, then\\
$\sqrt{bT}\left(\widehat{\mathbf{R}}_j-\mathbf{H}_R^{\top}\mathbf{R}_j\right)\xrightarrow{\mathcal{D}}N\left(\mathbf{0},D^{-1}Q^{-1}\left[\mathbf{\Gamma}_{R_j}^{obs}+\mathbf{\Gamma}_{R_j}^{miss}\right]\left(Q^{-1}\right)^{\top}D^{-1}\right)$;\\
For column loading matrix $\mathbf{C}$, if $\frac{\sqrt{aT}}{b}\rightarrow 0$, then\\
$\sqrt{aT}\left(\widehat{\mathbf{C}}_j-\mathbf{H}_C^{\top}\mathbf{C}_j\right)\xrightarrow{\mathcal{D}}N\left(\mathbf{0},E^{-1}P^{-1}\left[\mathbf{\Gamma}_{C_j}^{obs}+\mathbf{\Gamma}_{C_j}^{miss}\right]\left(P^{-1}\right)^{\top}E^{-1}\right)$.
Where the definitions of $D,E,Q,P$ are given in the Lemmas \ref{Le:4-1} and \ref{Le:4-2}.
\end{theorem}

 \section{Simulation}\label{s:sim}
In this section, we use Monte Carlo simulations to evaluate the accuracy of estimating the dimensions of latent factors. We also assess the validity of the asymptotic results in approximating the finite sample distributions of $\hat{\mathbf{R}_i}$ and $\hat{\mathbf{C}_j}$, as well as the convergence rate of $\mathbf{F}_t$.
\subsection{Settings}
Throughout the matrix $\mathbf{Y}_t$'s are generated according to model \eqref{eq:2-1}. We study two observation patterns which are illustrated in section \ref{sec:2}.

I. \textit{Missing at random.} Entries are observed independently with probability 0.75.

II. \textit{Simultaneous missing.} Once a unit can not be observed, it stays missing afterward. $25\%$ randomly selected units can not be observed from time $0.75\times T$ and the remaining $75\%$ units can be observed until the end.

The dimension of the latent factor matrix $\mathbf{F}_t$ is fixed at $k\times r=3\times 3$. The values of $a,b$, and $T$ can vary in different settings. The true loading matrices $\mathbf{R}$ and $\mathbf{C}$ are independently sampled from the normal distribution $N\left(1,1\right)$. Additionally, the latent factor and noise matrices are allowed to be dependent across rows, columns, or time, respectively.

We simulate data under three Settings as follows:
\begin{enumerate}[1.]
\item \textit{Uncorrelated.} The entries of both $\mathbf{F}_t$ and $\mathbf{E}_t$ are uncorrelated across time, rows and columns. Specifically, we simulate temporally independent matrices $\mathbf{F}_t\sim \mathcal{MN}_{3\times 3}\left(\mathbf{0},\mathbf{I},\mathbf{I}\right)$ and $\mathbf{E}_t\sim \mathcal{MN}_{a\times b}\left(\mathbf{0},\mathbf{I},\mathbf{I}\right)$.
\item \textit{Weakly correlated cross time.} The entries of $\mathbf{F}_t$ and $\mathbf{E}_t$ are uncorrelated across rows and columns, but weakly correlated temporally. Specifically, we simulate $VEC\left(\mathbf{F}_t\right)$ from a vector autoregressive model of order one (VAR(1) model):
\begin{equation*}
VEC\left(\mathbf{F}_t\right)=\Psi\cdot VEC\left(\mathbf{F}_{t-1}\right)+\epsilon_t,
\end{equation*}
where the VAR coefficient matrix $\Psi=\psi\cdot \mathbf{I}_9$ and $var[\epsilon_t]=1-\psi^2\cdot \mathbf{I}_9$. We choose $\psi=0.1$ and $\psi=0.5$ to examine how temporal dependence may affect our results. The noise matrix $\mathbf{E}_t$ is also simulated from a VAR(1) model:
\begin{equation*}
VEC\left(\mathbf{E}_t\right)=\Phi\cdot VEC\left(\mathbf{E}_{t-1}\right)+\mathbf{u}_t,
\end{equation*}
where $\Phi=0.1\cdot \mathbf{I}_{a \times b}$ and $var[\mathbf{u}_t]=0.99\cdot \mathbf{I}_{a\times b}$.
\item \textit{Weakly correlated across rows or columns.} The entries of $\mathbf{F}_t$ and $\mathbf{E}_t$ are temporally uncorrelated, but $\mathbf{E}_t$ is weakly correlated across rows and columns. Specifically, we simulate the temporally independent matrix $\mathbf{F}_t\sim \mathcal{MN}_{3\times 3}\left(\mathbf{0},\mathbf{I}, \mathbf{I}\right)$ and $\mathbf{E}_t\sim \mathcal{MN}_{a\times b}\left(\mathbf{0},\mathbf{U}_E, \mathbf{V}_E\right)$, where $\mathbf{U}_E$ and $\mathbf{V}_E$ both have ones on the diagonal, while have $1/a$ and $1/b$ off-diagonal, respectively.
\end{enumerate}
We present the following results under various Settings and missing patterns. We refer to our method (denoted as M) and the one proposed by \cite{Chen03042023} (denoted as F), where all missing observations are imputed with zero. Additionally, we will discuss the direct method (denoted as Z), with which the covariance matrices would be estimated by $\widehat{\mathbf{M}}_R=\frac{1}{abT}\sum_{t=1}^T \widetilde{\mathbf{Y}}_t \widetilde{\mathbf{Y}}_t^{\top}$, $\widehat{\mathbf{M}}_C=\frac{1}{abT}\sum_{t=1}^T \widetilde{\mathbf{Y}}_t^{\top} \widetilde{\mathbf{Y}}_t$.
\begin{enumerate}[1.]
\item \textit{Estimating latent dimensions.} The latent dimensions are estimated using the eigen-ratio method as described in Equation \eqref{eq:2-7}. The
results are presented in tables showing the frequencies of $\left(\hat{k},\hat{r}\right)$.
\item \textit{Convergence of $\widehat{\mathbf{R}}, \widehat{\mathbf{C}}$.} We report values in tables and boxplots of space distances $\mathcal{D}\left(\widehat{\mathbf{R}},\mathbf{R}\right)$, $\mathcal{D}\left(\widehat{\mathbf{C}},\mathbf{C}\right)$, which is defined as:
\begin{equation}
\mathcal{D}\left(\widehat{\mathbf{A}},\mathbf{A}\right)=\left\Vert \widehat{\mathbf{A}}\left( \widehat{\mathbf{A}}^{\top} \widehat{\mathbf{A}}\right)^{-1}\widehat{\mathbf{A}}^{\top}- \mathbf{A}\left(\mathbf{A}^{\top}\mathbf{A}\right)^{-1}\mathbf{A}^{\top}\right\Vert.
\end{equation}
\item \textit{Convergence of $\widehat{\mathbf{F}}_t$.} To demonstrate that $\widehat{\mathbf{F}}_t$ estimates a transformation of $\mathbf{F}_t$ for $t \in [T]$, we calculate $\mathbf{H}_R$ and $\mathbf{H}_C$ according to Equations \eqref{eq:3-1} and \eqref{eq:3-2}, respectively. We report values in tables and boxplots showing $\left\Vert \widehat{\mathbf{F}}_t-\mathbf{H}_R^{-1}\mathbf{F}_t\mathbf{H}_C^{-1\top}\right\Vert$.
\item \textit{Asymptotic normality of $\widehat{\mathbf{R}}$.} We present $3$-dimensional histograms and QQ plots of the first row of $\widehat{\mathbf{R}}-\mathbf{H}_R^{\top}\mathbf{R}$, as the true value of $k$ is 3.
\end{enumerate}
\subsection{Comparison of Convergence}
In this section, we examine the finite sample convergence of $\widehat{\mathbf{R}}_i, \widehat{\mathbf{C}}_j$ and $\mathbf{F}_t$. We choose the values for $\left(a,b\right)$ from the options $\left(50,50\right), \left(100,100\right), \left(200,200\right)$ and, while $T$ takes on the values $50, 100$, and $200$. The results in this section are based on 500 repetitions, which are deemed sufficient, as indicated by the reported standard deviations.
\subsubsection{Accuracy of Estimating Unknown Dimensions}
We present the frequencies of estimated $\left(\hat{k},\hat{r}\right)$ pairs for 3 different simulated data Settings and 2 different missing patterns. Specifically, for Setting 2, there are two scenarios with $\psi=0.1$ and $\psi=0.5$, respectively, resulting in a total of $4 \times 2=8$ distinct situations.

Table \ref{tab0} shows the frequencies of estimated $\left(\hat{k},\hat{r}\right)$ pairs for 3 different Settings under missing pattern I. Our method demonstrates the highest frequencies of correct estimations, they are all 1. Additionally, the  accuracy of other methods increases as $a, b,$ and $T$ increase. Here, we need to note that in all the tables, the first row represents the frequencies obtained from our method, while the second row shows the frequencies from the direct method. The third, fourth, and fifth rows present the frequencies from \cite{Chen03042023} for $\alpha=-1,\alpha=0,\alpha=1$ for each pair,  respectively.

Table \ref{tab1} in Appendix \ref{sec:A2} presents the frequencies of estimated $\left(\hat{k},\hat{r}\right)$ pairs for 3 different Settings under missing pattern II. Here, our method also shows the highest frequencies of correct estimations, with accuracy increasing as $a, b,$ and $T$ increase. In contrast, the other methods tend to estimate the dimensions as $6 \times 3$. This preference may arise from the fact that, under missing pattern II, for some units, all features are missing after $0.75\times T$, meaning that $25\%$ of the rows in $\mathbf{Y}_t$ are missing. Consequently, the row loading matrix $\mathbf{R}$ can not be estimated precisely, including $k$. However, the remaining $75\%$ of rows in $\mathbf{Y}_t$ can be fully observed, so none of the columns in $\mathbf{Y}_t$ are completely unobservable,  allowing for a more accurate estimation of the column matrix  $\mathbf{C}$ and the $r$.

\subsubsection{Errors of Loading Matrices and Factor Matrices Estimation}
Table \ref{tab2} in the main text and Table \ref{tab3} in Appendix \ref{sec:A2} display the numeric values of the space distances between estimators and the true values under missing patterns I and II, respectively. Specifically, we examine $\mathcal{D}\left(\widehat{\mathbf{R}},\mathbf{R}\right), \mathcal{D}\left(\widehat{\mathbf{C}},\mathbf{C}\right)$, and the $l_2$ norm of the discrepancy between the estimated $\widehat{\mathbf{F}}_t$ and the transformed true $\mathbf{F}_t$, given by the expression $\left\Vert \widehat{\mathbf{F}}_t-\mathbf{H}_R^{-1}\mathbf{F}_t\mathbf{H}_C^{-1\top}\right\Vert$. We can see that our method consistently yields the smallest errors, which decrease as $a, b,$ and $T$ increase across various settings under missing pattern I. Under missing pattern II, the errors of row loading matrix $\widehat{\mathbf{R}}$ are the smallest when using our method. For the errors of column loading matrix $\widehat{\mathbf{C}}$, the differences are negligible and the errors are small, similar to the accurate estimation of $\hat{r}$. Regarding the $l_2$ norm of the factor matrices, our method shows the smallest errors in most cases for Setting 1 and Setting 3. In other cases, the minimum values are very close to those obtained with our method. For Setting 2, when $T=50$, the method proposed by \cite{Chen03042023} performs best; however, as $T$ increases, our method exhibits the smallest errors.

Figures \ref{fig:4-1}, \ref{fig:4-2} in the main text, and Figures \ref{fig:4-3}, \ref{fig:4-4} in Appendix \ref{sec:A2} display the boxplots of estimation errors for loading matrices and factor matrices under missing pattern I and II, respectively. These figures correspond to the data presented in Table \ref{tab2} in the main text and Table \ref{tab3} in Appendix \ref{sec:A2}. In all the figures, "F" represents the method proposed by \cite{Chen03042023}, where $\alpha$ is set to -1. For the space distances $\mathcal{D}\left(\widehat{\mathbf{R}},\mathbf{R}\right), \mathcal{D}\left(\widehat{\mathbf{C}},\mathbf{C}\right)$, there is a tendency of higher convergence at higher $T$. Similarly, the convergence of $\widehat{\mathbf{F}}_t$ also enhances with larger values of $T$ and shows slight improvement as $a,b$ are increased.

\subsubsection{Asymptotic Normality}
 In this section, we consider the asymptotic normality of loading matrices. We only report the results for $\hat{\mathbf{R}_i}$ since $\hat{\mathbf{C}_j}$ exhibits similar properties. Specifically, we consider the first row of $\widehat{\mathbf{R}}-\mathbf{H}_R^{\top}\mathbf{R}$. According to Theorem \ref{Th4}, the asymptotic normality requires that either $\sqrt{bT}/a\rightarrow 0$ or $\sqrt{aT}/b\rightarrow 0$. For our analysis, we set the parameters $\left(a,b,T\right)$ to $\left(200,200,100\right)$.

 The results for asymptotic normality are based on 500 repetitions. Given that $k=3$, we report $3$-dimensional QQ plots and histograms of the first row of $\widehat{\mathbf{R}}-\mathbf{H}_R^{\top}\mathbf{R}$.
 All the results for missing pattern I are shown in the main text as the figures \ref{fig:4-5} $\sim$ \ref{fig:4-12}, while the results for missing pattern II are included in appendix \ref{sec:A2} as the figures \ref{fig:6-1} $\sim$ \ref{fig:6-8}.  And the frequencies highlighted in pink indicate the best results in all the talbes.
As $k=3$, each plot displays three entries per row. The first row presents the results from our method, the second row shows the results from the method "Z", and the third, fourth, and fifth rows illustrate results from the method proposed by \cite{Chen03042023} with $\alpha=-1, \alpha=0, \alpha=1$, respectively. In all cases, the QQ plots and histograms demonstrate that our method exhibits the best asymptotic normality, as expected from the theorem.
\begin{table}
\caption{\label{tab0}Table of frequencies of estimated $\left(\hat{k},\hat{r}\right)$ pairs under missing pattern I. The truth is $\left(3,3\right)$.}
\resizebox{0.7\linewidth}{!}{
\begin{tabular}{lccccccccc}
    \hline
    \multicolumn{10}{c}{Setting=1, Missing pattern=I}\\
      &  \multicolumn{3}{c}{a=50,b=50} & \multicolumn{3}{c}{a=100,b=100}& \multicolumn{3}{c}{a=200,b=200}\\
     \cline{2-4} \cline{5-7} \cline{8-10}
 $\left(\hat{k},\hat{r}\right)$ & T=50 & T=100 & T=200 & T=50 & T=100 & T=200 & T=50 & T=100 & T=200 \\ \hline
(3,3) &\sethlcolor{pink}\hl{1.0}&\sethlcolor{pink}\hl{1.0}&\sethlcolor{pink}\hl{1.0}&\sethlcolor{pink}\hl{1.0}&\sethlcolor{pink}\hl{1.0}&\sethlcolor{pink}\hl{1.0}&\sethlcolor{pink}\hl{1.0}&\sethlcolor{pink}\hl{1.0}&\sethlcolor{pink}\hl{1.0} \\
      &0.986&1.0&0.58&1.0&1.0&1.0&1.0&1.0&1.0 \\

 &0.76&0.91&1.0&1.0&1.0&1.0&1.0&1.0&1.0 \\

 &0.3&0.898&0.394&1.0&1.0&1.0&1.0&1.0&1.0 \\

  &0.866&1.0&1.0&1.0&1.0&1.0&1.0&1.0&1.0 \\  \hline

others
&0.0&0.0&0.0&0.0&0.0&0.0&0.0&0.0&0.0 \\

&0.014&0.0&0.42&0.0&0.0&0.0&0.0&0.0&0.0 \\

 &0.24&0.09&0.0&0.0&0.0&0.0&0.0&0.0&0.0 \\

  &0.7&0.102&0.606&0.0&0.0&0.0&0.0&0.0&0.0 \\

  &0.134&0.0&0.0&0.0&0.0&0.0&0.0&0.0&0.0 \\
  \hline
  \multicolumn{10}{c}{Setting=2 with $\psi=0.1$, Missing pattern=I}\\
      &  \multicolumn{3}{c}{a=50,b=50} & \multicolumn{3}{c}{a=100,b=100}& \multicolumn{3}{c}{a=200,b=200}\\
     \cline{2-4} \cline{5-7} \cline{8-10}
 $\left(\hat{k},\hat{r}\right)$   & T=50 & T=100 & T=200 & T=50 & T=100 & T=200 & T=50 & T=100 & T=200 \\ \hline
  (3,3)  &\sethlcolor{pink}\hl{1.0}&\sethlcolor{pink}\hl{1.0}&\sethlcolor{pink}\hl{1.0}&\sethlcolor{pink}\hl{1.0}&\sethlcolor{pink}\hl{1.0}&\sethlcolor{pink}\hl{1.0}&\sethlcolor{pink}\hl{1.0}&\sethlcolor{pink}\hl{1.0}&\sethlcolor{pink}\hl{1.0}\\
          &0.962&0.196&1.0&1.0&1.0&1.0&1.0&1.0&1.0\\

          &0.22&0.632&0.434&0.564&0.858&0.908&0.932&0.98&1.0\\

          &0.384&0.544&0.932&0.698&0.936&0.988&0.808&0.998&0.998\\

          &0.412&0.536&0.946&0.76&0.82&0.978&0.904 &0.968&1.0\\ \hline
  other   &0.0&0.0&0.0&0.0&0.0&0.0&0.0&0.0&0.0 \\

           &0.038&0.804&0.0&0.0&0.0&0.0&0.0&0.0&0.0 \\

           &0.78&0.368&0.6&0.436&0.142&0.092&0.068&0.02&0.0\\

           &0.616&0.456&0.068&0.302&0.064&0.012&0.192&0.002&0.002 \\

           &0.588&0.464&0.054&0.24&0.18&0.022&0.096&0.032&0.0 \\  \hline
\multicolumn{10}{c}{Setting=2 with $\psi=0.5$, Missing pattern=I}\\
      &  \multicolumn{3}{c}{a=50,b=50} & \multicolumn{3}{c}{a=100,b=100}& \multicolumn{3}{c}{a=200,b=200}\\
     \cline{2-4} \cline{5-7} \cline{8-10}
$\left(\hat{k},\hat{r}\right)$       & T=50 & T=100 & T=200 & T=50 & T=100 & T=200 & T=50 & T=100 & T=200 \\\hline

    (3,3) &\sethlcolor{pink}\hl{1.0}&\sethlcolor{pink}\hl{1.0}&\sethlcolor{pink}\hl{1.0}&\sethlcolor{pink}\hl{1.0}&\sethlcolor{pink}\hl{1.0}&\sethlcolor{pink}\hl{1.0}&\sethlcolor{pink}\hl{1.0}&\sethlcolor{pink}\hl{1.0}&\sethlcolor{pink}\hl{1.0} \\
        &0.85&1.0&1.0&0.996&0.99&1.0&1.0&1.0&1.0 \\

         &0.9&0.994&0.958&0.996&0.996&1.0&1.0&1.0&1.0 \\

          &0.718&1.0&0.644&0.974&0.998&1.0&1.0&1.0&1.0 \\

         &0.616&0.964&0.838&1.0&0.998&1.0&1.0&1.0&1.0 \\    \hline

other    &0.0&0.0&0.0&0.0&0.0&0.0&0.0&0.0&0.0 \\

         &0.15&0.0&0.0&0.004&0.01&0.0&0.0&0.0&0.0 \\

         &0.1&0.006&0.042&0.004&0.004&0.0&0.0&0.0&0.0 \\

         &0.282&0.0&0.356&0.026&0.002&0.0&0.0&0.0&0.0 \\

         &0.384&0.036&0.162&0.0&0.002&0.0&0.0&0.0&0.0 \\  \hline
    \multicolumn{10}{c}{Setting=3, Missing pattern=I}\\
    &  \multicolumn{3}{c}{a=50,b=50} & \multicolumn{3}{c}{a=100,b=100}& \multicolumn{3}{c}{a=200,b=200}\\
     \cline{2-4} \cline{5-7} \cline{8-10}
 $\left(\hat{k},\hat{r}\right)$     & T=50 & T=100 & T=200 & T=50 & T=100 & T=200 & T=50 & T=100 & T=200 \\ \hline
  (3,3) &\sethlcolor{pink}\hl{1.0}&\sethlcolor{pink}\hl{1.0}&\sethlcolor{pink}\hl{1.0}&\sethlcolor{pink}\hl{1.0}&\sethlcolor{pink}\hl{1.0}&\sethlcolor{pink}\hl{1.0}&\sethlcolor{pink}\hl{1.0}&\sethlcolor{pink}\hl{1.0}&\sethlcolor{pink}\hl{1.0} \\
        &0.964&0.952&1.0&0.998&1.0&1.0&1.0&1.0&1.0 \\

          &0.976&0.882&1.0&1.0&1.0&1.0&1.0&1.0&1.0 \\

          &0.998&1.0&0.992&0.996&1.0&1.0&1.0&1.0&1.0 \\

         &0.758&0.938&0.998&0.998&1.0&1.0&1.0&1.0&1.0 \\   \hline

other    &0.0&0.0&0.0&0.0&0.0&0.0&0.0&0.0&0.0 \\

         &0.036&0.048&0.0&0.002&0.0&0.0&0.0&0.0&0.0 \\

         &0.024&0.118&0.0&0.0&0.0&0.0&0.0&0.0&0.0 \\

         &0.002&0.0&0.008&0.004&0.0&0.0&0.0&0.0&0.0 \\

         &0.242&0.062&0.002&0.002&0.0&0.0&0.0&0.0&0.0 \\     \hline  \hline
  \end{tabular}
  }
\end{table}
\begin{table}
\caption{\label{tab2}Means and standard deviations (in parentheses) of $\mathcal{D}\left(\widehat{\mathbf{R}},\mathbf{R}\right), \mathcal{D}\left(\widehat{\mathbf{C}},\mathbf{C}\right),\left\Vert \widehat{\mathbf{F}}_t-\mathbf{H}_R^{-1}\mathbf{F}_t\mathbf{H}_C^{-1\top}\right\Vert$ with missing pattern I.}
\resizebox{.90\linewidth}{!}{
\begin{tabular}{lccccccccc}
    \hline
    \multicolumn{10}{c}{Setting=1, Missing pattern=I}\\
      &  \multicolumn{3}{c}{T=50} & \multicolumn{3}{c}{T=100}& \multicolumn{3}{c}{T=200}\\
     \cline{2-4} \cline{5-7} \cline{8-10}
 $\left(a,b\right)$ & $\mathcal{D}\left(\widehat{\mathbf{R}},\mathbf{R}\right)$ & $\mathcal{D}\left(\widehat{\mathbf{C}},\mathbf{C}\right)$ & $\left\Vert \widehat{\mathbf{F}}_t-\mathbf{H}_R^{-1}\mathbf{F}_t\mathbf{H}_C^{-1\top}\right\Vert$& $\mathcal{D}\left(\widehat{\mathbf{R}},\mathbf{R}\right)$ & $\mathcal{D}\left(\widehat{\mathbf{C}},\mathbf{C}\right)$ & $\left\Vert \widehat{\mathbf{F}}_t-\mathbf{H}_R^{-1}\mathbf{F}_t\mathbf{H}_C^{-1\top}\right\Vert$& $\mathcal{D}\left(\widehat{\mathbf{R}},\mathbf{R}\right)$ & $\mathcal{D}\left(\widehat{\mathbf{C}},\mathbf{C}\right)$ & $\left\Vert \widehat{\mathbf{F}}_t-\mathbf{H}_R^{-1}\mathbf{F}_t\mathbf{H}_C^{-1\top}\right\Vert$ \\ \hline
(50,50) &\sethlcolor{pink}\hl{0.044(0.0053)}&\sethlcolor{pink}\hl{0.039(0.005)}&\sethlcolor{pink}\hl{0.1(0.023)}&\sethlcolor{pink}\hl{0.031(0.0037)}&\sethlcolor{pink}\hl{0.032(0.0037)}&\sethlcolor{pink}\hl{0.11(0.028)}&\sethlcolor{pink}\hl{0.024(0.0027)}&\sethlcolor{pink}\hl{0.023(0.0026)}&\sethlcolor{pink}\hl{0.081(0.027)} \\
&0.053(0.0075)&0.055(0.0074)&0.32(0.14)&0.045(0.0059)&0.04(0.0049)&0.26(0.1)&0.063(0.0071)&0.033(0.0035)&0.16(0.067) \\
 &0.072(0.012)&0.056(0.0086)&0.31(0.13)&0.05(0.0066)&0.047(0.0062)&0.24(0.098)&0.038(0.0043)&0.032(0.0031)&0.14(0.053) \\

 &0.061(0.011)&0.08(0.015)&0.33(0.14)&0.049(0.0065)&0.039(0.005)&0.24(0.09)&0.057(0.0074)&0.042(0.0045)&0.16(0.067) \\
 &0.056(0.0072)&0.083(0.015)&0.32(0.13)&0.041(0.005)&0.053(0.0067)&0.22(0.077)&0.037(0.0041)&0.05(0.0055)&0.15(0.059) \\
 \hline
(100,100)&\sethlcolor{pink}\hl{0.029(0.0028)}&\sethlcolor{pink}\hl{0.029(0.0028)}&\sethlcolor{pink}\hl{0.1(0.021)}&\sethlcolor{pink}\hl{0.022(0.002)}&\sethlcolor{pink}\hl{0.021(0.002)}&\sethlcolor{pink}\hl{0.1(0.024)}&\sethlcolor{pink}\hl{0.014(0.0012)}&\sethlcolor{pink}\hl{0.015(0.0012)}&\sethlcolor{pink}\hl{0.081(0.025)} \\
 &0.038(0.0046)&0.039(0.0049)&0.33(0.14)&0.032(0.0034)&0.034(0.004)&0.25(0.11)&0.028(0.0028)&0.026(0.0027)&0.16(0.058) \\
 &0.036(0.0041)&0.035(0.0044)&0.33(0.12)&0.029(0.0031)&0.034(0.0042)&0.2(0.075)&0.021(0.0017)&0.02(0.0016)&0.13(0.045) \\

  &0.036(0.0042)&0.04(0.0055)&0.33(0.13)&0.029(0.0028)&0.025(0.0024)&0.24(0.094)&0.026(0.0027)&0.025(0.0023)&0.16(0.067) \\
  &0.04(0.0052)&0.037(0.004)&0.31(0.13)&0.028(0.0026)&0.027(0.0025)&0.21(0.078)&0.024(0.0019)&0.026(0.0027)&0.14(0.054) \\
  \hline
(200,200) &\sethlcolor{pink}\hl{0.022(0.0018)}&\sethlcolor{pink}\hl{0.021(0.0018)}&\sethlcolor{pink}\hl{0.091(0.017)}&\sethlcolor{pink}\hl{0.015(0.0011)}&\sethlcolor{pink}\hl{0.014(0.00095)}&\sethlcolor{pink}\hl{0.1(0.024)}&\sethlcolor{pink}\hl{0.0099(0.0006)}&\sethlcolor{pink}\hl{0.011(0.00069)}&\sethlcolor{pink}\hl{0.077(0.024)} \\
 &0.026(0.0026)&0.026(0.003)&0.34(0.14)&0.018(0.0015)&0.018(0.0014)&0.23(0.094)&0.014(0.001)&0.014(0.0011)&0.16(0.06) \\
 &0.024(0.0023)&0.025(0.0025)&0.32(0.12)&0.017(0.0014)&0.018(0.0014)&0.24(0.088)&0.015(0.0011)&0.013(0.00096)&0.16(0.059) \\

  &0.026(0.0027)&0.025(0.0023)&0.31(0.12)&0.02(0.0019)&0.02(0.0018)&0.24(0.093)&0.013(0.0011)&0.013(0.00094)&0.16(0.06) \\
  &0.025(0.0024)&0.025(0.0027)&0.32(0.13)&0.018(0.0014)&0.018(0.0014)&0.24(0.092)&0.013(0.00084)&0.013(0.0009)&0.15(0.058) \\
  \hline

  \multicolumn{10}{c}{Setting=2($\psi=0.1$), Missing pattern=I}\\
   &  \multicolumn{3}{c}{T=50} & \multicolumn{3}{c}{T=100}& \multicolumn{3}{c}{T=200}\\
     \cline{2-4} \cline{5-7} \cline{8-10}
 $\left(a,b\right)$ & $\mathcal{D}\left(\widehat{\mathbf{R}},\mathbf{R}\right)$ & $\mathcal{D}\left(\widehat{\mathbf{C}},\mathbf{C}\right)$ & $\left\Vert \widehat{\mathbf{F}}_t-\mathbf{H}_R^{-1}\mathbf{F}_t\mathbf{H}_C^{-1\top}\right\Vert$& $\mathcal{D}\left(\widehat{\mathbf{R}},\mathbf{R}\right)$ & $\mathcal{D}\left(\widehat{\mathbf{C}},\mathbf{C}\right)$ & $\left\Vert \widehat{\mathbf{F}}_t-\mathbf{H}_R^{-1}\mathbf{F}_t\mathbf{H}_C^{-1\top}\right\Vert$& $\mathcal{D}\left(\widehat{\mathbf{R}},\mathbf{R}\right)$ & $\mathcal{D}\left(\widehat{\mathbf{C}},\mathbf{C}\right)$ & $\left\Vert \widehat{\mathbf{F}}_t-\mathbf{H}_R^{-1}\mathbf{F}_t\mathbf{H}_C^{-1\top}\right\Vert$ \\ \hline
  (50,50) &\sethlcolor{pink}\hl{0.044(0.0051)}&\sethlcolor{pink}\hl{0.039(0.0046)}&\sethlcolor{pink}\hl{0.099(0.023)}&\sethlcolor{pink}\hl{0.032(0.0037)}&\sethlcolor{pink}\hl{0.036(0.0045)}&\sethlcolor{pink}\hl{0.11(0.027)}&\sethlcolor{pink}\hl{0.02(0.0022)}&\sethlcolor{pink}\hl{0.021(0.0022)}&\sethlcolor{pink}\hl{0.084(0.023)} \\
          &0.074(0.015)&0.053(0.007)&0.35(0.15)&0.077(0.012)&0.054(0.0081)&0.28(0.13)&0.051(0.0056)&0.038(0.0043)&0.18(0.069) \\
          &0.13(0.067)&0.12(0.07)&2.5(0.86)&0.064(0.021)&0.079(0.024)&1.8(0.65)&0.044(0.0096)&0.077(0.025)&1.6(0.63) \\
          &0.087(0.043)&0.083(0.041)&2.5(1.0)&0.068(0.024)&0.06(0.018)&1.9(0.79)&0.045(0.0097)&0.043(0.0098)&1.2(0.46) \\
          &0.079(0.032)&0.069(0.039)&2.7(1.1)&0.071(0.018)&0.06(0.02)&2.1(0.81)&0.041(0.011)&0.035(0.0067)&1.5(0.56) \\ \hline

(100,100)&\sethlcolor{pink}\hl{0.031(0.0034)}&\sethlcolor{pink}\hl{0.031(0.0031)}&\sethlcolor{pink}\hl{0.099(0.026)}&\sethlcolor{pink}\hl{0.021(0.002)}&\sethlcolor{pink}\hl{0.021(0.0019)}&\sethlcolor{pink}\hl{0.097(0.021)}&\sethlcolor{pink}\hl{0.015(0.0012)}&\sethlcolor{pink}\hl{0.016(0.0014)}&\sethlcolor{pink}\hl{0.081(0.023)} \\
         &0.039(0.0047)&0.038(0.0045)&0.4(0.18)&0.027(0.0029)&0.029(0.0031)&0.28(0.12)&0.021(0.0018)&0.023(0.0021)&0.18(0.069) \\
         &0.07(0.042)&0.061(0.022)&2.6(1.0)&0.036(0.011)&0.036(0.0096)&2.1(0.88)&0.037(0.01)&0.039(0.0084)&1.4(0.57) \\
         &0.046(0.015)&0.046(0.02)&2.7(1.0)&0.031(0.0076)&0.03(0.0071)&1.9(0.7)&0.026(0.0047)&0.023(0.0044)&1.4(0.48) \\
         &0.046(0.015)&0.049(0.017)&2.8(0.99)&0.032(0.008)&0.036(0.012)&2.0(0.75)&0.025(0.0054)&0.029(0.0066)&1.5(0.55) \\  \hline
(200,200) &\sethlcolor{pink}\hl{0.022(0.0019)}&\sethlcolor{pink}\hl{0.023(0.0022)}&\sethlcolor{pink}\hl{0.094(0.023)}&\sethlcolor{pink}\hl{0.015(0.0011)}&\sethlcolor{pink}\hl{0.016(0.0011)}&\sethlcolor{pink}\hl{0.1(0.023)}&\sethlcolor{pink}\hl{0.01(0.00066)}&\sethlcolor{pink}\hl{0.011(0.00062)}&\sethlcolor{pink}\hl{0.087(0.027)} \\
         &0.025(0.0026)&0.025(0.0028)&0.37(0.15)&0.017(0.0014)&0.017(0.0012)&0.26(0.1)&0.016(0.0013)&0.015(0.0013)&0.18(0.071) \\
         &0.032(0.0082)&0.033(0.0083)&2.6(0.96)&0.022(0.0048)&0.022(0.0046)&2.1(0.85)&0.017(0.0029)&0.016(0.0024)&1.6(0.6) \\
         &0.032(0.01)&0.032(0.0095)&2.9(1.1)&0.019(0.0032)&0.019(0.0039)&2.1(0.7)&0.014(0.002)&0.016(0.0025)&1.5(0.54) \\
         &0.029(0.0073)&0.029(0.0076)&2.7(0.98)&0.021(0.0038)&0.021(0.0048)&2.0(0.75)&0.015(0.0023)&0.014(0.0023)&1.5(0.56) \\ \hline
  \multicolumn{10}{c}{Setting=2($\psi=0.5$), Missing pattern=I}\\
   &  \multicolumn{3}{c}{T=50} & \multicolumn{3}{c}{T=100}& \multicolumn{3}{c}{T=200}\\
     \cline{2-4} \cline{5-7} \cline{8-10}
 $\left(a,b\right)$ & $\mathcal{D}\left(\widehat{\mathbf{R}},\mathbf{R}\right)$ & $\mathcal{D}\left(\widehat{\mathbf{C}},\mathbf{C}\right)$ & $\left\Vert \widehat{\mathbf{F}}_t-\mathbf{H}_R^{-1}\mathbf{F}_t\mathbf{H}_C^{-1\top}\right\Vert$& $\mathcal{D}\left(\widehat{\mathbf{R}},\mathbf{R}\right)$ & $\mathcal{D}\left(\widehat{\mathbf{C}},\mathbf{C}\right)$ & $\left\Vert \widehat{\mathbf{F}}_t-\mathbf{H}_R^{-1}\mathbf{F}_t\mathbf{H}_C^{-1\top}\right\Vert$& $\mathcal{D}\left(\widehat{\mathbf{R}},\mathbf{R}\right)$ & $\mathcal{D}\left(\widehat{\mathbf{C}},\mathbf{C}\right)$ & $\left\Vert \widehat{\mathbf{F}}_t-\mathbf{H}_R^{-1}\mathbf{F}_t\mathbf{H}_C^{-1\top}\right\Vert$ \\ \hline
(50,50)
&\sethlcolor{pink}\hl{0.041(0.0055)}&\sethlcolor{pink}\hl{0.045(0.0065)}&\sethlcolor{pink}\hl{0.16(0.16)}&\sethlcolor{pink}\hl{0.027(0.0032)}&\sethlcolor{pink}\hl{0.034(0.0044)}&\sethlcolor{pink}\hl{0.12(0.035)}&\sethlcolor{pink}\hl{0.022(0.0029)}&\sethlcolor{pink}\hl{0.025(0.0031)}&\sethlcolor{pink}\hl{0.12(0.029)} \\
&0.065(0.012)&0.077(0.016)&0.79(0.35)&0.036(0.0049)&0.048(0.0065)&0.48(0.2)&0.039(0.0054)&0.046(0.0051)&0.33(0.14) \\
&0.059(0.01)&0.057(0.0094)&0.67(0.27)&0.05(0.0066)&0.043(0.0063)&0.43(0.16)&0.037(0.0046)&0.046(0.0057)&0.28(0.1) \\
&0.061(0.012)&0.066(0.011)&0.56(0.21)&0.043(0.0066)&0.042(0.0055)&0.4(0.16)&0.046(0.0056)&0.063(0.0095)&0.33(0.14) \\
&0.057(0.01)&0.074(0.015)&0.6(0.24)&0.049(0.0081)&0.052(0.0067)&0.45(0.17)&0.06(0.0082)&0.05(0.0062)&0.32(0.13) \\ \hline

(100,100)
&\sethlcolor{pink}\hl{0.033(0.0042)}&\sethlcolor{pink}\hl{0.032(0.0039)}&\sethlcolor{pink}\hl{0.16(0.17)}&\sethlcolor{pink}\hl{0.02(0.0019)}&\sethlcolor{pink}\hl{0.021(0.002)}&\sethlcolor{pink}\hl{0.12(0.027)}&\sethlcolor{pink}\hl{0.014(0.0012)}&\sethlcolor{pink}\hl{0.014(0.0013)}&\sethlcolor{pink}\hl{0.12(0.024)} \\
&0.039(0.0049)&0.038(0.0053)&0.76(0.34)&0.032(0.0042)&0.028(0.0033)&0.52(0.22)&0.026(0.0027)&0.022(0.0021)&0.33(0.15) \\
&0.037(0.0048)&0.039(0.0052)&0.63(0.25)&0.027(0.003)&0.036(0.0053)&0.48(0.18)&0.025(0.0026)&0.028(0.0031)&0.3(0.11) \\
&0.044(0.0064)&0.038(0.0051)&0.66(0.25)&0.029(0.0034)&0.037(0.0055)&0.45(0.18)&0.024(0.0023)&0.024(0.0022)&0.33(0.14) \\
&0.038(0.0049)&0.034(0.004)&0.59(0.22)&0.027(0.0032)&0.031(0.0046)&0.44(0.16)&0.021(0.0021)&0.025(0.0029)&0.32(0.13) \\ \hline
(200,200)
 &\sethlcolor{pink}\hl{0.022(0.0023)}&\sethlcolor{pink}\hl{0.022(0.002)}&\sethlcolor{pink}\hl{0.15(0.13)}&\sethlcolor{pink}\hl{0.016(0.0013)}&\sethlcolor{pink}\hl{0.016(0.0014)}&\sethlcolor{pink}\hl{0.12(0.038)}&\sethlcolor{pink}\hl{0.01(0.00073)}&\sethlcolor{pink}\hl{0.011(0.00076)}&\sethlcolor{pink}\hl{0.12(0.026)} \\
&0.024(0.0025)&0.024(0.0025)&0.77(0.34)&0.017(0.0015)&0.017(0.0014)&0.5(0.21)&0.014(0.0012)&0.014(0.001)&0.35(0.14) \\
&0.024(0.0027)&0.025(0.0028)&0.61(0.22)&0.017(0.0014)&0.018(0.0017)&0.45(0.17)&0.014(0.0012)&0.015(0.0013)&0.31(0.12) \\
&0.024(0.0027)&0.025(0.003)&0.61(0.22)&0.018(0.0017)&0.018(0.0017)&0.43(0.15)&0.014(0.0013)&0.014(0.0011)&0.32(0.12) \\
 &0.024(0.0026)&0.024(0.0024)&0.59(0.22)&0.019(0.0018)&0.019(0.002)&0.44(0.17)&0.014(0.0012)&0.015(0.0013)&0.33(0.13) \\  \hline
   \multicolumn{10}{c}{Setting=3, Missing pattern=I}\\
    &  \multicolumn{3}{c}{T=50} & \multicolumn{3}{c}{T=100}& \multicolumn{3}{c}{T=200}\\
     \cline{2-4} \cline{5-7} \cline{8-10}
 $\left(a,b\right)$ & $\mathcal{D}\left(\widehat{\mathbf{R}},\mathbf{R}\right)$ & $\mathcal{D}\left(\widehat{\mathbf{C}},\mathbf{C}\right)$ & $\left\Vert \widehat{\mathbf{F}}_t-\mathbf{H}_R^{-1}\mathbf{F}_t\mathbf{H}_C^{-1\top}\right\Vert$& $\mathcal{D}\left(\widehat{\mathbf{R}},\mathbf{R}\right)$ & $\mathcal{D}\left(\widehat{\mathbf{C}},\mathbf{C}\right)$ & $\left\Vert \widehat{\mathbf{F}}_t-\mathbf{H}_R^{-1}\mathbf{F}_t\mathbf{H}_C^{-1\top}\right\Vert$& $\mathcal{D}\left(\widehat{\mathbf{R}},\mathbf{R}\right)$ & $\mathcal{D}\left(\widehat{\mathbf{C}},\mathbf{C}\right)$ & $\left\Vert \widehat{\mathbf{F}}_t-\mathbf{H}_R^{-1}\mathbf{F}_t\mathbf{H}_C^{-1\top}\right\Vert$ \\ \hline
  (50,50) &\sethlcolor{pink}\hl{0.049(0.0066)}&\sethlcolor{pink}\hl{0.045(0.0058)}&\sethlcolor{pink}\hl{0.091(0.023)}&\sethlcolor{pink}\hl{0.033(0.004)}&\sethlcolor{pink}\hl{0.03(0.0036)}&\sethlcolor{pink}\hl{0.1(0.026)}&\sethlcolor{pink}\hl{0.022(0.0022)}&\sethlcolor{pink}\hl{0.023(0.0025)}&\sethlcolor{pink}\hl{0.083(0.026)} \\
          &0.074(0.011)&0.058(0.0084)&0.37(0.17)&0.047(0.006)&0.052(0.006)&0.24(0.1)&0.033(0.0038)&0.032(0.0032)&0.16(0.064) \\
          &0.06(0.0093)&0.058(0.0078)&0.34(0.14)&0.05(0.0078)&0.045(0.0062)&0.22(0.092)&0.04(0.0046)&0.044(0.005)&0.16(0.06) \\
          &0.051(0.0067)&0.055(0.0074)&0.27(0.1)&0.037(0.0045)&0.047(0.0057)&0.23(0.085)&0.035(0.004)&0.063(0.0066)&0.15(0.056) \\
          &0.066(0.01)&0.058(0.0088)&0.34(0.14)&0.066(0.0094)&0.054(0.0079)&0.23(0.1)&0.048(0.0063)&0.038(0.0035)&0.14(0.058) \\  \hline
(100,100)
    &\sethlcolor{pink}\hl{0.03(0.003)}&\sethlcolor{pink}\hl{0.032(0.0036)}&\sethlcolor{pink}\hl{0.1(0.023)}&\sethlcolor{pink}\hl{0.022(0.002)}&\sethlcolor{pink}\hl{0.019(0.0016)}&\sethlcolor{pink}\hl{0.098(0.023)}&\sethlcolor{pink}\hl{0.015(0.0013)}&\sethlcolor{pink}\hl{0.016(0.0013)}&\sethlcolor{pink}\hl{0.075(0.022)} \\
         &0.037(0.0047)&0.037(0.0043)&0.35(0.15)&0.025(0.0024)&0.025(0.0027)&0.23(0.085)&0.021(0.0022)&0.018(0.0015)&0.17(0.059) \\
        &0.043(0.0062)&0.036(0.0042)&0.31(0.13)&0.029(0.0031)&0.028(0.0029)&0.23(0.098)&0.022(0.0021)&0.023(0.0022)&0.16(0.056) \\
        &0.042(0.0065)&0.036(0.0042)&0.3(0.11)&0.029(0.0031)&0.029(0.003)&0.22(0.08)&0.025(0.0025)&0.026(0.0026)&0.17(0.066) \\
        &0.038(0.0047)&0.035(0.0039)&0.33(0.13)&0.026(0.0025)&0.028(0.003)&0.22(0.084)&0.02(0.0016)&0.019(0.0017)&0.17(0.063) \\ \hline
(200,200) &\sethlcolor{pink}\hl{0.023(0.002)}&\sethlcolor{pink}\hl{0.023(0.002)}&\sethlcolor{pink}\hl{0.087(0.018)}&\sethlcolor{pink}\hl{0.015(0.0012)}&\sethlcolor{pink}\hl{0.015(0.0011)}&\sethlcolor{pink}\hl{0.098(0.024)}&\sethlcolor{pink}\hl{0.01(0.00064)}&\sethlcolor{pink}\hl{0.01(0.00064)}&\sethlcolor{pink}\hl{0.079(0.025)} \\
          &0.025(0.0024)&0.024(0.0022)&0.35(0.15)&0.019(0.0015)&0.019(0.0015)&0.25(0.1)&0.013(0.00093)&0.016(0.0014)&0.18(0.075) \\
          &0.024(0.0023)&0.023(0.0021)&0.33(0.12)&0.019(0.0016)&0.017(0.0013)&0.23(0.086)&0.015(0.0011)&0.014(0.001)&0.15(0.056) \\
           &0.024(0.002)&0.023(0.002)&0.32(0.12)&0.019(0.0017)&0.02(0.0017)&0.22(0.086)&0.014(0.00095)&0.012(0.00079)&0.16(0.062) \\
           &0.023(0.002)&0.023(0.002)&0.31(0.12)&0.017(0.0013)&0.016(0.0013)&0.22(0.075)&0.013(0.00095)&0.015(0.001)&0.16(0.058) \\ \hline
   \hline
  \end{tabular}
  }
\end{table}

\begin{figure}
\includegraphics[scale=0.3]{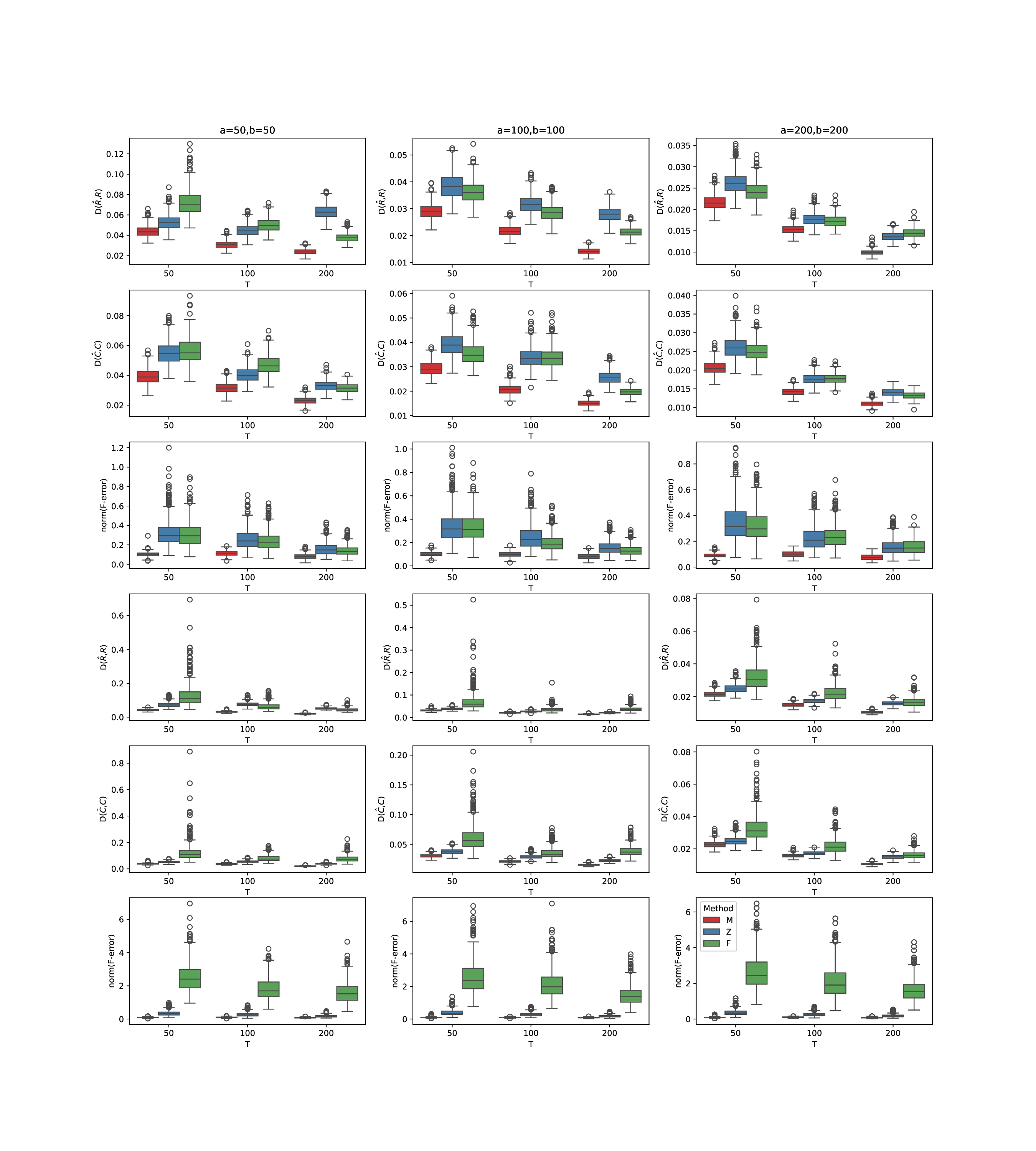}
\caption{Boxplots of errors of loading matrices estimation and factor matrices estimation for Setting 1 and Setting 2 with $\psi=0.1$ under missing pattern I.}
\label{fig:4-1}
\end{figure}
\begin{figure}
\includegraphics[scale=0.3]{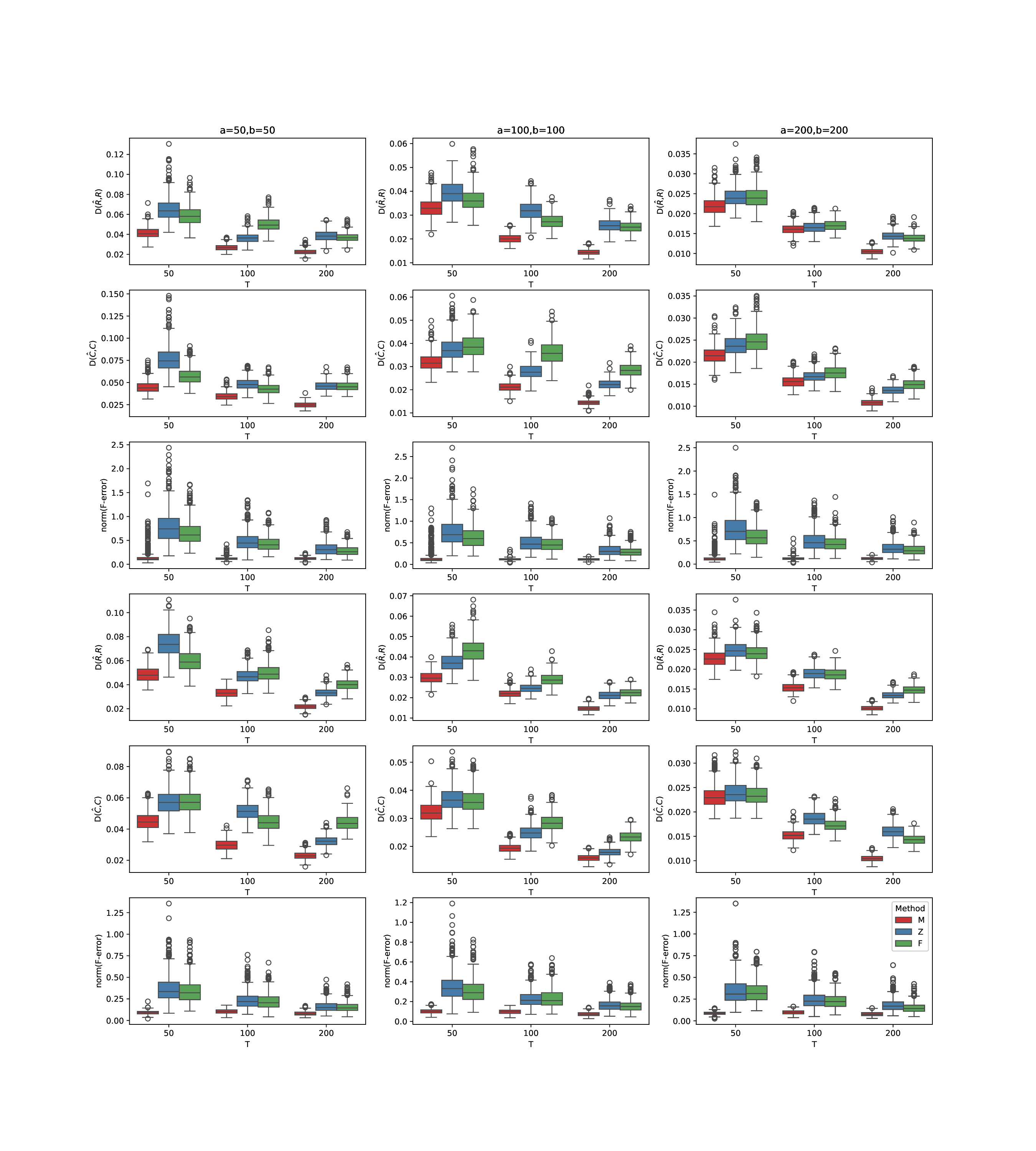}
\caption{Boxplots of errors of loading matrices estimation
and factor matrices estimation for Setting 2 with $\psi=0.5$ and Setting 3 under missing pattern I.}
\label{fig:4-2}
\end{figure}
\begin{figure}
\includegraphics[scale=0.3]{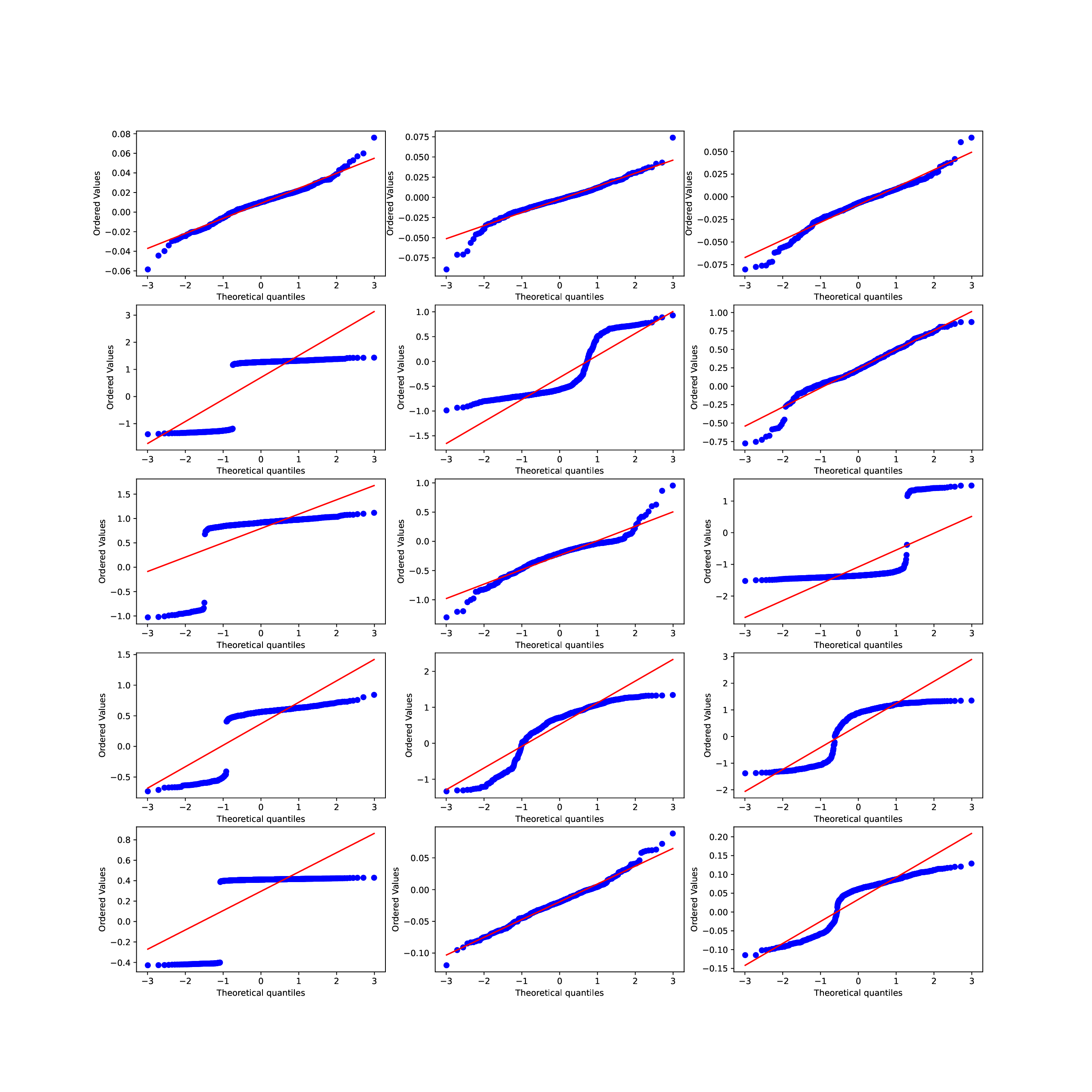}
\caption{The QQ plots for Setting 1, missing pattern I.}
\label{fig:4-5}
\end{figure}

\begin{figure}
\includegraphics[scale=0.3]{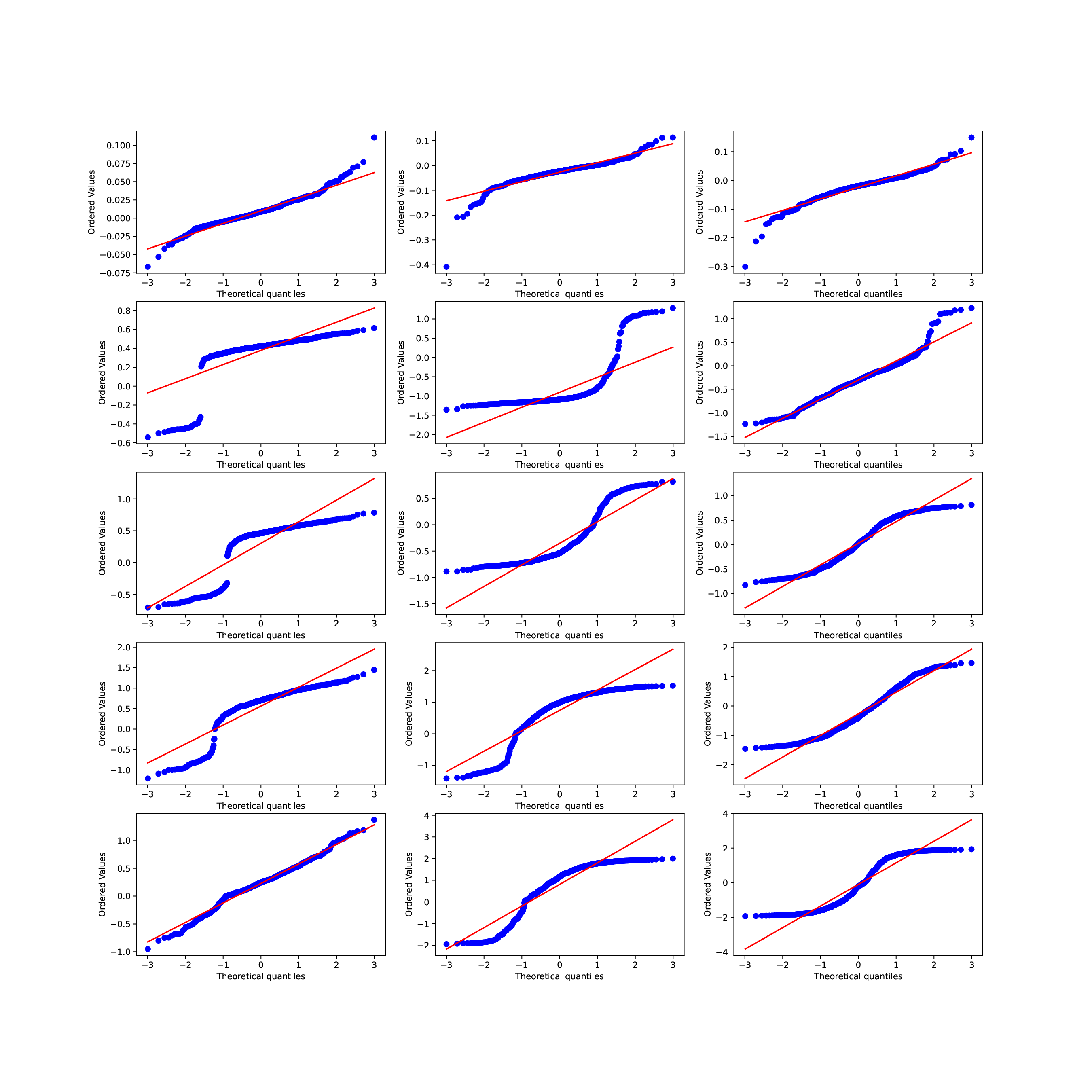}
\caption{The QQ plots for Setting 2 with $\psi=0.1$, missing pattern I. }
\label{fig:4-6}
\end{figure}

\begin{figure}
\includegraphics[scale=0.3]{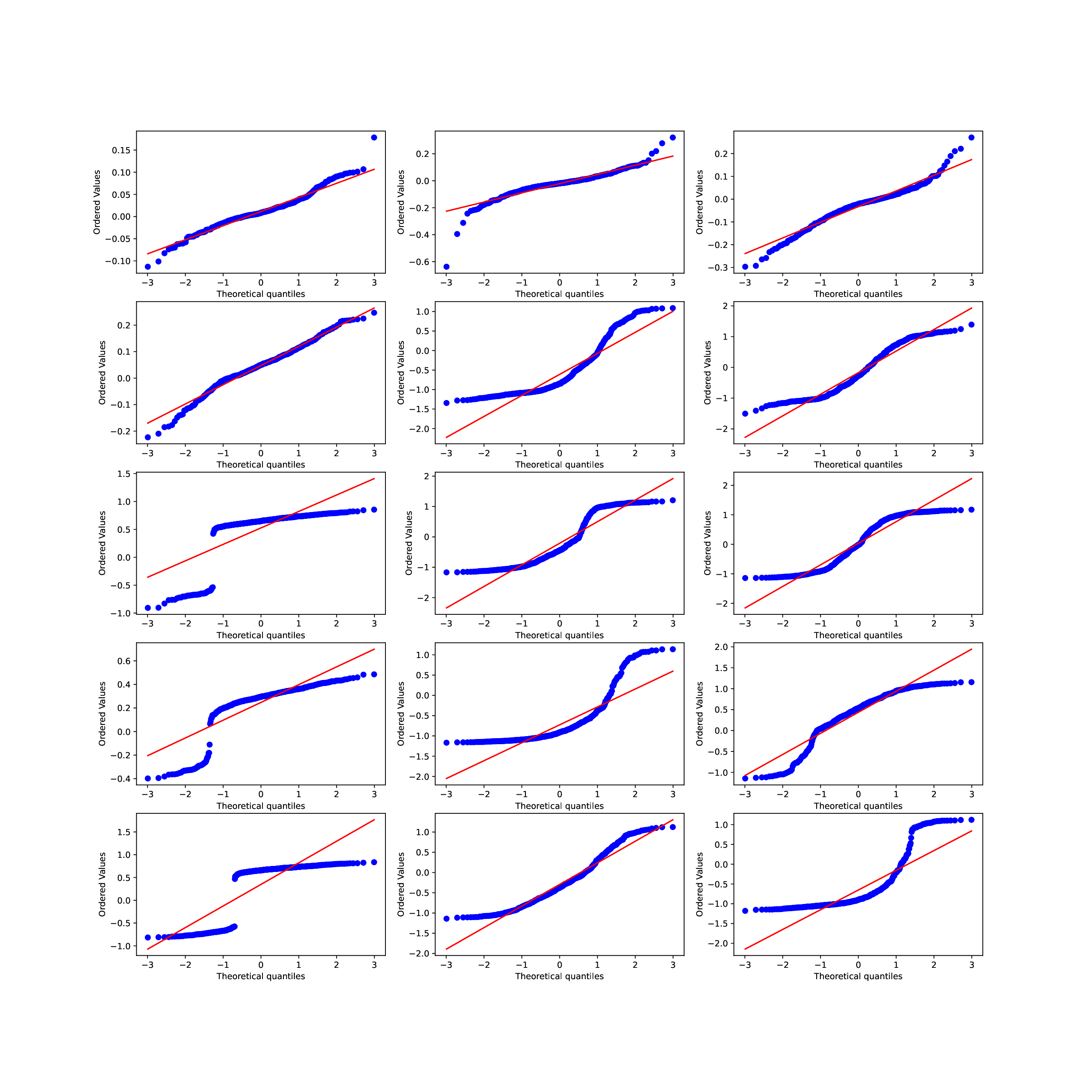}
\caption{The QQ plots for Setting 2 with $\psi=0.5$, missing pattern I. }
\label{fig:4-7}
\end{figure}

\begin{figure}
\includegraphics[scale=0.3]{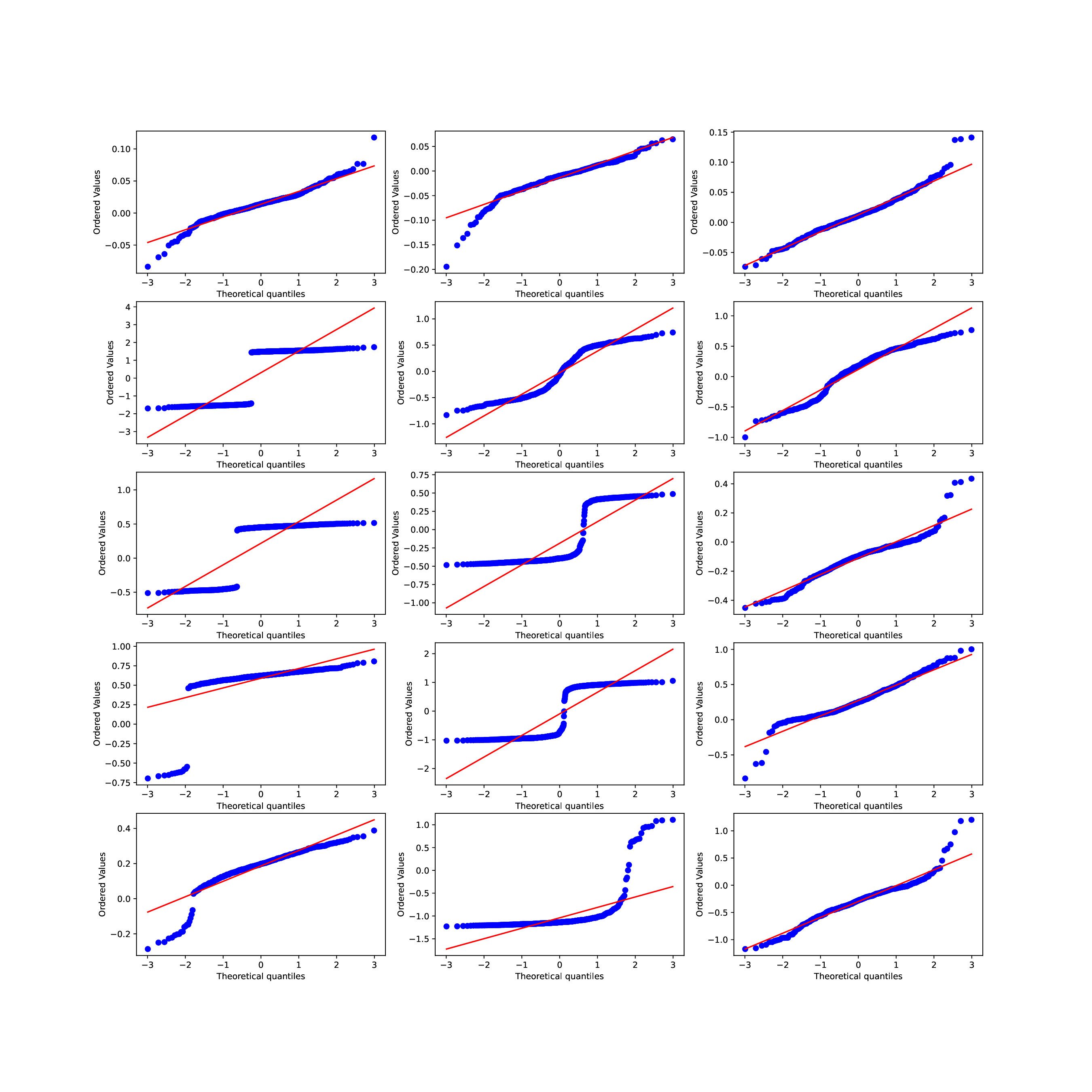}
\caption{The QQ plots for Setting 3, missing pattern I. }
\label{fig:4-8}
\end{figure}

\begin{figure}
\includegraphics[scale=0.4]{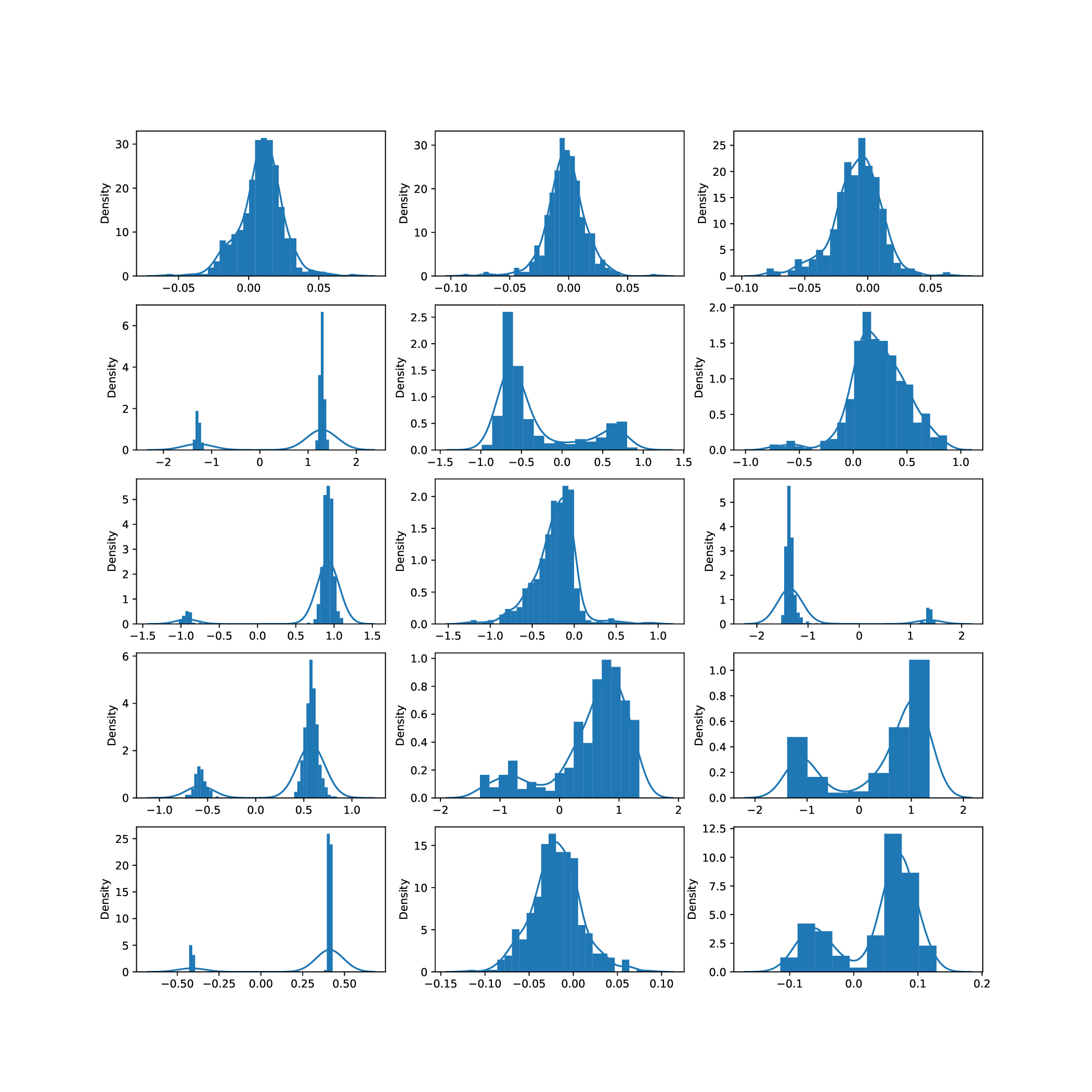}
\caption{The histograms for Setting 1, missing pattern I.}
\label{fig:4-9}
\end{figure}
\begin{figure}
\includegraphics[scale=0.4]{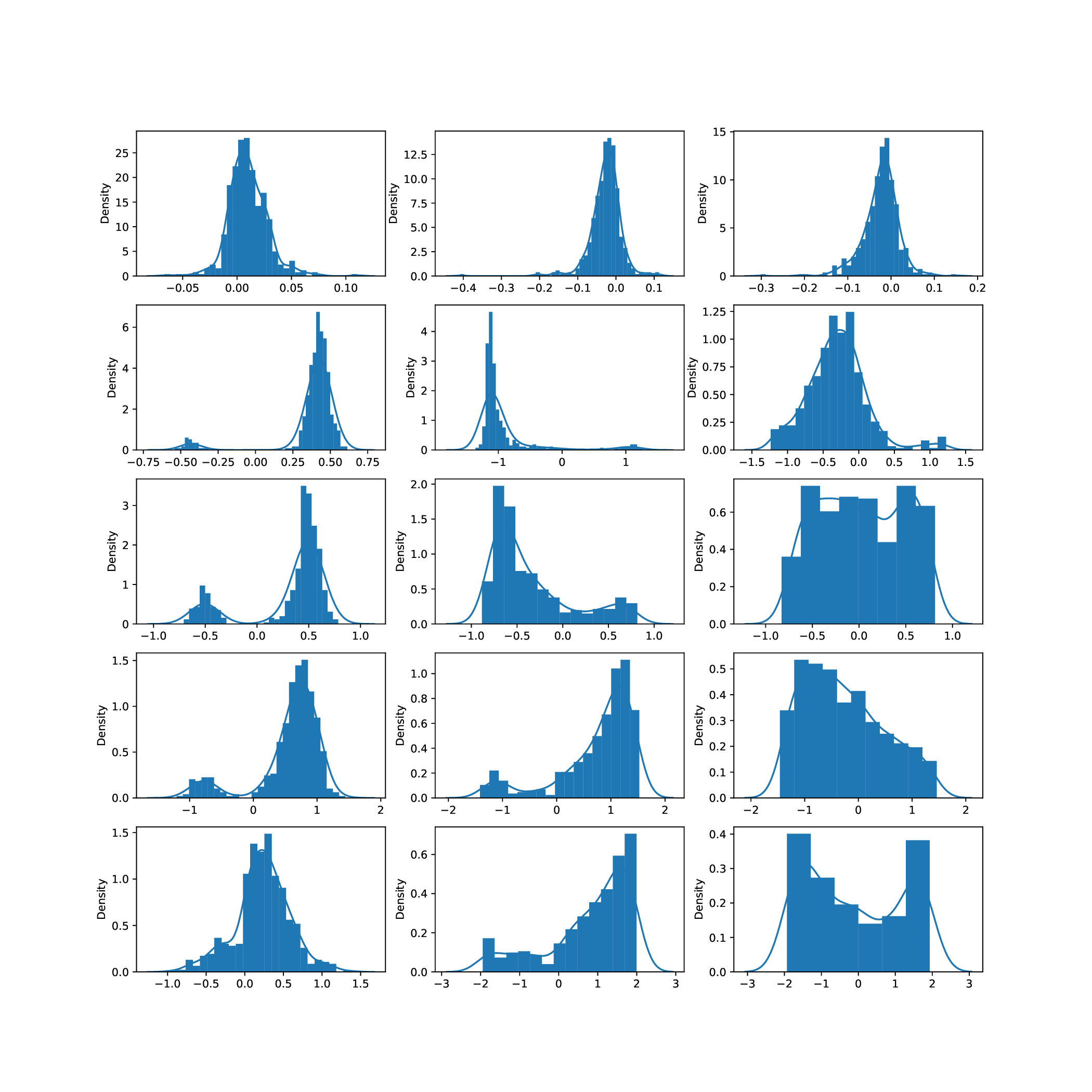}
\caption{The histograms for Setting 2 with $\psi=0.1$, missing pattern I.}
\label{fig:4-10}
\end{figure}

\begin{figure}
\includegraphics[scale=0.4]{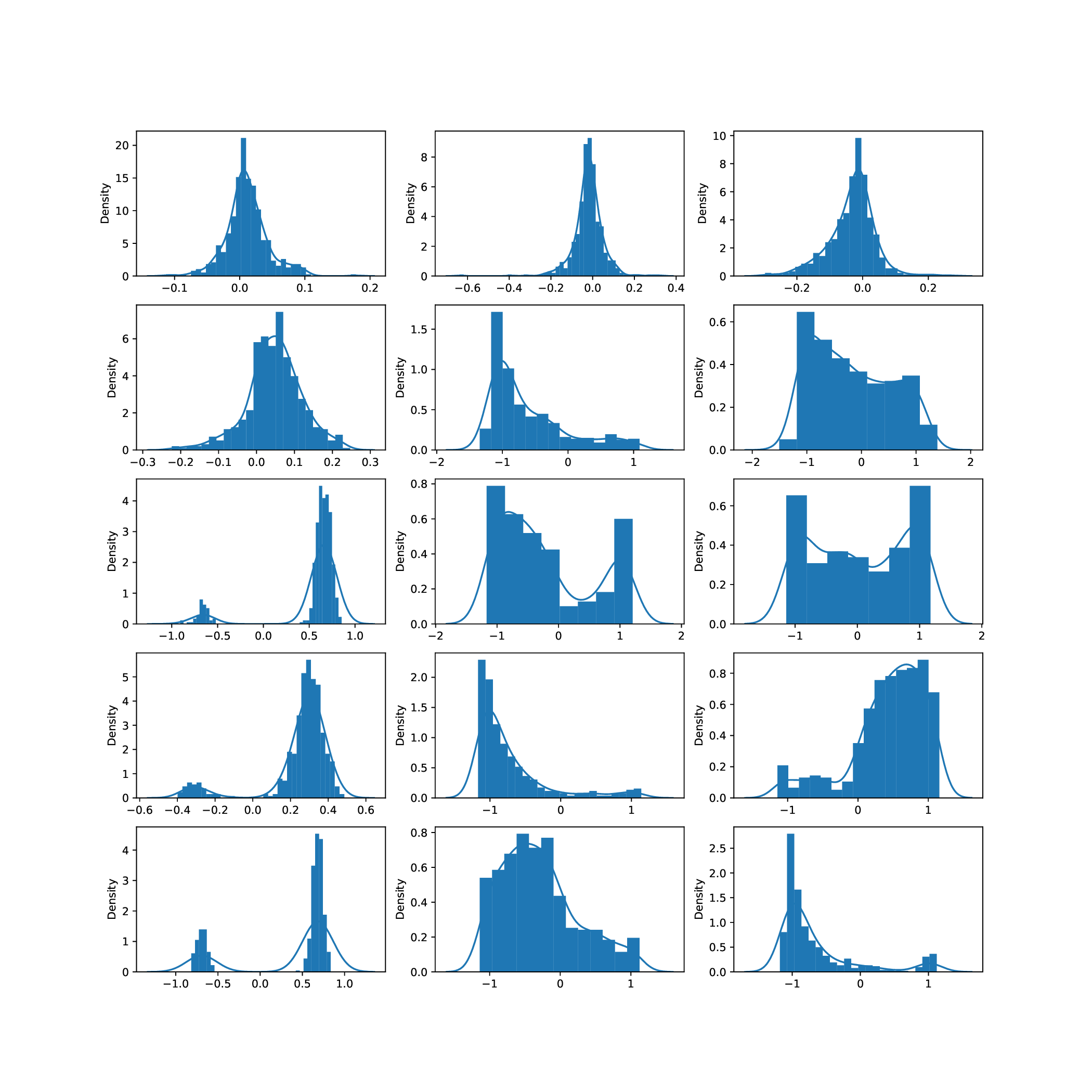}
\caption{The histograms for Setting 2 with $\psi=0.5$, missing pattern I.}
\label{fig:4-11}
\end{figure}

\begin{figure}
\includegraphics[scale=0.4]{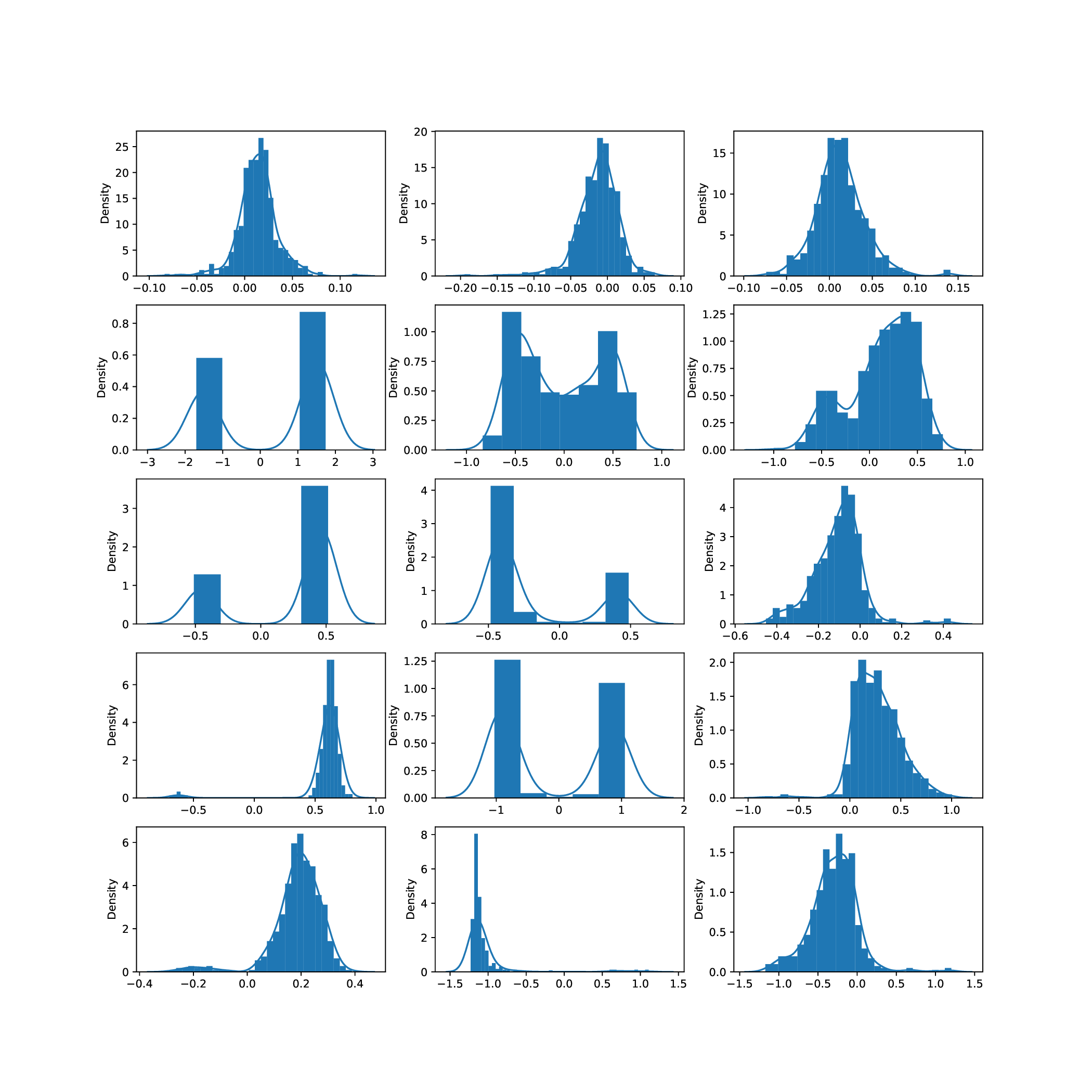}
\caption{The histograms for Setting 3, missing pattern I.}
\label{fig:4-12}
\end{figure}

\section{Application: Multinational Macroeconomic Indices}\label{sec:App}
 In this section, we apply our method to a dataset of multinational macroeconomic indices collected from OECD. This dataset includes 10 quarterly macroeconomic indices from 16 countries, covering the period from the second quarter of 1970 to the fourth quarter of 2015, resulting in a total of 181 quarters. Therefore, we have $T=181$ and $a \times b = 16 \times 10$ matrix-valued time series. The countries represented in the dataset are Australia (AU), Austria (AUT), Canada (CAN), Denmark (DK), Finland (FIN), France (FRA), Germany (GER), Greece (GR), Ireland (IE), Italy (ITA), Netherlands (NL), New Zealand (NZL), Norway (NOR), Sweden (SWE), the United Kingdom (UK), and the United States of America (USA).

 The indices fall into four major categories. Production: Industrial Production (P:IP), Total Manufacturing (P:TM), Gross Domestic Product (GDP). Consumer Prices: Food (CPI: Food), Energy (CPI: Energy), Total (CPI: Total). Money Market: Long-Term Rates (IR:L), Short-Term Rates (IR:S). International Trade: Trade Export (IT:EX), Trade Import (IT:IM).

 In the dataset, there are 14 countries and 8 macroeconomic indices with missing values up to 153 time points. There are a total of 3,152 missing values, resulting in a missing percentage of $10.88\%$.
  Each original univariate time series is transformed by applying either the first or second difference or taking the logarithm, in order to meet the mixing condition outlined in Assumption \ref{As:3-1}. For detailed descriptions of the dataset and the transformations used, refer to Table \ref{tabA3-1} in Appendix \ref{sec:A2}. Figure \ref{fig:5-1} displays the transformed time series of macroeconomic indicators from all countries. It is evident that similar patterns exist among the time series in the same row or column.

  We apply our method to the OECD data set and obtain estimates $\left(\hat{k},\hat{r}\right)=\left(2,2\right)$. However, using the method outlined in \cite{Chen03042023} with $\alpha=0.55$ (This value is recommended by \cite{Chen03042023}), we arrive at $\left(\hat{k},\hat{r}\right)=\left(6,2\right)$, which is consistent with the results from the direct method.

  Here, we present the results generated by our method. Figure \ref{fig:5-2} displays the eigenvalues and the eigenratios of $\left(\widehat{\mathbf{M}}_R,\widehat{\mathbf{M}}_C\right)$, calculated according to Equations \eqref{eq:2-3} and \eqref{eq:2-4}.  From these estimates of $\widehat{\mathbf{M}}$ with $\left(\hat{k},\hat{r}\right)=\left(2,2\right)$, we calculate the loading matrices $\widehat{\mathbf{R}}, \widehat{\mathbf{C}}$, respectively.

  Tables \ref{tab5-1} and \ref{tab5-2} present the estimates of the row and column loading matrices, respectively. These matrices have been VARIMAX-rotated to reveal a clear structure, and they are scaled and rounded for easier display. We can interpret the latent structure of the global macroeconomy by analyzing the estimated row and column loading matrices. Specifically, from the pairs of $\widehat{\mathbf{R}}, \widehat{\mathbf{C}}$, we can cluster countries or macroeconomic indices based on their loading matrices.

  Using the row loading matrix $\widehat{\mathbf{R}}$, we can form three groups: Group 1: AU, AUT, CAN, DK, FIN, FRA, GER, IE, ITA, NL, NOR, SWE, UK, USA; Group 2: GR; Group 3: NZL. In this example, Group 1 loads heavily on Row 2; Group 2 loads heavily on Row 1; and Group 3 loads heavily on Row 2 and lightly on Row 1. This analysis can help identify which countries exhibit stronger correlations in their macroeconomic characteristics.

  From the column loading matrix $\widehat{\mathbf{C}}$, we can identify three groups of variables. Group 1: IT:EX, IT:IM, GDP, R:TM, CPI:Food, CPI:Energy, CPI:Total, P:IP; Group 2: IR:L; Group 3: IR:S. Group 1 loads both heavily on two factors, while Group 2 and Group 3 are highly loaded on the first and second rows, respectively.

\begin{figure}
\hspace*{-2cm} \includegraphics[scale=0.25]{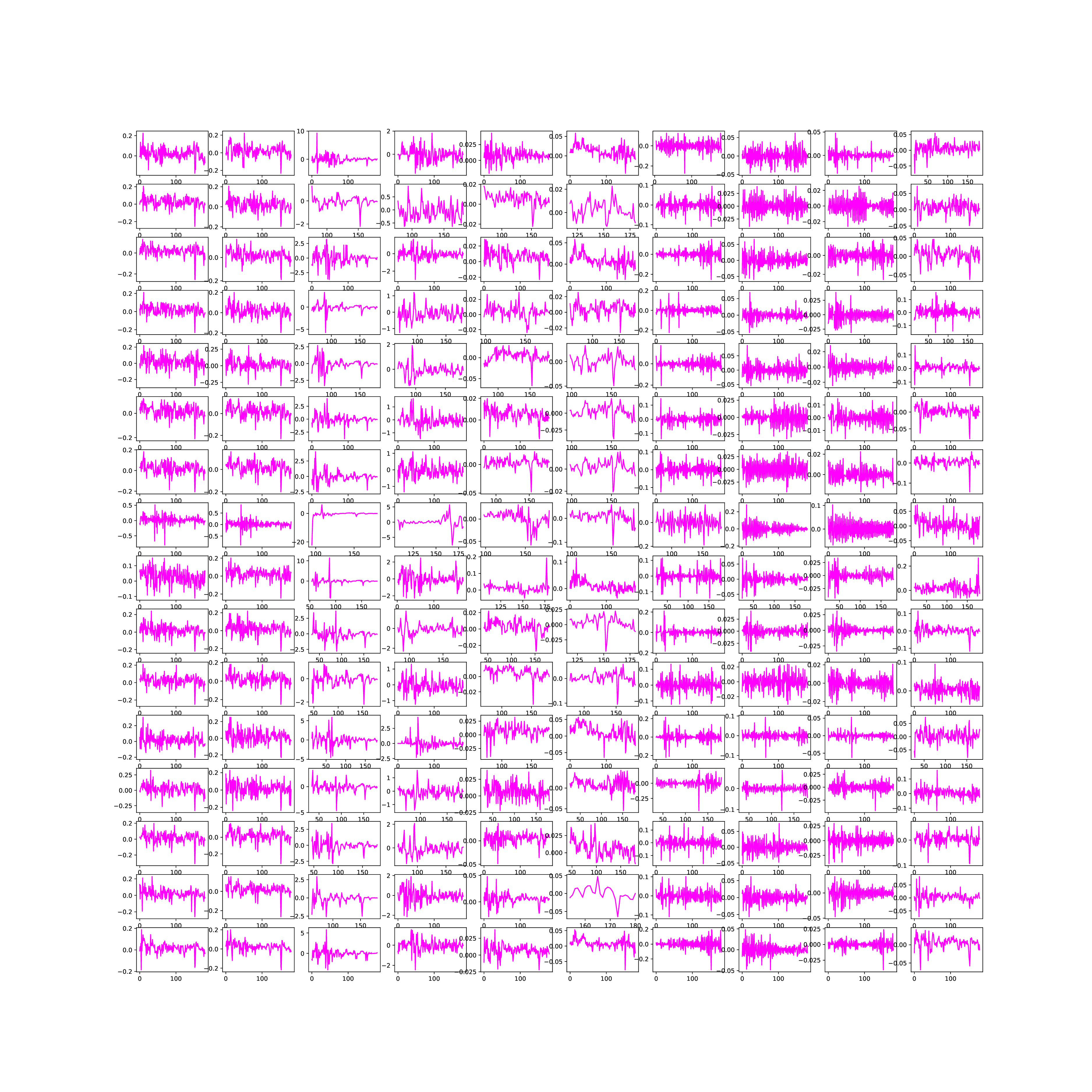}
\caption{Time series plots of macroeconomic indicators of all the countries (after data transformation).}
\label{fig:5-1}
\end{figure}

\begin{table}
\caption{Estimations of row loading matrices (VARIMAX rotated). The loading matrix is multiplied by 100 and rounded for ease indisplay.\label{tab5-1}}%
\resizebox{.9\linewidth}{!}{
\begin{tabular}{lllllllllllllllll}
 \hline
 Row  & AU &AUT &CAN& DK& FIN& FRA& GER&GR&IE&ITA& NL& NZL& NOR& SWE& UK & USA\\
 \hline
Row 1    & 0   &0  & 0 &0 &0& 0 &0&10&0&0&0&-1&0&0&0&0\\
Row 2    & -3  & -1& -3 &-2 &-2& -3&-2&0&-5&-2&-2&-3&-2&-2&-2&-2\\
 \hline
\end{tabular}
}
\end{table}

\begin{table}
\caption{Estimations of column loading matrices (VARIMAX rotated). The loading matrix is multiplied by 100 and rounded for ease indisplay.\label{tab5-2}}%
\resizebox{.9\linewidth}{!}{
\begin{tabular}{lllllllllll}
 \hline
Row  &IT:EX &IT:IM &IR:L& IR:S&GDP&R:TM&CPI:Energy&CPI:Food&CPI:Total&(P:IP)\\
 \hline
Row 1    &-5   &-9  & -1000 &9 &-1& -1&2&0.5&0&-2 \\
Row 2    & -9  & -12& -8 &-1000 &0&-5&-11&-0.3&-1&-2\\
 \hline
\end{tabular}
}
\end{table}
\section{Conclusion}\label{sec:Con}
 This paper develops an inferential theory for matrix-variate latent factor models with missing observations. We propose a simple all-purpose estimator. Both the simulation results and the real data application demonstrate that our method outperforms others. In the future, we plan to generalize the method to high-dimensional tensor factor models with missing observations.

\section{Competing interests}
No competing interest is declared.

\section{Author contributions statement}
M.T. and Y.Z. proposed the conceptualization and methodology. L.X. , K.Y. and J.L. analyzed the results of simulation and real data application. Y.Z. and J.L. wrote and reviewed the manuscript.
\section{Acknowledgments}
This work is sponsored by the Yuxiu Innovation Project of NCUT (Project No. 2024NCUTYXCX104)

\bibliographystyle{abbrvnat}
\bibliography{reference}
\appendix
\section{Proof of Theorem}\label{sec:A}
\subsection{Proof of Theorem \ref{Th1}}
From model \eqref{eq:2-1}, we can get that $\widetilde{\mathbf{Y}}_t=\mathbf{R}\mathbf{F}_t \mathbf{C}^{\top}\odot \mathbf{W}_t +\mathbf{E}_t\odot \mathbf{W}_t$, let $\tilde{q}_{R,ij}=\frac{|Q_{R,ij}|}{bT}$, then from formula \eqref{eq:3-3}, we can obtain:
\begin{equation*}
\frac{1}{abT}\sum_{t=1}^T\left(\mathbf{R} \mathbf{F}_t \mathbf{C}^{\top} \odot \mathbf{W}_t +\mathbf{E}_t\odot \mathbf{W}_t\right)\left(\mathbf{C} \mathbf{F}_t^{\top}\mathbf{R}^{\top}\odot \mathbf{W}_t^{\top}+ \mathbf{E}_t^{\top}\odot \mathbf{W}_t^{\top}\right)\odot\left[\frac{1}{\tilde{q}_{R,ij}}\right]\widehat{\mathbf{R}}V_{R,abT}^{-1}=\widehat{\mathbf{R}}
\end{equation*}
Note that the $\left(i,j\right)$-th entries in $\left(\mathbf{R} \mathbf{F}_t \mathbf{C}^{\top} \odot \mathbf{W}_t\right)\left(\mathbf{C} \mathbf{F}_t^{\top}\mathbf{R}^{\top}\odot \mathbf{W}_t^{\top}\right)$,\\
$\left(\mathbf{R} \mathbf{F}_t \mathbf{C}^{\top} \odot \mathbf{W}_t\right)\left(\mathbf{E}_t^{\top}\odot \mathbf{W}_t^{\top}\right)$,
$\left(\mathbf{E}_t\odot \mathbf{W}_t\right)\left(\mathbf{C} \mathbf{F}_t^{\top}\mathbf{R}^{\top}\odot \mathbf{W}_t^{\top}\right)$, and $\left(\mathbf{E}_t\odot \mathbf{W}_t\right)\left(\mathbf{E}_t^{\top}\odot \mathbf{W}_t^{\top}\right)$ have the following form:
\begin{align*}
&\left(\mathbf{R} \mathbf{F}_t \mathbf{C}^{\top} \odot \mathbf{W}_t\right)\left(\mathbf{C} \mathbf{F}_t^{\top}\mathbf{R}^{\top}\odot \mathbf{W}_t^{\top}\right)_{ij}=\left(\mathbf{R}_i^{\top}\mathbf{F}_t\mathbf{C}^{\top}\odot\mathbf{W}_{t,i}^{\top}\right)\left(\mathbf{C}\mathbf{F}_t^{\top}\mathbf{R}_j\odot\mathbf{W}_{t,j}\right)\\
&=\mathbf{R}_i^{\top}\mathbf{F}_t\mathbf{C}^{\top} diag\left(\mathbf{W}_{t,i}\odot \mathbf{W}_{t,j}\right)\mathbf{C}\mathbf{F}_t^{\top}\mathbf{R}_j\\
&\left(\mathbf{R} \mathbf{F}_t \mathbf{C}^{\top} \odot \mathbf{W}_t\right)\left(\mathbf{E}_t^{\top}\odot \mathbf{W}_t^{\top}\right)_{ij}=\left(\mathbf{R}_i^{\top}\mathbf{F}_t\mathbf{C}^{\top}\odot\mathbf{W}_{t,i}^{\top}\right)\left(\mathbf{E}_{t,j}\odot\mathbf{W}_{t,j}\right)\\
&=\mathbf{R}_i^{\top}\mathbf{F}_t\mathbf{C}^{\top} diag\left(\mathbf{W}_{t,i}\odot \mathbf{W}_{t,j}\right)\mathbf{E}_{t,j}\\
&\left(\mathbf{E}_t\odot \mathbf{W}_t\right)\left(\mathbf{C} \mathbf{F}_t^{\top}\mathbf{R}^{\top}\odot \mathbf{W}_t^{\top}\right)_{ij}=\left(\mathbf{E}_{t,i}^{\top}\odot\mathbf{W}_{t,i}^{\top}\right)\left(\mathbf{C} \mathbf{F}_t^{\top}\mathbf{R}_j\odot \mathbf{W}_{t,j} \right)\\
&=\mathbf{E}_{t,i}^{\top} diag\left(\mathbf{W}_{t,i}\odot \mathbf{W}_{t,j}\right) \mathbf{C} \mathbf{F}_t^{\top}\mathbf{R}_j\\
&\left(\mathbf{E}_t\odot \mathbf{W}_t\right)\left(\mathbf{E}_t^{\top}\odot \mathbf{W}_t^{\top}\right)_{ij}=\left(\mathbf{E}_{t,i}^{\top}\odot\mathbf{W}_{t,i}^{\top}\right)\left(\mathbf{E}_{t,j}\odot\mathbf{W}_{t,j}\right)\\
&=\mathbf{E}_{t,i}^{\top} diag\left(\mathbf{W}_{t,i}\odot \mathbf{W}_{t,j}\right)\mathbf{E}_{t,j}
\end{align*}
Let
\begin{equation*}
 \mathbf{H}_j=\frac{1}{a}V_{R,abT}^{-1}\sum_{i=1}^a \widehat{\mathbf{R}}_i \mathbf{R}_i^{\top} \frac{1}{|Q_{R,ij}|}\sum_{t,m \in Q_{R,ij}}\mathbf{F}_t \mathbf{C}_m \mathbf{C}_m^{\top} \mathbf{F}_t^{\top}\in \mathbb{R}^{k\times k},
 \end{equation*}
Then,
\begin{equation*}
\begin{split}
&\widehat{\mathbf{R}}_j-\mathbf{H}_j\mathbf{R}_j\\
&=\frac{1}{abT}V_{R,abT}^{-1}\left[\sum_{t=1}^T\sum_{i=1}^a \widehat{\mathbf{R}}_i \mathbf{R}_i^{\top}\mathbf{F}_t \mathbf{C}^{\top} diag \left(\mathbf{W}_{t,i}\odot\mathbf{W}_{t,j}\right) \mathbf{E}_{t,j}/\tilde{q}_{R,ij}\right]\\
&+\frac{1}{abT}V_{R,abT}^{-1}\left[\sum_{t=1}^T\sum_{i=1}^a \widehat{\mathbf{R}}_i \mathbf{E}_{t,i}^{\top} diag\left(\mathbf{W}_{t,i}\odot \mathbf{W}_{t,j}\right)\mathbf{C} \mathbf{F}_t^{\top}\mathbf{R}_j/ \tilde{q}_{R,ij}\right]\\
&+\frac{1}{abT}V_{R,abT}^{-1}\left[\sum_{t=1}^T\sum_{i=1}^a \widehat{\mathbf{R}}_i \mathbf{E}_{t,i}^{\top} diag\left(\mathbf{W}_{t,i}\odot \mathbf{W}_{t,j}\right)\mathbf{E}_{t,j}/ \tilde{q}_{R,ij} \right]\\
&=\frac{1}{a}V_{R,abT}^{-1}\left[\sum_{i=1}^a \widehat{\mathbf{R}}_i \mathbf{R}_i^{\top} \frac{1}{|Q_{R,ij}|} \sum_{t,m \in Q_{R,ij}}\mathbf{F}_t \mathbf{C}_m e_{t,jm}\right]\\
&+\frac{1}{a}V_{R,abT}^{-1}\left[\sum_{i=1}^a \widehat{\mathbf{R}}_i\frac{1}{|Q_{R,ij}|} \sum_{t,m \in Q_{R,ij}}e_{t,im}\mathbf{C}_m^{\top}\mathbf{F}_t^{\top}\mathbf{R}_j\right] \\
&+\frac{1}{a}V_{R,abT}^{-1}\left[\sum_{i=1}^a \widehat{\mathbf{R}}_i\frac{1}{|Q_{R,ij}|} \sum_{t,m \in Q_{R,ij}}e_{t,im}e_{t,jm}\right]\\
&=V_{R,abT}^{-1}\left[\frac{1}{a}\sum_{i=1}^a \widehat{\mathbf{R}}_i \mathbf{R}_i^{\top}\frac{1}{|Q_{R,ij}|}\sum_{t,m \in Q_{R,ij}} \mathbf{F}_t \mathbf{C}_m e_{t,jm}\right]\\
&+V_{R,abT}^{-1}\left[\frac{1}{a}\sum_{i=1}^a \widehat{\mathbf{R}}_i\frac{1}{|Q_{R,ij}|} \sum_{t,m \in Q_{R,ij}}e_{t,im}\mathbf{C}_m^{\top}\mathbf{F}_t^{\top}\mathbf{R}_j\right]\\
&+V_{R,abT}^{-1}\left[\frac{1}{a}\sum_{i=1}^a \widehat{\mathbf{R}}_i\frac{1}{|Q_{R,ij}|} \sum_{t,m \in Q_{R,ij}}e_{t,im}e_{t,jm}\right]
\end{split}
\end{equation*}
Let
\begin{align*}
&\gamma\left(i,j\right)=\frac{1}{|Q_{R,ij}|}\sum_{t,m \in Q_{R,ij}}\mathbb{E}\left(e_{t,im}e_{t,jm}\right),\\
&\zeta_{ij}=\frac{1}{|Q_{R,ij}|}\sum_{t,m \in Q_{R,ij}}e_{t,im}e_{t,jm}-\gamma\left(i,j\right),\\ &\eta_{ij}=\frac{1}{|Q_{R,ij}|}\mathbf{R}_i^{\top}\sum_{t,m \in Q_{R,ij}}\mathbf{F}_t \mathbf{C}_m e_{t,jm},\\
&\xi_{ij}=\frac{1}{|Q_{R,ij}|}\sum_{t,m \in Q_{R,ij}} \mathbf{R}_j^{\top} \mathbf{F}_t \mathbf{C}_m e_{t,im}.
\end{align*}
Then we can get:
\begin{equation*}
\widehat{\mathbf{R}}_j-\mathbf{H}_j\mathbf{R}_j=V_{R,abT}^{-1}\left[\frac{1}{a}\sum_{i=1}^a \widehat{\mathbf{R}}_i\gamma\left(i,j\right)+\frac{1}{a}\sum_{i=1}^a \widehat{\mathbf{R}}_i\zeta_{ij} +\frac{1}{a}\sum_{i=1}^a \widehat{\mathbf{R}}_i \eta_{ij}   +\frac{1}{a}\sum_{i=1}^a \widehat{\mathbf{R}}_i \xi_{ij}\right]
\end{equation*}
\begin{lemma}\label{Le:1}
Under Assumptions \ref{As:2}, \ref{As:3},  and \ref{As:4}, we have for some $M<\infty$, and for all $a, b,T$:\\
1. $\sum_{i=1}^a \gamma\left(i,j\right)^2 \leq M$ and $\frac{1}{a}\sum_{i=1}^a \sum_{j=1}^b\gamma\left(i,j\right)^2 \leq M$;\\
2. $\mathbb{E}\left[\left(\frac{1}{\sqrt{Q_{R,ij}}}\mathbf{R}_i^{\top}\sum_{t,m \in Q_{R,ij}}\mathbf{F}_t \mathbf{C}_m e_{t,jm} \right)^2 \right]\leq M$.
\end{lemma}
\begin{proof}
    1. Let
\begin{align*}
\rho\left(i,j\right)&=\gamma\left(i,j\right)/\left[\left(\frac{1}{|Q_{R,ij}|}\sum_{t,m \in Q_{R,ij}}\mathbb{E}[e_{t,im}^2]\right)\left(\frac{1}{|Q_{R,ij}|}\sum_{t,m \in Q_{R,ij}}\mathbb{E}[e_{t,jm}^2]\right)\right]^{1/2}\\
&=\gamma\left(i,j\right)/\left[\gamma\left(i,i\right)\gamma\left(j,j\right)\right]^{1/2}
\end{align*}
So, $\lvert \rho\left(i,j\right) \rvert\leq 1$, and $\rho\left(i,j\right)^2 \leq \lvert \rho\left(i,j\right) \rvert$. \\
From Assumption \ref{As:3}, $\lvert\gamma\left(i,i\right) \rvert\leq M$,  $\lvert\gamma\left(j,j\right) \rvert\leq M$. We then have for all $i$ and $j$,
\begin{align*}
\gamma\left(i,j\right)^2= \gamma\left(i,i\right) \gamma\left(j,j\right)\rho\left(i,j\right)^2\leq M \lvert\gamma\left(i,i\right)\gamma\left(j,j\right) \rvert^{1/2}\lvert \rho\left(i,j\right) \rvert= M \lvert\gamma\left(i,j\right) \rvert.
\end{align*}
Then,
\begin{align*}
&\sum_{i=1}^a \gamma\left(i,j\right)^2 \leq M\sum_{i=1}^a\lvert\gamma\left(i,j\right) \rvert\leq M^2\\
&\frac{1}{a}\sum_{i=1}^a \sum_{j=1}^a\gamma\left(i,j\right)^2 \leq \frac{M}{a}\sum_{i=1}^a \sum_{j=1}^a\lvert\gamma\left(i,j\right) \rvert \leq M^2
\end{align*}
2. From Assumptions \ref{As:2} and \ref{As:4}
\begin{align*}
&\mathbb{E}\left[\left(\frac{1}{\sqrt{|Q_{R,ij}|}}\mathbf{R}_i^{\top}\sum_{t,m \in Q_{R,ij}}\mathbf{F}_t \mathbf{C}_m e_{t,jm} \right)^2 \right]\\
&\leq \mathbb{E}\left[\left\Vert\mathbf{R}_i\right\Vert^2\right]\mathbb{E}\left\Vert\frac{1}{\sqrt{|Q_{R,ij}|}}\sum_{t,m \in Q_{R,ij}}\mathbf{F}_t \mathbf{C}_m e_{t,jm}\right\Vert^2\leq \bar{\mathbf{R}} M
\end{align*}
\end{proof}
\begin{lemma}\label{Le:2}
Under Assumptions \ref{As:C1}, \ref{As:2}, and \ref{As:4}, let $\delta_{a,bT}=min(a,bT)$, we have
\begin{equation*}
\delta_{a,bT}\left(\frac{1}{a}\sum_{j=1}^a \left\Vert \widehat{\mathbf{R}}_j- \mathbf{H}_j \mathbf{R}_j\right\Vert^2\right)=\mathcal{O}_p\left(1\right)
\end{equation*}
\end{lemma}
\begin{proof}

From the Cauchy-Schwartz inequality, we have $\left\Vert \widehat{\mathbf{R}}_j- \mathbf{H}_j \mathbf{R}_j\right\Vert^2 \leq 4\left\Vert V_{R,abT}^{-1} \right\Vert^2\left(e_j+b_j+c_j+d_j\right)$, where
\begin{align*}
&e_j=\frac{1}{a^2}\left\Vert \sum_{i=1}^a \widehat{\mathbf{R}}_i \gamma\left(i,j\right) \right\Vert^2 \\
&b_j=\frac{1}{a^2}\left\Vert \sum_{i=1}^a \widehat{\mathbf{R}}_i \zeta_{ij} \right\Vert^2  \\
&c_j=\frac{1}{a^2}\left\Vert \sum_{i=1}^a \widehat{\mathbf{R}}_i \eta_{ij} \right\Vert^2  \\
&d_j=\frac{1}{a^2}\left\Vert \sum_{i=1}^a \widehat{\mathbf{R}}_i \xi_{ij} \right\Vert^2
\end{align*}
(1) From $\frac{1}{a^2}\left\Vert \sum_{i=1}^a \widehat{\mathbf{R}}_i \gamma\left(i,j\right) \right\Vert^2 \leq \left(\frac{1}{a}\sum_{i=1}^a \left\Vert \widehat{\mathbf{R}}_i \right\Vert^2 \right)\left(\frac{1}{a}\sum_{i=1}^a \gamma\left(i,j\right)^2\right)$, then
\begin{equation*}
\frac{1}{a}\sum_{j=1}^a e_j\leq \frac{1}{a} \left(\frac{1}{a}\sum_{i=1}^a \left\Vert \widehat{\mathbf{R}}_i \right\Vert^2\right)\left(\frac{1}{a}\sum_{j=1}^a\sum_{i=1}^a \gamma\left(i,j\right)^2\right)=\mathcal{O}_p(\frac{1}{a})
\end{equation*}
where $\frac{1}{a}\sum_{i=1}^a \left\Vert \widehat{\mathbf{R}}_i \right\Vert^2 =\mathcal{O}_p\left(1\right)$ which follows from $\frac{1}{a}\widehat{\mathbf{R}}^{\top}\widehat{\mathbf{R}}=\mathbf{I}_r$.\\
(2)
\begin{align*}
\frac{1}{a}\sum_{j=1}^a b_j&= \frac{1}{a}\sum_{j=1}^a \frac{1}{a^2}\left\Vert \sum_{i=1}^a \widehat{\mathbf{R}}_i \zeta_{ij} \right\Vert^2\\
&=\frac{1}{a^3}\sum_{j=1}^a\sum_{i=1}^a\sum_{l=1}^a\left\Vert \widehat{\mathbf{R}}_i^{\top} \widehat{\mathbf{R}}_l \right\Vert \zeta_{ij}\zeta_{lj}\\
&\leq \frac{1}{a^3}\left(\sum_{i=1}^a\sum_{l=1}^a \left\Vert\widehat{\mathbf{R}}_i^{\top} \widehat{\mathbf{R}}_l \right\Vert^2\right)^{1/2}\left[\sum_{i=1}^a\sum_{l=1}^a\left(\sum_{j=1}^a \zeta_{ij} \zeta_{lj}\right)^2\right]^{1/2}\\
&\leq \frac{1}{a^3}\sum_{i=1}^a \left\Vert \widehat{\mathbf{R}}_i \right\Vert^2\left[\sum_{i=1}^a\sum_{l=1}^a\left(\sum_{j=1}^a \zeta_{ij} \zeta_{lj}\right)^2\right]^{1/2}\\
&=\left(\frac{1}{a}\sum_{i=1}^a \left\Vert \widehat{\mathbf{R}}_i \right\Vert^2 \right)\frac{1}{a^2}\left[\sum_{i=1}^a\sum_{l=1}^a\left(\sum_{j=1}^a \zeta_{ij} \zeta_{lj}\right)^2 \right]^{1/2}
\end{align*}
From Assumption \ref{As:3},
$\mathbb{E}\left[\sum_{j=1}^a \zeta_{ij}\zeta_{lj}\right]^2\leq a^2 max_{i,j} E|\zeta_{ij}|^4\\
\mathbb{E}|\zeta_{ij}|^4=\frac{1}{|Q_{R,ij}|^2}\mathbb{E}\lvert \frac{1}{|Q_{R,ij}|^{1/2}}\sum_{t,m\in Q_{R,ij} }\left(e_{t,im}e_{t,jm}-\mathbb{E}\left[e_{t,im}e_{t,jm}\right]\right)\rvert^4\leq \frac{M}{|Q_{R,ij}|^2}=\mathcal{O}_p\left(\frac{1}{b^2T^2}\right)$\\
$\therefore  \frac{1}{a}\sum_{j=1}^a b_j=\mathcal{O}_p\left(\frac{1}{bT}\right)$.\\
(3)\begin{align*}
c_j&=\frac{1}{a^2}\left\Vert \sum_{i=1}^a \widehat{\mathbf{R}}_i \eta_{ij}\right\Vert^2=\frac{1}{a^2}\left\Vert \sum_{i=1}^a \widehat{\mathbf{R}}_i\mathbf{R}_i^{\top} \left(\frac{1}{|Q_{R,ij}|}\sum_{t,m \in Q_{R,ij}}\mathbf{F}_t \mathbf{C}_m e_{t,jm}\right)\right\Vert^2\\
&\leq \left(\frac{1}{a}\sum_{i=1}^a\left\Vert \widehat{\mathbf{R}}_i \right\Vert^2\right)\left(\frac{1}{a}\sum_{i=1}^a\frac{1}{|Q_{R,ij}|}\left(\frac{1}{\sqrt{|Q_{R,ij}|}}\mathbf{R}_i^{\top} \sum_{t,m \in Q_{R,ij}}\mathbf{F}_t \mathbf{C}_m e_{t,jm}\right)^2\right)\\
&=\mathcal{O}_p\left(\frac{1}{bT}\right)
\end{align*}
From Lemma \ref{Le:1}.2, we can get:
\begin{align*}
&\frac{1}{a}\sum_{i=1}^a\frac{1}{|Q_{R,ij}|}\left(\frac{1}{\sqrt{|Q_{R,ij}|}}\mathbf{R}_i^{\top} \sum_{t,m \in Q_{R,ij}}\mathbf{F}_t \mathbf{C}_m e_{t,jm}\right)^2\\
&\leq max_i \frac{1}{|Q_{R,ij}|}\frac{1}{a}\sum_{i=1}^a\left(\frac{1}{\sqrt{|Q_{R,ij}|}}\mathbf{R}_i^{\top} \sum_{t,m \in Q_{R,ij}}\mathbf{F}_t \mathbf{C}_m e_{t,jm}\right)^2=\mathcal{O}_p\left(\frac{1}{bT}\right)\mathcal{O}_p\left(1\right).
\end{align*}
$\therefore \frac{1}{a}\sum_{j=1}^a c_j=\mathcal{O}_p\left(\frac{1}{bT}\right)$.\\
(4) Similarly, we can show
\begin{align*}
&d_j=\frac{1}{a^2}\left\Vert \sum_{i=1}^a \widehat{\mathbf{R}}_i \xi_{ij} \right\Vert^2\\
&\frac{1}{a}\sum_{j=1}^a d_j=\mathcal{O}_p\left(\frac{1}{bT}\right)\\
&\left\Vert V_{R,abT}^{-1} \right\Vert=\mathcal{O}_p(1)\\
\therefore \frac{1}{a}\sum_{j=1}^a \left\Vert \widehat{\mathbf{R}}_j - \mathbf{H}_j \mathbf{R}_j \right\Vert^2 &\leq 4 \left\Vert V_{R,abT}^{-1} \right\Vert^2 \frac{1}{a}\sum_{j=1}^a\left(e_j+b_j+c_j+d_j\right)\\
&=\mathcal{O}_p\left(\frac{1}{a}\right)+ \mathcal{O}_p\left(\frac{1}{bT}\right)=\mathcal{O}_p\left(\frac{1}{\delta_{a,bT}}\right)
\end{align*}
\end{proof}
\begin{lemma}\label{Le:3}
Under Assumptions \ref{As:C1}, and \ref{As:3}, let $\delta_{a,bT}=min\left(a,bT\right)$, then we have $\mathbf{H}_j=\mathcal{O}_p(1), \mathbf{H}_R=\mathcal{O}_p(1)$ and for all $j$, $\left\Vert\mathbf{H}_j - \mathbf{H}_R^{\top}\right\Vert=\mathcal{O}_p\left(1/\sqrt{\delta_{a,bT}}\right)$.
\end{lemma}
\begin{proof}
\begin{align*}
&\left\Vert\mathbf{H}_j - \mathbf{H}_R^{\top}\right\Vert = \Vert \frac{1}{a}V_{R,abT}^{-1}\sum_{i=1}^a \widehat{\mathbf{R}}_i \mathbf{R}_i^{\top} \frac{1}{|Q_{R,ij}|}\sum_{t,m \in Q_{R,ij}}\mathbf{F}_t \mathbf{C}_m \mathbf{C}_m^{\top} \mathbf{F}_t^{\top}\\
&- \frac{1}{abT}V_{R,abT}^{-1} \sum_{i=1}^a \widehat{\mathbf{R}}_i \mathbf{R}_i^{\top}\sum_{t=1}^T\sum_{m=1}^b\mathbf{F}_t \mathbf{C}_m \mathbf{C}_m^{\top}\mathbf{F}_t^{\top} \Vert \\
&=\left\Vert \frac{1}{a}V_{R,abT}^{-1}\sum_{i=1}^a \widehat{\mathbf{R}}_i \mathbf{R}_i^{\top}\left(\frac{1}{|Q_{R,ij}|}\sum_{t,m \in Q_{R,ij}}\mathbf{F}_t \mathbf{C}_m \mathbf{C}_m^{\top} \mathbf{F}_t^{\top} - \frac{1}{bT}\sum_{t=1}^T\sum_{m=1}^b\mathbf{F}_t \mathbf{C}_m \mathbf{C}_m^{\top}\mathbf{F}_t^{\top}\right) \right\Vert \\
&\leq \left\Vert V_{R,abT}^{-1} \right\Vert \left[\frac{1}{a}\sum_{i=1}^a \left\Vert \widehat{\mathbf{R}}_i \right\Vert  \left\Vert \mathbf{R}_i \right\Vert \left\Vert \frac{1}{|Q_{R,ij}|} \sum_{t,m \in Q_{R,ij}} \mathbf{F}_t \mathbf{C}_m \mathbf{C}_m^{\top} \mathbf{F}_t^{\top}- \frac{1}{bT}\sum_{t=1}^T\sum_{m=1}^b\mathbf{F}_t \mathbf{C}_m \mathbf{C}_m^{\top}\mathbf{F}_t^{\top}\right\Vert\right]\\
&\leq \left\Vert V_{R,abT}^{-1} \right\Vert \left[\frac{1}{a}\sum_{i=1}^a \left\Vert\widehat{\mathbf{R}}_i \right\Vert^2\right]^{1/2}\\
&\cdot\left[\frac{1}{a}\sum_{i=1}^a\left\Vert\mathbf{R}_i\right\Vert^2 \left\Vert \frac{1}{|Q_{R,ij}|}\sum_{t,m \in Q_{R,ij}}\mathbf{F}_t \mathbf{C}_m \mathbf{C}_m^{\top} \mathbf{F}_t^{\top}- \frac{1}{bT}\sum_{t=1}^T\sum_{m=1}^b\mathbf{F}_t \mathbf{C}_m \mathbf{C}_m^{\top}\mathbf{F}_t^{\top}\right\Vert^2\right]^{1/2}
\end{align*}
Let $\triangle \triangleq \frac{1}{a}\sum_{i=1}^a\left\Vert\mathbf{R}_i\right\Vert^2 \left\Vert \frac{1}{|Q_{R,ij}|}\sum_{t,m \in Q_{R,ij}}\mathbf{F}_t \mathbf{C}_m \mathbf{C}_m^{\top} \mathbf{F}_t^{\top}- \frac{1}{bT}\sum_{t=1}^T\sum_{m=1}^b\mathbf{F}_t \mathbf{C}_m \mathbf{C}_m^{\top}\mathbf{F}_t^{\top}\right\Vert^2$,  as $\mathbf{R}_i$ is independent of $\mathbf{F}_t, \mathbf{C}$, from Assumption \ref{As:1}, we have
\begin{align*}
\mathbb{E}\left[\triangle\right]&=\frac{1}{a}\sum_{i=1}^a \mathbb{E}\left\Vert \mathbf{R}_i \right\Vert^2 \mathbb{E} \left\Vert \frac{1}{|Q_{R,ij}|}\sum_{t,m \in Q_{R,ij}}\mathbf{F}_t \mathbf{C}_m \mathbf{C}_m^{\top} \mathbf{F}_t^{\top}- \frac{1}{bT}\sum_{t=1}^T\sum_{m=1}^b\mathbf{F}_t \mathbf{C}_m \mathbf{C}_m^{\top}\mathbf{F}_t^{\top} \right\Vert^2 \\
&\leq max \frac{1}{|Q_{R,ij}|} \frac{M}{a}\sum_{i=1}^a \mathbb{E}\left\Vert \mathbf{R}_i \right\Vert^2=\mathcal{O}_p\left(\frac{1}{bT}\right)
\end{align*}
where
\begin{align*}
&\mathbb{E}\left\Vert \frac{1}{|Q_{R,ij}|}\sum_{t,m \in Q_{R,ij}}\mathbf{F}_t \mathbf{C}_m \mathbf{C}_m^{\top} \mathbf{F}_t^{\top}- \frac{1}{bT}\sum_{t=1}^T\sum_{m=1}^b\mathbf{F}_t \mathbf{C}_m \mathbf{C}_m^{\top}\mathbf{F}_t^{\top} \right\Vert^2\\
&=\mathbb{E}\left\Vert \left(\frac{1}{|Q_{R,ij}|}\sum_{t,m \in Q_{R,ij}}\mathbf{F}_t \mathbf{C}_m \mathbf{C}_m^{\top} \mathbf{F}_t^{\top}-\Sigma_{FC}\right)-\left(\frac{1}{bT}\sum_{t=1}^T\sum_{m=1}^b\mathbf{F}_t \mathbf{C}_m \mathbf{C}_m^{\top}\mathbf{F}_t^{\top}-\Sigma_{FC}\right) \right\Vert^2\\
&\leq \mathbb{E} \left\Vert \frac{1}{|Q_{R,ij}|} \sum_{t,m \in Q_{R,ij}}\mathbf{F}_t \mathbf{C}_m \mathbf{C}_m^{\top} \mathbf{F}_t^{\top}-\Sigma_{FC}\right\Vert^2+\mathbb{E} \left\Vert \frac{1}{bT}\sum_{t=1}^T\sum_{m=1}^b\mathbf{F}_t \mathbf{C}_m \mathbf{C}_m^{\top}\mathbf{F}_t^{\top}-\Sigma_{FC} \right\Vert^2\\
&=\frac{M}{|Q_{R,ij}|}+\frac{M}{bT}=\mathcal{O}_p\left(\frac{1}{bT}\right)
\end{align*}
$\therefore \left\Vert\mathbf{H}_j - \mathbf{H}_R^{\top}\right\Vert=\mathcal{O}_p\left(1/\sqrt{\delta_{a,bT}}\right)$
\begin{equation*}
\left\Vert \mathbf{H}_j\right\Vert^2\leq \left\Vert V_{R,abT}^{-1} \right\Vert^2\left(\frac{1}{a}\sum_{i=1}^a \left\Vert \widehat{\mathbf{R}}_i \right\Vert^2\ \right)\left(\frac{1}{a}\sum_{i=1}^a \left\Vert \mathbf{R}_i^{\top}\frac{1}{|Q_{R,ij}|}\sum_{t,m \in Q_{R,ij}}\mathbf{F}_t \mathbf{C}_m \mathbf{C}_m^{\top}\mathbf{F}_t^{\top}\right\Vert^2\right)
\end{equation*}
Let $\triangle_1\triangleq\frac{1}{a}\sum_{i=1}^a \left\Vert \mathbf{R}_i^{\top}\frac{1}{|Q_{R,ij}|}\sum_{t,m \in Q_{R,ij}}\mathbf{F}_t \mathbf{C}_m \mathbf{C}_m^{\top}\mathbf{F}_t^{\top}\right\Vert^2$, then from Assumption \ref{As:1}\\
\begin{align*}
\mathbb{E}\left[\triangle_1\right]&=\mathbb{E}\left[\frac{1}{a}\sum_{i=1}^a \left\Vert \mathbf{R}_i^{\top}\frac{1}{|Q_{R,ij}|}\sum_{t,m \in Q_{R,ij}}\mathbf{F}_t \mathbf{C}_m \mathbf{C}_m^{\top}\mathbf{F}_t^{\top}\right\Vert^2\right]\\
&=\frac{1}{a}\sum_{i=1}^a\mathbb{E}\left[\left\Vert \mathbf{R}_i\right\Vert^2 \right]\mathbb{E}\left\Vert \frac{1}{|Q_{R,ij}|}\sum_{t,m \in Q_{R,ij}}\mathbf{F}_t \mathbf{C}_m \mathbf{C}_m^{\top}\mathbf{F}_t^{\top}  \right\Vert^2\\
&\leq \bar{R}\bar{F}_C
\end{align*}
$\therefore \triangle_1=\mathcal{O}_p(1), \left\Vert\mathbf{H}_j\right\Vert =\mathcal{O}_p(1), \left\Vert\mathbf{H}\right\Vert=\left\Vert\mathbf{H}_j-\mathcal{O}_p\left(\frac{1}{bT}\right)\right\Vert=\mathcal{O}_p(1)$.
\end{proof}
\textit{Proof of Theorem \ref{Th1}}
\begin{proof}
\begin{align*}
\frac{1}{a}\sum_{j=1}^a \Vert \widehat{\mathbf{R}}_j-\mathbf{H}_R^
{\top}\mathbf{R}_j \Vert^2 \leq \frac{1}{a}\sum_{j=1}^a\Vert \widehat{\mathbf{R}}_j-\mathbf{H}_j\mathbf{R}_j \Vert^2+\frac{1}{a}\sum_{j=1}^a\Vert \left(\mathbf{H}_j-\mathbf{H}_R^
{\top}\right)\mathbf{R}_j \Vert^2
\end{align*}
From Lemma \ref{Le:2}, the first term satisfies:
\begin{align*}
&\frac{1}{a}\sum_{j=1}^a\left\Vert \widehat{\mathbf{R}}_j-\mathbf{H}_j\mathbf{R}_j \right\Vert^2=\mathcal{O}_p\left(\frac{1}{\delta_{a,bT}}\right)
\end{align*}
For the second term:
\small{
\begin{align*}
&\frac{1}{a}\sum_{j=1}^a\left\Vert \left(\mathbf{H}_j-\mathbf{H}_R^
{\top}\right)\mathbf{R}_j \right\Vert^2\\
&=\frac{1}{a}\sum_{j=1}^a \left\Vert \frac{1}{a}V_{R,abT}^{-1} \sum_{i=1}^a \widehat{\mathbf{R}}_i\mathbf{R}_i^{\top}\left(\frac{1}{|Q_{R,ij}|}\sum_{t,m \in Q_{R,ij}}\mathbf{F}_t \mathbf{C}_m \mathbf{C}_m^{\top} \mathbf{F}_t^{\top}-\frac{1}{bT}\sum_{t=1}^T\sum_{m=1}^b\mathbf{F}_t \mathbf{C}_m \mathbf{C}_m^{\top}\mathbf{F}_t^{\top}\right) \mathbf{R}_j\right\Vert^2\\
&\leq \left\Vert V_{R,abT}^{-1}\right\Vert^2 \left[\frac{1}{a}\sum_{i=1}^a\left\Vert \widehat{\mathbf{R}}_i \right\Vert^2\right]\\
 &\cdot\left[\frac{1}{a^2}\sum_{i=1}^a \sum_{j=1}^a \left\Vert \mathbf{R}_i\right\Vert^2 \left\Vert \mathbf{R}_j\right\Vert^2 \left\Vert \frac{1}{|Q_{R,ij}|}\sum_{t,m \in Q_{R,ij}}\mathbf{F}_t \mathbf{C}_m \mathbf{C}_m^{\top} \mathbf{F}_t^{\top}- \frac{1}{bT}\sum_{t=1}^T\sum_{m=1}^b\mathbf{F}_t \mathbf{C}_m \mathbf{C}_m^{\top}\mathbf{F}_t^{\top}  \right\Vert^2\right]
\end{align*}
}
\small{ Let\\
$\triangle_2 \triangleq \frac{1}{a^2}\sum_{i=1}^a \sum_{j=1}^a \Vert \mathbf{R}_i\Vert^2 \Vert \mathbf{R}_j\Vert^2\Vert \frac{1}{|Q_{R,ij}|}\sum_{t,m \in Q_{R,ij}}\mathbf{F}_t \mathbf{C}_m \mathbf{C}_m^{\top} \mathbf{F}_t^{\top}- \frac{1}{bT}\sum_{t=1}^T\sum_{m=1}^b\mathbf{F}_t \mathbf{C}_m \mathbf{C}_m^{\top}\mathbf{F}_t^{\top}  \Vert^2$}.
From Assumption \ref{As:1}, we can get
\begin{align*}
\mathbb{E}\left[\triangle_2\right]&=\frac{1}{a^2}\sum_{i=1}^a \sum_{j=1}^a \mathbb{E}\left[\left\Vert \mathbf{R}_i \right\Vert^2 \left\Vert \mathbf{R}_j \right\Vert^2\right]\mathbb{E}\left\Vert \frac{1}{|Q_{R,ij}|}\sum_{t,m \in Q_{R,ij}}\mathbf{F}_t \mathbf{C}_m \mathbf{C}_m^{\top} \mathbf{F}_t^{\top}- \frac{1}{bT}\sum_{t=1}^T\sum_{m=1}^b\mathbf{F}_t \mathbf{C}_m \mathbf{C}_m^{\top}\mathbf{F}_t^{\top} \right\Vert^2\\
&\leq  \frac{1}{a^2} \sum_{i=1}^a \sum_{j=1}^a \left(\mathbb{E}\left[\left\Vert \mathbf{R}_i \right\Vert^4\right]\mathbb{E}\left[\left\Vert \mathbf{R}_j \right\Vert^4\right]\right)^{1/2}\left(\frac{M}{|Q_{R,ij}|}+\frac{M}{bT}\right)\\
&=\mathcal{O}_p\left(\frac{1}{bT}\right)\\
&\therefore \frac{1}{a}\sum_{j=1}^a\left\Vert \left(\mathbf{H}_j-\mathbf{H}_R^
{\top}\right)\mathbf{R}_j \right\Vert^2=\mathcal{O}_p\left(\frac{1}{bT}\right)\\
&\therefore \frac{1}{a}\sum_{j=1}^a\left\Vert \widehat{\mathbf{R}}_j-\mathbf{H}_R^
{\top}\mathbf{R}_j \right\Vert^2 = \mathcal{O}_p\left(\frac{1}{\delta_{a,bT}}\right)
\end{align*}
Then, $\frac{1}{a}\left\Vert \widehat{\mathbf{R}} - \mathbf{R} \mathbf{H}_R\right\Vert_{F}^2=\frac{1}{a}\sum_{j=1}^a\left\Vert \widehat{\mathbf{R}}_j-\mathbf{H}_R^
{\top}\mathbf{R}_j \right\Vert^2=\mathcal{O}_p\left(\frac{1}{\delta_{a,bT}}\right)$, and $\frac{1}{a}\left\Vert \widehat{\mathbf{R}} - \mathbf{R} \mathbf{H}_R\right\Vert^2\leq \frac{1}{a}\left\Vert \widehat{\mathbf{R}} - \mathbf{R} \mathbf{H}_R\right\Vert_{F}^2=\mathcal{O}_p
\left(\frac{1}{\delta_{a,bT}}\right
)$.
Similarly, we can get the corresponding results about $\widehat{\mathbf{C}}$.
\end{proof}
\subsection{Proof of Theorem \ref{Th3}}
\begin{proof}
\begin{align*}
\widehat{\mathbf{F}}_t&=\frac{1}{ab}\widehat{\mathbf{R}}^{\top}\widetilde{\mathbf{Y}}_t\widehat{\mathbf{C}}=\frac{1}{ab}\widehat{\mathbf{R}}^{\top}\left(\mathbf{Y}_t\odot \mathbf{W}\right)\widehat{\mathbf{C}}\\
&=\frac{1}{ab}\widehat{\mathbf{R}}^{\top}\left[\left(\mathbf{R}\mathbf{F}_t \mathbf{C}^{\top}\right)\odot \mathbf{W}_t +\mathbf{E}_t\odot \mathbf{W}_t\right]\widehat{\mathbf{C}}\\
&=\frac{1}{ab}\widehat{\mathbf{R}}^{\top}\left[\left(\mathbf{R}-\widehat{\mathbf{R}}\mathbf{H}_R^{-1}+\widehat{\mathbf{R}}\mathbf{H}_R^{-1}\right) \mathbf{F}_t  \left(\mathbf{C}-\widehat{\mathbf{C}}\mathbf{H}_C^{-1}+\widehat{\mathbf{C}}\mathbf{H}_C^{-1}\right)^{\top}\odot \mathbf{W}_t +\mathbf{E}_t\odot \mathbf{W}_t \right]\widehat{\mathbf{C}}\\
&=\frac{1}{ab}\widehat{\mathbf{R}}^{\top}\left[\left(\mathbf{R}-\widehat{\mathbf{R}}\mathbf{H}_R^{-1}\right)\mathbf{F}_t \left(\mathbf{C}-\widehat{\mathbf{C}}\mathbf{H}_C^{-1}\right)^{\top}\right]\odot\mathbf{W}_t\widehat{\mathbf{C}}\\
&+\frac{1}{ab}\widehat{\mathbf{R}}^{\top}\left[\left(\mathbf{R}-\widehat{\mathbf{R}}\mathbf{H}_R^{-1}\right)\mathbf{F}_t \left(\widehat{\mathbf{C}}\mathbf{H}_C^{-1}\right)^{\top} \right]\odot \mathbf{W}_t \widehat{\mathbf{C}}\\
&+\frac{1}{ab}\widehat{\mathbf{R}}^{\top}\left(\widehat{\mathbf{R}}\mathbf{H}_R^{-1}\mathbf{F}_t\left(\mathbf{C}-\widehat{\mathbf{C}}\mathbf{H}_C^{-1}\right)^{\top} \right)\odot \mathbf{W}_t \widehat{\mathbf{C}}\\
&+\frac{1}{ab}\widehat{\mathbf{R}}^{\top}\left(\widehat{\mathbf{R}}\mathbf{H}_R^{-1}\mathbf{F}_t\mathbf{H}_C^{-1\top}\widehat{\mathbf{C}}^{\top}\right)\odot \mathbf{W}_t \widehat{\mathbf{C}}\\
&+\frac{1}{ab}\widehat{\mathbf{R}}^{\top}\mathbf{E}_t\odot \mathbf{W}_t\widehat{\mathbf{C}}
\end{align*}
then,
\begin{align*}
\widehat{\mathbf{F}}_t-\mathbf{H}_R^{-1}\mathbf{F}_t\mathbf{H}_C^{-1\top}&=\frac{1}{ab}\widehat{\mathbf{R}}^{\top}\left[\left(\mathbf{R}-\widehat{\mathbf{R}}\mathbf{H}_R^{-1}\right)\mathbf{F}_t \left(\mathbf{C}-\widehat{\mathbf{C}}\mathbf{H}_C^{-1}\right)^{\top}\right]\odot\mathbf{W}_t\widehat{\mathbf{C}}\\
&+\frac{1}{ab}\widehat{\mathbf{R}}^{\top}\left[\left(\mathbf{R}-\widehat{\mathbf{R}}\mathbf{H}_R^{-1}\right)\mathbf{F}_t \left(\widehat{\mathbf{C}}\mathbf{H}_C^{-1}\right)^{\top} \right]\odot \mathbf{W}_t \widehat{\mathbf{C}}\\
&+\frac{1}{ab}\widehat{\mathbf{R}}^{\top}\left(\widehat{\mathbf{R}}\mathbf{H}_R^{-1}\mathbf{F}_t\left(\mathbf{C}-\widehat{\mathbf{C}}\mathbf{H}_C^{-1}\right)^{\top} \right)\odot \mathbf{W}_t \widehat{\mathbf{C}}\\
&+\frac{1}{ab}\widehat{\mathbf{R}}^{\top}\left(\widehat{\mathbf{R}}\mathbf{H}_R^{-1}\mathbf{F}_t\mathbf{H}_C^{-1\top}\widehat{\mathbf{C}}^{\top}\right)\odot \mathbf{W}_t \widehat{\mathbf{C}}\\
&+\frac{1}{ab}\left(\widehat{\mathbf{R}}-\mathbf{R}\mathbf{H}_R+\mathbf{R}\mathbf{H}_R\right)^{\top}\mathbf{E}_t\odot \mathbf{W}_t\left(\widehat{\mathbf{C}}-\mathbf{C}\mathbf{H}_C+\mathbf{C}\mathbf{H}_C\right)\\
&-\frac{1}{ab}\widehat{\mathbf{R}}^{\top}\widehat{\mathbf{R}}\mathbf{H}_R^{-1}\mathbf{F}_t\mathbf{H}_C^{-1\top}\widehat{\mathbf{C}}^{\top}\widehat{\mathbf{C}}\\
&=I_1+I_2+I_3+I_4+I_5+I_6+I_7+I_8-I_9
\end{align*}
\small{Where $I_1=\frac{1}{ab}\widehat{\mathbf{R}}^{\top}\left[\left(\mathbf{R}-\widehat{\mathbf{R}}\mathbf{H}_R^{-1}\right)\mathbf{F}_t \left(\mathbf{C}-\widehat{\mathbf{C}}\mathbf{H}_C^{-1}\right)^{\top}\right]\odot\mathbf{W}_t\widehat{\mathbf{C}}, \\
I_2=\frac{1}{ab}\widehat{\mathbf{R}}^{\top}\left[\left(\mathbf{R}-\widehat{\mathbf{R}}\mathbf{H}_R^{-1}\right)\mathbf{F}_t \left(\widehat{\mathbf{C}}\mathbf{H}_C^{-1}\right)^{\top} \right]\odot \mathbf{W}_t \widehat{\mathbf{C}}, I_3=\frac{1}{ab}\widehat{\mathbf{R}}^{\top}\left(\widehat{\mathbf{R}}\mathbf{H}_R^{-1}\mathbf{F}_t\left(\mathbf{C}-\widehat{\mathbf{C}}\mathbf{H}_C^{-1}\right)^{\top} \right)\odot \mathbf{W}_t \widehat{\mathbf{C}}, I_4=\frac{1}{ab}\left(\widehat{\mathbf{R}}-\mathbf{R}\mathbf{H}_R\right)^{\top}\mathbf{E}_t\odot \mathbf{W}_t\left(\widehat{\mathbf{C}}-\mathbf{C}\mathbf{H}_C\right), I_5=\frac{1}{ab}\left(\widehat{\mathbf{R}}-\mathbf{R}\mathbf{H}_R\right)^{\top}\mathbf{E}_t\odot \mathbf{W}_t\mathbf{C}\mathbf{H}_C, I_6= \frac{1}{ab}\mathbf{H}_R^{\top}\mathbf{R}^{\top}\mathbf{E}_t\odot \mathbf{W}_t\left(\widehat{\mathbf{C}}-\mathbf{C}\mathbf{H}_C\right),I_7= \frac{1}{ab}\mathbf{H}_R^{\top}\mathbf{R}^{\top}\mathbf{E}_t\odot \mathbf{W}_t \mathbf{C}\mathbf{H}_C, I_8=\frac{1}{ab}\widehat{\mathbf{R}}^{\top}\left(\widehat{\mathbf{R}}\mathbf{H}_R^{-1}\mathbf{F}_t\mathbf{H}_C^{-1\top}\widehat{\mathbf{C}}^{\top}\right)\odot \mathbf{W}_t \widehat{\mathbf{C}},I_9=\frac{1}{ab}\widehat{\mathbf{R}}^{\top}\widehat{\mathbf{R}}\mathbf{H}_R^{-1}\mathbf{F}_t\mathbf{H}_C^{-1\top}\widehat{\mathbf{C}}^{\top}\widehat{\mathbf{C}}$.} We can see that $I_1$ is controlled by $I_2$ and $I_3$, $I_4$ is controlled by $I_5$ and $I_6$, so we just need to study the norms of $I_2$, $I_3$, $I_5$, $I_6$, $I_7$ and $I_8-I_9$.

\begin{align*}
\left\Vert I_2\right\Vert &= \frac{1}{ab}\left\Vert \widehat{\mathbf{R}}^{\top}\left[\left(\mathbf{R}-\widehat{\mathbf{R}}\mathbf{H}_R^{-1}\right)\mathbf{F}_t \left(\widehat{\mathbf{C}}\mathbf{H}_C^{-1}\right)^{\top} \right]\odot \mathbf{W}_t \widehat{\mathbf{C}}\right\Vert \\
&\leq \frac{1}{ab}\left\Vert \widehat{\mathbf{R}}^{\top} \right\Vert \left\Vert \left[\left(\mathbf{R}-\widehat{\mathbf{R}}\mathbf{H}_R^{-1}\right)\mathbf{F}_t \left(\widehat{\mathbf{C}}\mathbf{H}_C^{-1}\right)^{\top} \right]\odot \mathbf{W}_t \right\Vert \left\Vert \widehat{\mathbf{C}} \right\Vert\\
&\leq \frac{1}{ab}\left\Vert \widehat{\mathbf{R}}^{\top} \right\Vert  \left\Vert \left(\mathbf{R}-\widehat{\mathbf{R}}\mathbf{H}_R^{-1}\right)\mathbf{F}_t \left(\widehat{\mathbf{C}}\mathbf{H}_C^{-1}\right)^{\top}\right\Vert_F \left\Vert \widehat{\mathbf{C}} \right\Vert\\
&\leq \frac{1}{ab}\left\Vert \widehat{\mathbf{R}}^{\top} \right\Vert \left\Vert \mathbf{R}-\widehat{\mathbf{R}}\mathbf{H}_R^{-1} \right\Vert_F \left\Vert \mathbf{F}_t\right\Vert \left\Vert \widehat{\mathbf{C}}^{\top} \right\Vert \left\Vert\mathbf{H}_C^{-1^{\top}} \right\Vert \left\Vert \widehat{\mathbf{C}} \right\Vert\\
&=\frac{1}{a}\left\Vert \widehat{\mathbf{R}}^{\top} \right\Vert \left\Vert \mathbf{R}-\widehat{\mathbf{R}}\mathbf{H}_R^{-1} \right\Vert_F
\end{align*}
$\therefore \left\Vert I_2\right\Vert ^2=\mathcal{O}_p\left(\frac{1}{\delta_{a,bT}}\right)$.
\begin{align*}
\left\Vert I_3\right\Vert &= \left\Vert\frac{1}{ab}\widehat{\mathbf{R}}^{\top}\left(\widehat{\mathbf{R}}\mathbf{H}_R^{-1}\mathbf{F}_t\left(\mathbf{C}-\widehat{\mathbf{C}}\mathbf{H}_C^{-1}\right)^{\top} \right)\odot \mathbf{W}_t \widehat{\mathbf{C}}\right\Vert\\
&\leq \frac{1}{ab} \left\Vert\widehat{\mathbf{R}}^{\top}\right\Vert \left\Vert\mathbf{C} \right\Vert
\left\Vert \widehat{\mathbf{R}}\mathbf{H}_R^{-1}\mathbf{F}_t\left(\mathbf{C}-\widehat{\mathbf{C}}\mathbf{H}_C^{-1}\right)^{\top}\odot \mathbf{W}_t \right\Vert _F\\
&\leq \frac{1}{ab}  \left\Vert\widehat{\mathbf{R}}^{\top}\right\Vert \left\Vert \mathbf{C} \right\Vert \left\Vert \widehat{\mathbf{R}}\mathbf{H}_R^{-1}\mathbf{F}_t\left(\mathbf{C}-\widehat{\mathbf{C}}\mathbf{H}_C^{-1}\right)^{\top} \right\Vert _F\\
&\leq \frac{1}{ab}\left\Vert\widehat{\mathbf{R}}^{\top}\right\Vert \left\Vert \mathbf{C} \right\Vert \left\Vert\widehat{\mathbf{R}}\right\Vert \left\Vert \mathbf{H}_R^{-1} \right\Vert \left\Vert \mathbf{F}_t \right\Vert \left\Vert \mathbf{C}-\widehat{\mathbf{C}}\mathbf{H}_C^{-1} \right\Vert _F\\
&=\frac{1}{b}\left\Vert \mathbf{C} \right\Vert \left\Vert \mathbf{C}-\widehat{\mathbf{C}}\mathbf{H}_C^{-1} \right\Vert _F
\end{align*}
$\therefore \left\Vert I_3\right\Vert^2=\mathcal{O}_p\left(\frac{1}{\delta_{b,aT}}\right)$.
\begin{align*}
\left\Vert I_5\right\Vert &= \frac{1}{ab}\left\Vert\left(\widehat{\mathbf{R}}-\mathbf{R}\mathbf{H}_R\right)^{\top}\mathbf{E}_t\odot \mathbf{W}_t\mathbf{C}\mathbf{H}_C\right\Vert\\
&\leq \frac{1}{ab}\left\Vert\left(\widehat{\mathbf{R}}-\mathbf{R}\mathbf{H}_R\right)^{\top}\right\Vert \left\Vert\mathbf{E}_t \right\Vert_F \left\Vert \mathbf{C}\mathbf{H}_C \right\Vert\\
&= \frac{1}{ab}\left\Vert\left(\widehat{\mathbf{R}}-\mathbf{R}\mathbf{H}_R\right)^{\top}\right\Vert
\end{align*}
$\therefore \Vert I_5\Vert^2=\frac{1}{ab}\mathcal{O}_p\left(\frac{1}{\delta_{a,bT}}\right)$.\\
\begin{align*}
\left\Vert I_6\right\Vert &= \frac{1}{ab}\left\Vert \mathbf{H}_R^{\top} \mathbf{R}^{\top}\mathbf{E}_t\odot \mathbf{W}_t\left(\widehat{\mathbf{C}}-\mathbf{C}\mathbf{H}_C\right)\right\Vert\\
&\leq \frac{1}{ab}\left\Vert \mathbf{R}^{\top} \right\Vert \left\Vert \mathbf{E}_t \right\Vert_F \left\Vert \widehat{\mathbf{C}}-\mathbf{C}\mathbf{H}_C \right\Vert\\
&\leq \frac{1}{ab}\left\Vert \mathbf{R}^{\top} \right\Vert\left\Vert \widehat{\mathbf{C}}-\mathbf{C}\mathbf{H}_C \right\Vert
\end{align*}
$\therefore \left\Vert I_6\right\Vert^2=\frac{1}{ab}\mathcal{O}_p\left(\frac{1}{\delta_{b,aT}}\right)$.\\
\begin{align*}
\left\Vert I_7 \right\Vert &= \frac{1}{ab}\left\Vert \mathbf{H}_R^{\top} \mathbf{R}^{\top}\mathbf{E}_t\odot \mathbf{W}_t\mathbf{C}\mathbf{H}_C\right\Vert\\
&\leq \frac{1}{ab}\left\Vert \mathbf{H}_R^{\top} \right\Vert \left\Vert\mathbf{R}^{\top} \right\Vert \left\Vert\mathbf{E}_t \right\Vert_F \left\Vert\mathbf{C} \right\Vert \left\Vert\mathbf{H}_C \right\Vert
\end{align*}
$\therefore \left\Vert I_7\right\Vert^2=\frac{1}{ab}\mathcal{O}_p(1)=\mathcal{O}_p\left(\frac{1}{ab}\right)$.\\
\begin{align*}
\left\Vert I_8-I_9 \right\Vert &=\frac{1}{ab}\left\Vert \widehat{\mathbf{R}}^{\top}\left[\widehat{\mathbf{R}}\mathbf{H}_R^{-1}\mathbf{F}_t\left(\widehat{\mathbf{C}}\mathbf{H}_C^{-1}\right)^{\top} \right]\odot \mathbf{W}_t\widehat{\mathbf{C}}- \widehat{\mathbf{R}}^{\top}\widehat{\mathbf{R}}\mathbf{H}_R^{-1}\mathbf{F}_t\left(\widehat{\mathbf{C}}\mathbf{H}_C^{-1}\right)^{\top}\widehat{\mathbf{C}}\right\Vert\\
&\leq\frac{1}{ab}\left\Vert\widehat{\mathbf{R}}^{\top}\right\Vert \left\Vert \widehat{\mathbf{C}} \right\Vert  \left\Vert \widehat{\mathbf{R}}\mathbf{H}_R^{-1}\mathbf{F}_t\left(\widehat{\mathbf{C}}\mathbf{H}_C^{-1}\right)^{\top}\right\Vert_F\\
&=\mathcal{O}_p\left( \frac{1}{ab}\right)
\end{align*}
\begin{align*}
&\therefore \left\Vert \widehat{\mathbf{F}}_t-\mathbf{H}_R^{-1}\mathbf{F}_t\mathbf{H}_C^{-1\top} \right\Vert = \mathcal{O}_p\left(\frac{1}{min\left(\sqrt{a},\sqrt{bT}\right)}\right)+\mathcal{O}_p\left(\frac{1}{min\left(\sqrt{b},\sqrt{aT}\right)}\right)\\
&+\frac{1}{\sqrt{ab}}\mathcal{O}_p\left(\frac{1}{min\left(\sqrt{a},\sqrt{bT}\right)}\right)+\frac{1}{\sqrt{ab}}\mathcal{O}_p\left(\frac{1}{min\left(\sqrt{b},\sqrt{aT}\right)}\right)+\mathcal{O}_p\left(\frac{1}{\sqrt{ab}}\right)\\
&=\mathcal{O}_p\left(\frac{1}{min\left(\sqrt{a},\sqrt{bT}\right)}\right)+\left(\frac{1}{min\left(\sqrt{b},\sqrt{aT}\right)}\right)+\mathcal{O}_p\left(\frac{1}{\sqrt{ab}}\right)\\
&=\mathcal{O}_p\left(\frac{1}{min\left(\sqrt{a},\sqrt{b}\right)}\right)
\end{align*}
\end{proof}
\subsection{Proof of Theorem \ref{Th4}}
\begin{lemma}\label{Le:4-1}
$V_{R,abT}\xrightarrow{p}D, D=diag(d_1,d_2,\cdots,d_k)$ are the eigenvalues of $\Sigma_R\Sigma_F$.\\
$V_{C,abT}\xrightarrow{p}E, E=diag(e_1,e_2,\cdots,e_r)$ are the eigenvalues of $\Sigma_{F^{\top}}\Sigma_C$
\end{lemma}
\begin{lemma}\label{Le:4-2}
Under Assumptions \ref{As:C1}$\sim$\ref{As:4}, it holds that:\\
1. $\frac{1}{a}\mathbf{R}^{\top}\mathbf{R}\xrightarrow{p} Q^{\top}$, where $Q$ is invertible, $Q=D^{1/2}\bf{\gamma}\Sigma_F^{1/2}$, The diagonal entries of $D=diag(d_1,d_2,\cdots,d_r)$ are the eigenvalues of $\Sigma_F^{1/2}\Sigma_R\Sigma_F^{1/2}$, and $\bf{\gamma}$ is the corresponding eigenvector matrix such that $\bf{\gamma}^{\top}\bf{\gamma}=\mathbf{I}$.\\
$\frac{1}{b}\mathbf{C}^{\top}\mathbf{C}\xrightarrow{p} P^{\top}$, where $P$ is invertible, $P=E^{1/2}\bf{\tau}\Sigma_{F^{\top}}^{1/2}$, The diagonal entries of $E=diag(e_1,e_2,\cdots,e_r)$ are the eigenvalues of $\Sigma_{F^{\top}}^{1/2}\Sigma_C\Sigma_{F^{\top}}^{1/2}$, and $\bf{\tau}$ is the corresponding eigenvector matrix such that $\bf{\tau}^{\top}\bf{\tau}=\mathbf{I}$.\\
2. $\mathbf{H}_R\xrightarrow{p} Q^{-1}, \mathbf{H}_C\xrightarrow{p} P^{-1}$.
\end{lemma}
\begin{proof}
    The proof of Lemmas \ref{Le:4-1} and \ref{Le:4-2} are similar as \cite{XIONG2023271}.
\end{proof}
\begin{lemma}\label{Le:5}
\begin{align*}
&1. \frac{1}{a}\sum_{i=1}^a\widehat{\mathbf{R}}_i \gamma(i,j) =\mathcal{O}_p\left(\frac{1}{\sqrt{a\delta_{a,bT}}}\right) \\
&2. \frac{1}{a}\sum_{i=1}^a\widehat{\mathbf{R}}_i \zeta_{ij}=\mathcal{O}_p\left(\frac{1}{\sqrt{bT\delta_{a,bT}}}\right) \\
&3. \frac{1}{a}\sum_{i=1}^a\widehat{\mathbf{R}}_i \eta_{ij}=\mathcal{O}_p\left(\frac{1}{\sqrt{bT}}\right) \\
&4. \frac{1}{a}\sum_{i=1}^a\widehat{\mathbf{R}}_i \xi_{ij}=\mathcal{O}_p\left(\frac{1}{\sqrt{bT\delta_{a,bT}}}\right)
\end{align*}
\end{lemma}
\begin{proof}
1.
\begin{equation*}
\frac{1}{a}\sum_{i=1}^a\widehat{\mathbf{R}}_i \gamma(i,j) = \frac{1}{a}\sum_{i=1}^a\left(\widehat{\mathbf{R}}_i - \mathbf{H}_R^{\top}\mathbf{R}_i\right)\gamma(i,j)+\frac{1}{a}\sum_{i=1}^a\mathbf{H}_R^{\top}\mathbf{R}_i\gamma(i,j)
\end{equation*}
For the second term $\frac{1}{a}\sum_{i=1}^a\mathbf{H}_R^{\top}\mathbf{R}_i\gamma(i,j)$, we have\\
$\mathbb{E}\left[\left\Vert \sum_{i=1}^a\mathbf{R}_i\gamma(i,j) \right\Vert\right]\leq \sum_{i=1}^a\mathbb{E}\left[\left\Vert \mathbf{R}_i\right\Vert\right]|\gamma(i,j)|\leq \bar{R}\sum_{i=1}^a|\gamma(i,j)|\leq \bar{R}\cdot M$
Together with $\mathbf{H}_R=\mathcal{O}_p\left(1\right)$, we have $\frac{1}{a}\sum_{i=1}^a\mathbf{H}_R^{\top}\mathbf{R}_i\gamma(i,j)=\mathcal{O}_p\left(\frac{1}{a}\right)$.
Next we consider the first term:
\begin{align*}
\left\Vert \frac{1}{a}\sum_{i=1}^a\left(\widehat{\mathbf{R}}_i - \mathbf{H}_R^{\top}\mathbf{R}_i\right)\gamma(i,j) \right\Vert &\leq \left(\frac{1}{a}\sum_{i=1}^a\left\Vert \widehat{\mathbf{R}}_i - \mathbf{H}_R^{\top}\mathbf{R}_i \right\Vert^2\right)^{1/2}\left(\frac{1}{a}\sum_{i=1}^a\gamma(i,j)^2 \right)^{1/2}\\
&\leq \mathcal{O}_p\left(\frac{1}{\sqrt{\delta_{a,bT}}}\right)\cdot \frac{1}{\sqrt{a}}\cdot M\\
&=\mathcal{O}_p\left(\frac{1}{\sqrt{a\delta_{a,bT}}}\right)\\
\end{align*}
\begin{align*}
\therefore \frac{1}{a}\sum_{i=1}^a\widehat{\mathbf{R}}_i \gamma(i,j) &= \frac{1}{a}\sum_{i=1}^a\left(\widehat{\mathbf{R}}_i - \mathbf{H}_R^{\top}\mathbf{R}_i\right)\gamma(i,j)+\frac{1}{a}\sum_{i=1}^a\mathbf{H}_R^{\top}\mathbf{R}_i\gamma(i,j)\\
&=\mathcal{O}_p\left(\frac{1}{\sqrt{a\delta_{a,bT}}}\right)+\mathcal{O}_p\left(\frac{1}{a}\right)=\mathcal{O}_p\left(\frac{1}{\sqrt{a\delta_{a,bT}}}\right)
\end{align*}
2. Let us decompose $\frac{1}{a}\sum_{i=1}^a\widehat{\mathbf{R}}_i \zeta_{ij}=\frac{1}{a}\sum_{i=1}^a\left(\widehat{\mathbf{R}}_i - \mathbf{H}_R^{\top}\mathbf{R}_i\right)\zeta_{ij}+\frac{1}{a}\sum_{i=1}^a\mathbf{H}_R^{\top}\mathbf{R}_i\zeta_{ij}$. First, we consider $\frac{1}{a}\sum_{i=1}^a \zeta_{ij}^2$.
\begin{align*}
\mathbb{E}\left[\frac{1}{a}\sum_{i=1}^a\zeta_{ij}^2\right]&=\mathbb{E}\left[\frac{1}{a}\sum_{i=1}^a\left(\frac{1}{|Q_{R,ij}|}\sum_{t,m\in Q_{R,ij}}e_{t,im}e_{t,jm}-\gamma(i,j)\right)^2\right]\\
&=\frac{1}{a}\sum_{i=1}^a\frac{1}{|Q_{R,ij}|}\mathbb{E}\left[\frac{1}{\sqrt{|Q_{R,ij}|}}\sum_{t,m\in Q_{R,ij}}\left(e_{t,im}e_{t,jm}-\mathbb{E}\left(e_{t,im}e_{t,jm}\right)\right)\right]^2\\
&\leq max_i\frac{1}{|Q_{R,ij}|}\cdot M=O\left(\frac{1}{bT}\right)\\
\therefore \frac{1}{a}\sum_{i=1}^a \zeta_{ij}^2=\mathcal{O}_p\left(\frac{1}{bT}\right)
\end{align*}
For the first term $\frac{1}{a}\sum_{i=1}^a\left(\widehat{\mathbf{R}}_i - \mathbf{H}_R^{\top}\mathbf{R}_i\right)\zeta_{ij}$, we obtain
\begin{align*}
\left\Vert \frac{1}{a}\sum_{i=1}^a\left(\widehat{\mathbf{R}}_i - \mathbf{H}_R^{\top}\mathbf{R}_i\right)\zeta_{ij}\right\Vert & \leq \left(\frac{1}{a}\sum_{i=1}^a\left\Vert \widehat{\mathbf{R}}_i - \mathbf{H}_R^{\top}\mathbf{R}_i \right\Vert^2\right)^{1/2}\left(\frac{1}{a}\sum_{i=1}^a\zeta_{ij}^2\right)^{1/2}\\
&=\mathcal{O}_p\left(\frac{1}{\sqrt{\delta_{a,bT}}}\right) \cdot \mathcal{O}_p\left(\frac{1}{\sqrt{bT}}\right)=\mathcal{O}_p\left(\frac{1}{\sqrt{bT\delta_{a,bT}}}\right)
\end{align*}
For the second term $\frac{1}{a}\sum_{i=1}^a\mathbf{H}_R^{\top}\mathbf{R}_i\zeta_{ij}$, let us consider $\frac{1}{a}\sum_{i=1}^a\mathbf{R}_i\zeta_{ij}$
\begin{align*}
\mathbb{E}\left[\left\Vert\frac{1}{a}\sum_{i=1}^a \mathbf{R}_i\zeta_{ij}\right\Vert^2\right]&=\mathbb{E}\left[\left\Vert\frac{1}{a}\sum_{i=1}^a \mathbf{R}_i \frac{1}{|Q_{R,ij}|}\sum_{t,m\in Q_{R,ij}}\left(e_{t,im}e_{t,jm}-\mathbb{E}\left(e_{t,im}e_{t,jm}\right)\right)\right\Vert^2\right]\\
&=O\left(\frac{1}{abT}\right)\\
\therefore \frac{1}{a}\sum_{i=1}^a\mathbf{H}_R^{\top}\mathbf{R}_i\zeta_{ij}=\mathcal{O}_p\left(\frac{1}{abT}\right)
\end{align*}
$\therefore \frac{1}{a}\sum_{i=1}^a\widehat{\mathbf{R}}_i \zeta_{ij}=\frac{1}{a}\sum_{i=1}^a\left(\widehat{\mathbf{R}}_i - \mathbf{H}_R^{\top}\mathbf{R}_i\right)\zeta_{ij}+\frac{1}{a}\sum_{i=1}^a\mathbf{H}_R^{\top}\mathbf{R}_i\zeta_{ij}=\mathcal{O}_p\left(\frac{1}{\sqrt{bT\delta_{a,bT}}}\right)$\\
3. We show that $\frac{1}{a}\sum_{i=1}^a\widehat{\mathbf{R}}_i \eta_{ij}=\mathcal{O}_p\left(\frac{1}{\sqrt{bT}}\right)$. We decompose $\frac{1}{a}\sum_{i=1}^a\widehat{\mathbf{R}}_i \eta_{ij}=\frac{1}{a}\sum_{i=1}^a\left(\widehat{\mathbf{R}}_i - \mathbf{H}_R^{\top}\mathbf{R}_i\right)\eta_{ij}+\frac{1}{a}\sum_{i=1}^a\mathbf{H}_R^{\top}\mathbf{R}_i\eta_{ij}$. Let us first consider $\frac{1}{a}\sum_{i=1}^a\mathbf{R}_i\eta_{ij}$ in the second term:
\begin{align*}
\left\Vert \frac{1}{a}\sum_{i=1}^a\mathbf{R}_i\eta_{ij}\right\Vert^2 &= \left\Vert\frac{1}{a}\sum_{i=1}^a\mathbf{R}_i \mathbf{R}_i^{\top}\frac{1}{|Q_{R,ij}|}\sum_{t,m\in Q_{R,ij}}\mathbf{F}_t \mathbf{C}_m e_{t,jm}\right\Vert^2\\
&\leq \left(\frac{1}{a}\sum_{i=1}^a\left\Vert \mathbf{R}_i \right\Vert^4 \right)\left(\frac{1}{a}\sum_{i=1}^a\left\Vert \frac{1}{|Q_{R,ij}|}\sum_{t,m\in Q_{R,ij}}\mathbf{F}_t \mathbf{C}_m e_{t,jm} \right\Vert^2\right)\\
&=\mathcal{O}_p\left(1\right)\cdot \mathcal{O}_p\left(\frac{1}{bT}\right)=\mathcal{O}_p\left(\frac{1}{bT}\right)
\end{align*}
where $\frac{1}{a}\sum_{i=1}^a\left\Vert \mathbf{R}_i \right\Vert^4=\mathcal{O}_p\left(1\right)$ follows from $\mathbb{E}\left[\frac{1}{a}\sum_{i=1}^a\left\Vert \mathbf{R}_i \right\Vert^4\right]=\frac{1}{a}\sum_{i=1}^a\mathbb{E}\left[\left\Vert \mathbf{R}_i \right\Vert^4\right]\leq M$, and $\frac{1}{a}\sum_{i=1}^a\left\Vert \frac{1}{|Q_{R,ij}|}\sum_{t,m\in Q_{R,ij}}\mathbf{F}_t \mathbf{C}_m e_{t,jm} \right\Vert^2=\mathcal{O}_p\left(\frac{1}{bT}\right)$ holds because of \small{$\mathbb{E}\left[\frac{1}{a}\sum_{i=1}^a\left\Vert \frac{1}{|Q_{R,ij}|}\sum_{t,m\in Q_{R,ij}}\mathbf{F}_t \mathbf{C}_m e_{t,jm}\right\Vert^2\right]=\frac{1}{a}\sum_{i=1}^a\frac{1}{|Q_{R,ij}|} \mathbb{E} \left[\left\Vert\frac{1}{\sqrt{|Q_{R,ij}|}}\sum_{t,m\in Q_{R,ij}}\mathbf{F}_t \mathbf{C}_m e_{t,jm}\right\Vert^2\right]\\
=O\left(\frac{1}{bT}\right)$}. Since $\mathbf{H}_R=\mathcal{O}_p\left(1\right)$, the second term satisfies $\frac{1}{a}\sum_{i=1}^a\mathbf{H}_R^{\top}\mathbf{R}_i\eta_{ij}=\mathcal{O}_p\left(\frac{1}{\sqrt{bT}}\right)$. Next, we consider the first term:
\begin{align*}
\left\Vert \frac{1}{a}\sum_{i=1}^a\left(\widehat{\mathbf{R}}_i - \mathbf{H}_R^{\top}\mathbf{R}_i\right)\eta_{ij}\right\Vert^2&\leq \left(\frac{1}{a}\sum_{i=1}^a\left\Vert \widehat{\mathbf{R}}_i - \mathbf{H}_R^{\top}\mathbf{R}_i \right\Vert^2\right)\left(\frac{1}{a}\sum_{i=1}^a \eta_{ij}^2\right)\\
&=\mathcal{O}_p\left(\frac{1}{bT\delta_{a,bT}}\right)
\end{align*}
where $\frac{1}{a}\sum_{i=1}^a \eta_{ij}^2=\mathcal{O}_p\left(\frac{1}{bT}\right)$ follows from \\$\mathbb{E}\left[\frac{1}{a}\sum_{i=1}^a \eta_{ij}^2\right]=\frac{1}{a}\sum_{i=1}^a\frac{1}{|Q_{R,ij}|}\mathbb{E}\left[\left(\frac{1}{\sqrt{|Q_{R,ij}|}}\mathbf{R}_i^{\top}\sum_{t,m\in Q_{R,ij}}\mathbf{F}_t \mathbf{C}_m e_{t,jm}\right)^2\right]=O\left(\frac{1}{bT}\right)$. Putting these together we conclude\\
$\frac{1}{a}\sum_{i=1}^a\widehat{\mathbf{R}}_i \eta_{ij}=\mathcal{O}_p\left(\frac{1}{\sqrt{bT\delta_{a,bT}}}\right)+\mathcal{O}_p\left(\frac{1}{\sqrt{bT}}\right)=\mathcal{O}_p\left(\frac{1}{\sqrt{bT}}\right)$.\\
4. Show that $ \frac{1}{a}\sum_{i=1}^a\widehat{\mathbf{R}}_i \xi_{ij}=\mathcal{O}_p\left(\frac{1}{\sqrt{bT\delta_{a,bT}}}\right)$. We start by decomposing $\frac{1}{a}\sum_{i=1}^a\widehat{\mathbf{R}}_i \xi_{ij}=\frac{1}{a}\sum_{i=1}^a\left(\widehat{\mathbf{R}}_i - \mathbf{H}_R^{\top}\mathbf{R}_i\right)\xi_{ij}+\frac{1}{a}\sum_{i=1}^a\mathbf{H}_R^{\top}\mathbf{R}_i\xi_{ij}$. Let us first consider the first term:
\begin{align*}
&\left\Vert \frac{1}{a}\sum_{i=1}^a\left(\widehat{\mathbf{R}}_i - \mathbf{H}_R^{\top}\mathbf{R}_i\right)\xi_{ij}\right\Vert = \left\Vert \frac{1}{a}\sum_{i=1}^a\left(\widehat{\mathbf{R}}_i - \mathbf{H}_R^{\top}\mathbf{R}_i\right)\frac{1}{|Q_{R,ij}|}\sum_{t,m\in Q_{R,ij}}\mathbf{R}_j^{\top}\mathbf{F}_t \mathbf{C}_m e_{t,im}\right\Vert\\
&\leq max_i\frac{1}{\sqrt{|Q_{R,ij}|}}\left(\frac{1}{a}\sum_{i=1}^a \left\Vert \widehat{\mathbf{R}}_i - \mathbf{H}_R^{\top}\mathbf{R}_i \right\Vert^2\right)^{1/2}\left(\frac{1}{a}\sum_{i=1}^a \left\Vert \frac{1}{\sqrt{|Q_{R,ij}|}}\sum_{t,m\in Q_{R,ij}}\mathbf{F}_t \mathbf{C}_m e_{t,im} \right\Vert^2\right)^{1/2}\left\Vert \mathbf{R}_j\right\Vert\\
&=\mathcal{O}_p\left(\frac{1}{\sqrt{bT\delta_{a,bT}}}\right).
\end{align*}
Let us consider the second term $\frac{1}{a}\sum_{i=1}^a\mathbf{H}_R^{\top}\mathbf{R}_i\xi_{ij}$.
\begin{align*}
\left\Vert \frac{1}{a}\sum_{i=1}^a\mathbf{H}_R^{\top}\mathbf{R}_i\xi_{ij}\right\Vert &= \left\Vert\frac{1}{a}\sum_{i=1}^a\mathbf{H}_R^{\top}\mathbf{R}_i\frac{1}{|Q_{R,ij}|}\sum_{t,m\in Q_{R,ij}}\mathbf{R}_j^{\top}\mathbf{F}_t \mathbf{C}_m e_{t,im}  \right\Vert\\
&\leq  \left\Vert \mathbf{H}_R \right\Vert\left(\frac{1}{\sqrt{a}}\right) \left\Vert \frac{1}{\sqrt{a}}\sum_{i=1}^a\frac{1}{|Q_{R,ij}|}\sum_{t,m\in Q_{R,ij}}\mathbf{R}_i\mathbf{C}_m^{\top}\mathbf{F}_t^{\top} e_{t,im} \right\Vert \left\Vert \mathbf{R}_j \right\Vert\\
&=\mathcal{O}_p\left(\frac{1}{\sqrt{abT}}\right)
\end{align*}
$\therefore \frac{1}{a}\sum_{i=1}^a\widehat{\mathbf{R}}_i \xi_{ij}=\mathcal{O}_p\left(\frac{1}{\sqrt{bT\delta_{a,bT}}}\right)+\mathcal{O}_p\left(\frac{1}{\sqrt{abT}}\right)=\mathcal{O}_p\left(\frac{1}{\sqrt{bT\delta_{a,bT}}}\right)$.
\end{proof}
\begin{lemma}\label{Le:6}
Under Assumptions \ref{As:C1} $\sim$ \ref{As:6}, it holds that:\\
$\mathbf{H}_R V_{R,abT}^{-1}\mathbf{H}_R^{-1}=\left(\frac{\mathbf{R}^{\top}\mathbf{R}}{a}\right)^{-1}\left(\frac{1}{bT}\sum_{t=1}^T\mathbf{F}_t\mathbf{C}^{\top}\mathbf{C}\mathbf{F}_t^{\top}\right)^{-1}+\mathcal{O}_p\left(\frac{1}{\sqrt{\delta_{a,bT}}}\right)$
\end{lemma}
\begin{proof}
From the definition of $\mathbf{H}_R=\frac{1}{abT}\sum_{t=1}^T\mathbf{F}_t\mathbf{C}^{\top}\mathbf{C}\mathbf{F}_t^{\top}\mathbf{R}^{\top}\widehat{\mathbf{R}}V_{R,abT}^{-1}$,
\begin{align*}
\mathbf{H}_R V_{R,abT}^{-1}\mathbf{H}_R^{-1}&=\mathbf{H}_R V_{R,abT}^{-1}V_{R,abT}\left(\frac{\mathbf{R}^{\top}\widehat{\mathbf{R}}}{a}\right)^{-1}\left(\frac{1}{bT}\sum_{t=1}^T\mathbf{F}_t\mathbf{C}^{\top}\mathbf{C}\mathbf{F}_t^{\top}\right)^{-1}\\
&=\mathbf{H}_R\mathbf{H}_R^{\top}\left(\frac{1}{bT}\sum_{t=1}^T\mathbf{F}_t\mathbf{C}^{\top}\mathbf{C}\mathbf{F}_t^{\top}\right)^{-1}+\mathcal{O}_p\left(\frac{1}{\sqrt{\delta_{a,bT}}}\right)\\
&=\left(\frac{\mathbf{R}^{\top}\mathbf{R}}{a}\right)^{-1}\left(\frac{1}{bT}\sum_{t=1}^T\mathbf{F}_t\mathbf{C}^{\top}\mathbf{C}\mathbf{F}_t^{\top}\right)^{-1}+\mathcal{O}_p\left(\frac{1}{\sqrt{\delta_{a,bT}}}\right)
\end{align*}
 where the last equality follows from $\mathbf{H}_R\mathbf{H}_R^{\top}=\left(\frac{\mathbf{R}^{\top}\mathbf{R}}{a}\right)^{-1}+\mathcal{O}_p\left(\frac{1}{\sqrt{\delta_{a,bT}}}\right)$
\begin{align*}
\mathbf{H}_R\mathbf{H}_R^{\top}&=\left(\frac{\widehat{\mathbf{R}}^{\top}\mathbf{R}}{a}\right)^{-1}\left(\frac{\mathbf{R}^{\top}\widehat{\mathbf{R}}}{a}\right)^{-1}+\mathcal{O}_p\left(\frac{1}{\sqrt{\delta_{a,bT}}}\right)\\
&=\left(\frac{\mathbf{R}^{\top}\widehat{\mathbf{R}}\widehat{\mathbf{R}}^{\top}\mathbf{R}}{a^2}\right)^{-1}+\mathcal{O}_p\left(\frac{1}{\sqrt{\delta_{a,bT}}}\right)\\
&=\left(\frac{\mathbf{R}^{\top}\mathbf{R}}{a}\right)^{-1}+\mathcal{O}_p\left(\frac{1}{\sqrt{\delta_{a,bT}}}\right)
\end{align*}
\end{proof}
\textit{Proof of Theorem \ref{Th4}}
\begin{proof}
$\sqrt{bT}\left(\widehat{\mathbf{R}}_j-\mathbf{H}_R^{\top}\mathbf{R}_j\right)=\sqrt{bT}\left(\widehat{\mathbf{R}}_j-\mathbf{H}_j\mathbf{R}_j\right)+\sqrt{bT}\left(\mathbf{H}_j-\mathbf{H}_R^{\top}\right)\mathbf{R}_j$.
First we consider $\sqrt{bT}\left(\widehat{\mathbf{R}}_j-\mathbf{H}_j\mathbf{R}_j\right)$, by Lemma \ref{Le:5}, the decomposition of $\widehat{\mathbf{R}}_j-\mathbf{H}_j\mathbf{R}_j$ is :
\begin{align*}
\widehat{\mathbf{R}}_j-\mathbf{H}_j\mathbf{R}_j= V_{R,abT}^{-1}\left[\frac{1}{a}\sum_{i=1}^a\widehat{\mathbf{R}}_i \gamma(i,j)+\frac{1}{a}\sum_{i=1}^a\widehat{\mathbf{R}}_i \zeta_{ij}+\frac{1}{a}\sum_{i=1}^a\widehat{\mathbf{R}}_i \eta_{ij}+\frac{1}{a}\sum_{i=1}^a\widehat{\mathbf{R}}_i \xi_{ij}\right]
\end{align*}
When $\sqrt{bT}/a\rightarrow 0$, the limiting distribution is determined by $\frac{1}{a}\sum_{i=1}^a\widehat{\mathbf{R}}_i \eta_{ij}$, i.e.,
\begin{align*}
\sqrt{bT}\left(\widehat{\mathbf{R}}_j-\mathbf{H}_j\mathbf{R}_j\right)&=\sqrt{bT}V_{R,abT}^{-1}\frac{1}{a}\sum_{i=1}^a\widehat{\mathbf{R}}_i\frac{1}{|Q_{R,ij}|}\mathbf{R}_i^{\top}\sum_{t,m\in Q_{R,ij}}\mathbf{F}_t \mathbf{C}_m e_{t,jm}+o_p\left(1\right)\\
&=\sqrt{bT}V_{R,abT}^{-1}\frac{1}{a}\sum_{i=1}^a\mathbf{H}_R^{\top}\mathbf{R}_i\mathbf{R}_i^{\top}\frac{1}{|Q_{R,ij}|}\sum_{t,m\in Q_{R,ij}}\mathbf{F}_t \mathbf{C}_m e_{t,jm}+o_p\left(1\right)\\
&=V_{R,abT}^{-1}\frac{1}{a}\sum_{i=1}^a\sqrt{\frac{bT}{|Q_{R,ij}|}}\mathbf{H}_R^{\top}\mathbf{R}_i\mathbf{R}_i^{\top}\frac{1}{\sqrt{|Q_{R,ij}|}}\sum_{t,m\in Q_{R,ij}}\mathbf{F}_t \mathbf{C}_m e_{t,jm}+o_p\left(1\right)
\end{align*}
Assumption \ref{As:6} yields $\frac{1}{a}\sum_{i=1}^a\sqrt{\frac{bT}{|Q_{R,ij}|}}\mathbf{R}_i\mathbf{R}_i^{\top}\frac{1}{\sqrt{|Q_{R,ij}|}}\sum_{t,m\in Q_{R,ij}}\mathbf{F}_t \mathbf{C}_m e_{t,jm}\xrightarrow{d}N\left(\mathbf{0},\mathbf{\Gamma}_{R_j}^{obs}\right)$.\\
Lemma \ref{Le:4-2} implies $\mathbf{H}_R\xrightarrow{p}Q^{-1}$, and Lemma \ref{Le:4-1} implies $V_{R,abT}^{-1} \xrightarrow{p}D^{-1}$. Combined with the Slutsky's Theorem, we conclude that\\
$V_{R,abT}^{-1}\frac{\sqrt{bT}}{a}\mathbf{H}_R^{\top}\sum_{i=1}^a\frac{1}{|Q_{R,ij}|}\mathbf{R}_i\mathbf{R}_i^{\top}\sum_{t,m\in Q_{R,ij}}\mathbf{F}_t \mathbf{C}_m e_{t,jm}\xrightarrow{d}N\left(\mathbf{0},D^{-1}Q^{-1}\mathbf{\Gamma}_{R_j}^{obs}\left(Q^{-1}\right)^{\top}D^{-1}\right)$\\
Next, we consider $\sqrt{bT}\left(\mathbf{H}_j-\mathbf{H}_R^{\top}\right)\mathbf{R}_j$. Lemma \ref{Le:3} implies, $\mathbf{H}_j-\mathbf{H}_R^{\top}=\mathcal{O}_p\left(\frac{1}{\sqrt{bT}}\right)$, and therefore $\sqrt{bT}\left(\mathbf{H}_j-\mathbf{H}_R^{\top}\right)\mathbf{R}_j=\mathcal{O}_p\left(1\right)$, recall the definition
\begin{align*}
&\mathbf{H}_R^{\top}=\frac{1}{abT}V_{R,abT}^{-1}\widehat{\mathbf{R}}^{\top}\mathbf{R}\sum_{t=1}^T\mathbf{F}_t \mathbf{C}^{\top}\mathbf{C}\mathbf{F}_t^{\top}=\frac{1}{abT}V_{R,abT}^{-1}\widehat{\mathbf{R}}^{\top}\mathbf{R}\sum_{t=1}^T\sum_{m=1}^b\mathbf{F}_t \mathbf{C}_m\mathbf{C}_m^{\top}\mathbf{F}_t^{\top}\\
&\mathbf{H}_j=\frac{1}{a}V_{R,abT}^{-1}\widehat{\mathbf{R}}^{\top}\mathbf{R}\frac{1}{|Q_{R,ij}|}\sum_{t,m\in Q_{R,ij}}\mathbf{F}_t \mathbf{C}_m\mathbf{C}_m^{\top}\mathbf{F}_t^{\top}\\
&\mathbf{H}_j-\mathbf{H}_R^{\top}=V_{R,abT}^{-1}\frac{1}{a}\sum_{i=1}^a\widehat{\mathbf{R}}_i\mathbf{R}_i^{\top}\left(\frac{1}{|Q_{R,ij}|}\sum_{t,m\in Q_{R,ij}}\mathbf{F}_t \mathbf{C}_m\mathbf{C}_m^{\top}\mathbf{F}_t^{\top}-\frac{1}{bT}\sum_{t=1}^T\sum_{m=1}^b\mathbf{F}_t \mathbf{C}_m\mathbf{C}_m^{\top}\mathbf{F}_t^{\top}\right)\\
&=V_{R,abT}^{-1}\frac{1}{a}\sum_{i=1}^a\mathbf{H}_R^{\top}\mathbf{R}_i\mathbf{R}_i^{\top}\Delta_{F,ij}+V_{R,abT}^{-1}\frac{1}{a}\sum_{i=1}^a\left(\widehat{\mathbf{R}}_i-\mathbf{H}_R^{\top}\mathbf{R}_i\right)\mathbf{R}_i^{\top}\Delta_{F,ij}
\end{align*}
where $\Delta_{F,ij}=\frac{1}{|Q_{R,ij}|}\sum_{t,m\in Q_{R,ij}}\mathbf{F}_t \mathbf{C}_m\mathbf{C}_m^{\top}\mathbf{F}_t^{\top}-\frac{1}{bT}\sum_{t=1}^T\sum_{m=1}^b\mathbf{F}_t \mathbf{C}_m\mathbf{C}_m^{\top}\mathbf{F}_t^{\top}$, let $\Delta_{H,1}=\frac{1}{a}\sum_{i=1}^a\mathbf{H}_R^{\top}\mathbf{R}_i\mathbf{R}_i^{\top}\Delta_{F,ij}, \Delta_{H,2}=\frac{1}{a}\sum_{i=1}^a\left(\widehat{\mathbf{R}}_i-\mathbf{H}_R^{\top}\mathbf{R}_i\right)\mathbf{R}_i^{\top}\Delta_{F,ij}$. For the term $\Delta_{H,2}$,
\begin{align*}
\left\Vert \Delta_{H,2} \right\Vert &=\left\Vert\frac{1}{a}\sum_{i=1}^a\left(\widehat{\mathbf{R}}_i-\mathbf{H}_R^{\top}\mathbf{R}_i\right)\mathbf{R}_i^{\top}\Delta_{F,ij} \right\Vert\\
&\leq \underbrace{\left(\frac{1}{a}\sum_{i=1}^a\left\Vert \widehat{\mathbf{R}}_i-\mathbf{H}_R^{\top}\mathbf{R}_i \right\Vert^2\right)^{1/2}}_{\mathcal{O}_p\left(\frac{1}{\sqrt{\delta_{a,bT}}}\right)}\left(\frac{1}{a}\sum_{i=1}^a\left\Vert \mathbf{R}_i \right\Vert^2 \left\Vert \Delta_{F,ij} \right\Vert^2\right)^{1/2}
\end{align*}
and the second term $\frac{1}{a}\sum_{i=1}^a\left\Vert \mathbf{R}_i \right\Vert^2 \left\Vert \Delta_{F,ij} \right\Vert^2$ satisfies
\begin{align*}
&\mathbb{E}\left[\frac{1}{a}\sum_{i=1}^a\left\Vert \mathbf{R}_i \right\Vert^2 \left\Vert \Delta_{F,ij} \right\Vert^2\right]=\frac{1}{a}\sum_{i=1}^a\mathbb{E}\left\Vert \mathbf{R}_i \right\Vert^2 \cdot \mathbb{E}\left\Vert \Delta_{F,ij} \right\Vert^2=O\left(\frac{1}{bT}\right)\\
&\therefore \left\Vert \Delta_{H,2} \right\Vert=\mathcal{O}_p\left(\frac{1}{\sqrt{bT\delta_{a,bT}}}\right)
\end{align*}

From Assumption \ref{As:6} and Slutsky's Theorem, we have
\small{
\begin{align*}
&\sqrt{bT}\left(\mathbf{H}_j-\mathbf{H}_R^{\top}\right)\mathbf{R}_j\\
&=\sqrt{bT}V_{R,abT}^{-1}\frac{1}{a}\sum_{i=1}^a\widehat{\mathbf{R}}_i\mathbf{R}_i^{\top}\left(\frac{1}{|Q_{R,ij}|}\sum_{t,m\in Q_{R,ij}}\mathbf{F}_t \mathbf{C}_m\mathbf{C}_m^{\top}\mathbf{F}_t^{\top}-\frac{1}{bT}\sum_{t=1}^T\sum_{m=1}^b\mathbf{F}_t \mathbf{C}_m\mathbf{C}_m^{\top}\mathbf{F}_t^{\top}\right)\mathbf{R}_j\\
&\sqrt{bT}V_{R,abT}^{-1}\frac{1}{a}\sum_{i=1}^a\mathbf{H}_R^{\top}\mathbf{R}_i\mathbf{R}_i^{\top}\left(\frac{1}{|Q_{R,ij}|}\sum_{t,m\in Q_{R,ij}}\mathbf{F}_t \mathbf{C}_m\mathbf{C}_m^{\top}\mathbf{F}_t^{\top}-\frac{1}{bT}\sum_{t=1}^T\sum_{m=1}^b\mathbf{F}_t \mathbf{C}_m\mathbf{C}_m^{\top}\mathbf{F}_t^{\top}\right)\mathbf{R}_j\\
&=V_{R,abT}^{-1}\mathbf{H}_R^{\top}\frac{\sqrt{bT}}{a}\sum_{i=1}^a\mathbf{R}_i\mathbf{R}_i^{\top}\left(\frac{1}{|Q_{R,ij}|}\sum_{t,m\in Q_{R,ij}}\mathbf{F}_t \mathbf{C}_m\mathbf{C}_m^{\top}\mathbf{F}_t^{\top}-\frac{1}{bT}\sum_{t=1}^T\sum_{m=1}^b\mathbf{F}_t \mathbf{C}_m\mathbf{C}_m^{\top}\mathbf{F}_t^{\top}\right)\mathbf{R}_j\\
&\xrightarrow{\mathcal{D}}N\left(\mathbf{0},D^{-1}Q^{-1}\left(\mathbf{\Gamma}_{R_j}^{miss}\right)\left(Q^{-1}\right)^{\top}D^{-1}\right)\mathcal{G}^t-Stably\\
&\therefore \sqrt{bT}\left(\widehat{\mathbf{R}}_j-\mathbf{H}_R^{\top}\mathbf{R}_j\right)\xrightarrow{\mathcal{D}}N\left(\mathbf{0},D^{-1}Q^{-1}\left[\mathbf{\Gamma}_{R_j}^{obs}+\mathbf{\Gamma}_{R_j}^{miss}\right]\left(Q^{-1}\right)^{\top}D^{-1}\right)\mathcal{G}^t-Stably
\end{align*}
}
Similarly, we can get the corresponding results about $\widehat{\mathbf{C}}_j$.
\end{proof}

\section{More Simulation Results}\label{sec:A2}%
 In this section, we present additional simulation results for missing pattern II, which includes the estimation of latent dimensions, convergence results, and the asymptotic normality of $\mathbf{R}_i$.
 \subsection{Latent Dimension Estimation}
 This subsection presents the frequencies of estimated $\left(\hat{k},\hat{r}\right)$ pairs under missing pattern II. The true values are $\left(3,3\right)$.
\begin{table}
\caption{\label{tab1}Table of frequencies of estimated $\left(\hat{k},\hat{r}\right)$ pairs under missing pattern II. The truth is $\left(3,3\right)$.}
\resizebox{.55\linewidth}{!}{
\begin{tabular}{lccccccccc}
    \hline
    \multicolumn{10}{c}{Setting=1, Missing pattern=II}\\
      &  \multicolumn{3}{c}{a=50,b=50} & \multicolumn{3}{c}{a=100,b=100}& \multicolumn{3}{c}{a=200,b=200}\\
     \cline{2-4} \cline{5-7} \cline{8-10}
 $\left(\hat{k},\hat{r}\right)$ & T=50 & T=100 & T=200 & T=50 & T=100 & T=200 & T=50 & T=100 & T=200 \\ \hline
(3,3) &\sethlcolor{pink}\hl{0.862}&\sethlcolor{pink}\hl{0.958}&\sethlcolor{pink}\hl{0.98}&\sethlcolor{pink}\hl{0.768}&\sethlcolor{pink}\hl{0.896}&\sethlcolor{pink}\hl{0.978}&\sethlcolor{pink}\hl{0.676}&\sethlcolor{pink}\hl{0.746}&\sethlcolor{pink}\hl{0.914} \\
 &0.64&0.156&0.16&0.11&0.022&0.034&0.0&0.0&0.0 \\
&0.28&0.39&0.238&0.096&0.016&0.024&0.0&0.0&0.0 \\
&0.45&0.316&0.308&0.112&0.068&0.028&0.0&0.0&0.0 \\
&0.536&0.424&0.142&0.138&0.032&0.004&0.0&0.0&0.0 \\
\hline
(6,3)&0.0&0.0&0.0&0.0&0.0&0.0&0.0&0.0&0.0 \\
&0.332&0.788&0.554&0.864&0.956&0.966&1.0&1.0&1.0 \\
&0.374&0.452&0.566&0.888&0.984&0.976&1.0&1.0&1.0 \\
&0.338&0.646&0.484&0.886&0.928&0.964&1.0&1.0&1.0 \\
&0.444&0.516&0.588&0.856&0.964&0.994&1.0&1.0&1.0 \\
\hline
other &0.138&0.042&0.02&0.232&0.104&0.022&0.324&0.254&0.086 \\
&0.028&0.056&0.286&0.026&0.022&0.0&0.0&0.0&0.0 \\
 &0.346&0.158&0.196&0.016&0.0&0.0&0.0&0.0&0.0 \\
&0.212&0.038&0.208&0.002&0.004&0.008&0.0&0.0&0.0 \\
&0.02&0.06&0.27&0.006&0.004&0.002&0.0&0.0&0.0 \\ \hline
  \multicolumn{10}{c}{Setting=2 with $\psi=0.1$, Missing pattern=II}\\
      &  \multicolumn{3}{c}{a=50,b=50} & \multicolumn{3}{c}{a=100,b=100}& \multicolumn{3}{c}{a=200,b=200}\\
     \cline{2-4} \cline{5-7} \cline{8-10}
  $\left(\hat{k},\hat{r}\right)$  & T=50 & T=100 & T=200 & T=50 & T=100 & T=200 & T=50 & T=100 & T=200 \\ \hline
  (3,3) &\sethlcolor{pink}\hl{0.782}&\sethlcolor{pink}\hl{0.93}&\sethlcolor{pink}\hl{0.95}&\sethlcolor{pink}\hl{0.758}&\sethlcolor{pink}\hl{0.87}&\sethlcolor{pink}\hl{0.962}&\sethlcolor{pink}\hl{0.594}&\sethlcolor{pink}\hl{0.768}&\sethlcolor{pink}\hl{0.884} \\

        &0.546&0.222&0.328&0.048&0.038&0.012&0.002&0.0&0.0 \\

        &0.564&0.412&0.328&0.156&0.032&0.026&0.002&0.0&0.0 \\

       &0.478&0.13&0.1&0.104&0.016&0.01&0.0&0.0&0.0 \\

       &0.428&0.264&0.18&0.09&0.024&0.018&0.0&0.0&0.0 \\           \hline
  (6,3)  &0.0&0.0&0.0&0.0&0.0&0.0&0.0&0.002&0.0 \\

       &0.418&0.558&0.454&0.916&0.96&0.984&0.998&1.0&1.0 \\

       &0.376&0.476&0.664&0.812&0.968&0.974&0.998&1.0&1.0 \\

       &0.32&0.558&0.536&0.892&0.974&0.98&1.0&1.0&1.0 \\

       &0.316&0.6&0.744&0.906&0.974&0.98&1.0&1.0&1.0 \\   \hline
  other &0.218&0.07&0.05&0.242&0.13&0.038&0.406&0.23&0.116 \\

       &0.036&0.22&0.218&0.036&0.002&0.004&0.0&0.0&0.0 \\

        &0.436&0.112&0.008&0.032&0.0&0.0&0.0&0.0&0.0 \\

        &0.202&0.312&0.364&0.004&0.01&0.01&0.0&0.0&0.0 \\

        &0.256&0.136&0.076&0.004&0.002&0.002&0.0&0.0&0.0 \\    \hline
 \multicolumn{10}{c}{Setting=2 with $\psi=0.5$, Missing pattern=II}\\
      &  \multicolumn{3}{c}{a=50,b=50} & \multicolumn{3}{c}{a=100,b=100}& \multicolumn{3}{c}{a=200,b=200}\\
     \cline{2-4} \cline{5-7} \cline{8-10}
  $\left(\hat{k},\hat{r}\right)$ & T=50 & T=100 & T=200 & T=50 & T=100 & T=200 & T=50 & T=100 & T=200 \\ \hline
(3,3)  &\sethlcolor{pink}\hl{0.688}&\sethlcolor{pink}\hl{0.812}&\sethlcolor{pink}\hl{0.936}&\sethlcolor{pink}\hl{0.564}&\sethlcolor{pink}\hl{0.708}&\sethlcolor{pink}\hl{0.87}&\sethlcolor{pink}\hl{0.426}&\sethlcolor{pink}\hl{0.592}&\sethlcolor{pink}\hl{0.758} \\
       &0.226&0.286&0.186&0.022&0.036&0.008&0.0&0.0&0.0 \\
       &0.362&0.19&0.2&0.028&0.026&0.014&0.0&0.0&0.0 \\
       &0.244&0.404&0.156&0.038&0.014&0.002&0.0&0.0&0.0 \\
       &0.208&0.222&0.194&0.034&0.016&0.008&0.0&0.0&0.0 \\  \hline
(6,3)  &0.0&0.0&0.0&0.004&0.0&0.0&0.006&0.006&0.002 \\
       &0.63&0.678&0.758&0.948&0.962&0.99&1.0&1.0&1.0 \\
       &0.558&0.646&0.766&0.882&0.972&0.986&1.0&1.0&1.0 \\
       &0.536&0.556&0.78&0.95&0.986&0.998&1.0&1.0&1.0 \\
       &0.36&0.692&0.792&0.91&0.982&0.992&1.0&1.0&1.0 \\ \hline
 other &0.172&0.038&0.024&0.176&0.106&0.074&0.362&0.232&0.076 \\
       &0.144&0.036&0.056&0.03&0.002&0.002&0.0&0.0&0.0 \\
       &0.08&0.164&0.034&0.09&0.002&0.0&0.0&0.0&0.0 \\
       &0.22&0.04&0.064&0.012&0.0&0.0&0.0&0.0&0.0 \\
       &0.432&0.086&0.014&0.056&0.002&0.0&0.0&0.0&0.0 \\ \hline
  \multicolumn{10}{c}{Setting=3, Missing pattern=II}\\
    &  \multicolumn{3}{c}{a=50,b=50} & \multicolumn{3}{c}{a=100,b=100}& \multicolumn{3}{c}{a=200,b=200}\\
     \cline{2-4} \cline{5-7} \cline{8-10}
  $\left(\hat{k},\hat{r}\right)$   & T=50 & T=100 & T=200 & T=50 & T=100 & T=200 & T=50 & T=100 & T=200 \\ \hline
  (3,3)  &\sethlcolor{pink}\hl{0.828}&\sethlcolor{pink}\hl{0.962}&\sethlcolor{pink}\hl{0.976}&\sethlcolor{pink}\hl{0.824}&\sethlcolor{pink}\hl{0.894}&\sethlcolor{pink}\hl{0.926}&\sethlcolor{pink}\hl{0.638}&\sethlcolor{pink}\hl{0.768}&\sethlcolor{pink}\hl{0.924} \\
         &0.558&0.602&0.402&0.09&0.016&0.026&0.0&0.0&0.0 \\
          &0.298&0.286&0.246&0.074&0.026&0.006&0.002&0.0&0.0 \\
          &0.128&0.42&0.636&0.18&0.074&0.038&0.002&0.0&0.0 \\
         &0.53&0.298&0.312&0.076&0.052&0.018&0.0&0.0&0.0 \\ \hline
   (6,3) &0.0&0.0&0.0&0.0&0.0&0.0&0.0&0.0&0.0 \\
         &0.368&0.39&0.568&0.898&0.982&0.97&1.0&1.0&1.0 \\
         &0.564&0.61&0.63&0.892&0.97&0.982&0.998&1.0&1.0 \\
         &0.232&0.52&0.364&0.82&0.922&0.962&0.998&1.0&1.0 \\
         &0.306&0.526&0.674&0.922&0.946&0.982&1.0&1.0&1.0 \\ \hline
  other   &0.172&0.038&0.024&0.176&0.106&0.074&0.362&0.232&0.076 \\
         &0.074&0.008&0.03&0.012&0.002&0.004&0.0&0.0&0.0 \\
         &0.138&0.104&0.124&0.034&0.004&0.012&0.0&0.0&0.0 \\
         &0.64&0.06&0.0&0.0&0.004&0.0&0.0&0.0&0.0 \\
         &0.164&0.176&0.014&0.002&0.002&0.0&0.0&0.0&0.0 \\   \hline \hline
  \end{tabular}
  }
\end{table}
\subsection{Errors of Loading Matrices and Factor Matrices Estimation}
This section presents the table of the estimation errors of loading matrices and factor matrices, along with the responding boxplots.
\begin{table}
\caption{\label{tab3} Means and standard deviations (in parentheses) of $\mathcal{D}\left(\widehat{\mathbf{R}},\mathbf{R}\right), \mathcal{D}\left(\widehat{\mathbf{C}},\mathbf{C}\right),\left\Vert \widehat{\mathbf{F}}_t-\mathbf{H}_R^{-1}\mathbf{F}_t\mathbf{H}_C^{-1\top}\right\Vert$ with missing pattern II.}
\resizebox{.90\linewidth}{!}{
\begin{tabular}{lccccccccc}
    \hline
    \multicolumn{10}{c}{Setting=1, Missing pattern=II}\\
      &  \multicolumn{3}{c}{T=50} & \multicolumn{3}{c}{T=100}& \multicolumn{3}{c}{T=200}\\
     \cline{2-4} \cline{5-7} \cline{8-10}
 $\left(a,b\right)$ & $\mathcal{D}\left(\widehat{\mathbf{R}},\mathbf{R}\right)$ & $\mathcal{D}\left(\widehat{\mathbf{C}},\mathbf{C}\right)$ & $\left\Vert \widehat{\mathbf{F}}_t-\mathbf{H}_R^{-1}\mathbf{F}_t\mathbf{H}_C^{-1\top}\right\Vert$& $\mathcal{D}\left(\widehat{\mathbf{R}},\mathbf{R}\right)$ & $\mathcal{D}\left(\widehat{\mathbf{C}},\mathbf{C}\right)$ & $\left\Vert \widehat{\mathbf{F}}_t-\mathbf{H}_R^{-1}\mathbf{F}_t\mathbf{H}_C^{-1\top}\right\Vert$& $\mathcal{D}\left(\widehat{\mathbf{R}},\mathbf{R}\right)$ & $\mathcal{D}\left(\widehat{\mathbf{C}},\mathbf{C}\right)$ & $\left\Vert \widehat{\mathbf{F}}_t-\mathbf{H}_R^{-1}\mathbf{F}_t\mathbf{H}_C^{-1\top}\right\Vert$ \\ \hline
(50,50) &\sethlcolor{pink}\hl{0.068(0.03)}&0.01(0.0012)&0.11(0.064)&\sethlcolor{pink}\hl{0.044(0.019)}&0.0063(0.0006)&0.069(0.033)&\sethlcolor{pink}\hl{0.034(0.013)}&0.0046(0.00042)&0.049(0.025) \\
 &0.13(0.029)&\sethlcolor{pink}\hl{0.0097(0.001)}&0.14(0.08)&0.13(0.027)&\sethlcolor{pink}\hl{0.0057(0.00052)}&0.11(0.057)&0.13(0.03)&0.0057(0.00057)&0.06(0.033) \\
&0.15(0.046)&0.012(0.0015)&0.1(0.056)&0.13(0.028)&0.0083(0.00081)&\sethlcolor{pink}\hl{0.065(0.035)}&0.12(0.023)&0.0043(0.00039)&\sethlcolor{pink}\hl{0.048(0.023)} \\
&0.14(0.036)&0.011(0.0012)&\sethlcolor{pink}\hl{0.092(0.049)}&0.12(0.024)&0.0066(0.00066)&0.072(0.036)&0.11(0.02)&\sethlcolor{pink}\hl{0.0043(0.00038)}&0.049(0.025) \\
&0.13(0.031)&0.01(0.0012)&0.1(0.053)&0.12(0.025)&0.0073(0.00077)&0.069(0.035)&0.12(0.027)&0.0045(0.00045)&0.057(0.031) \\
\hline
(100,100)
&\sethlcolor{pink}\hl{0.065(0.027)}&\sethlcolor{pink}\hl{0.0061(0.00053)}&0.11(0.061)&\sethlcolor{pink}\hl{0.044(0.018)}&0.0044(0.00036)&0.069(0.03)&\sethlcolor{pink}\hl{0.03(0.011)}&\sethlcolor{pink}\hl{0.0032(0.00024)}&\sethlcolor{pink}\hl{0.046(0.019)} \\
&0.14(0.031)&0.0075(0.00067)&0.16(0.1)&0.13(0.029)&0.0043(0.00032)&0.099(0.055)&0.12(0.015)&0.0037(0.00027)&0.052(0.023) \\
 &0.14(0.036)&0.0072(0.00074)&0.094(0.045)&0.13(0.021)&\sethlcolor{pink}\hl{0.0042(0.00034)}&0.07(0.032)&0.12(0.016)&0.0035(0.00022)&0.055(0.026) \\

 &0.13(0.03)&0.007(0.00058)&\sethlcolor{pink}\hl{0.088(0.039)}&0.13(0.021)&0.005(0.00036)&\sethlcolor{pink}\hl{0.063(0.028)}&0.12(0.016)&0.0033(0.00024)&0.048(0.021) \\
 &0.13(0.031)&0.0074(0.00071)&0.089(0.036)&0.13(0.022)&0.0051(0.00045)&0.073(0.031)&0.12(0.019)&0.0034(0.00026)&0.052(0.023) \\
 \hline
(200,200) &\sethlcolor{pink}\hl{0.064(0.028)}&0.0048(0.00038)&0.097(0.054)&\sethlcolor{pink}\hl{0.049(0.019)}&0.0031(0.0002)&0.068(0.028)&\sethlcolor{pink}\hl{0.031(0.011)}&\sethlcolor{pink}\hl{0.0022(0.00012)}&\sethlcolor{pink}\hl{0.046(0.018)} \\
&0.14(0.032)&\sethlcolor{pink}\hl{0.0047(0.00035)}&0.15(0.088)&0.13(0.022)&\sethlcolor{pink}\hl{0.0029(0.00017)}&0.096(0.046)&0.12(0.015)&0.0024(0.00013)&0.058(0.028) \\
&0.14(0.034)&0.0048(0.00039)&0.096(0.044)&0.13(0.024)&0.0034(0.00021)&\sethlcolor{pink}\hl{0.067(0.029)}&0.12(0.013)&0.0022(0.00013)&0.052(0.021) \\

 &0.14(0.034)&0.0048(0.00036)&0.095(0.042)&0.13(0.02)&0.0034(0.00022)&0.068(0.027)&0.12(0.014)&0.0023(0.00012)&0.05(0.022) \\
 &0.14(0.034)&0.0048(0.00035)&\sethlcolor{pink}\hl{0.091(0.035)}&0.13(0.025)&0.0031(0.00019)&0.075(0.03)&0.12(0.017)&0.0023(0.00013)&0.053(0.023) \\
 \hline

  \multicolumn{10}{c}{Setting=2($\psi=0.1$), Missing pattern=II}\\
   &  \multicolumn{3}{c}{T=50} & \multicolumn{3}{c}{T=100}& \multicolumn{3}{c}{T=200}\\
     \cline{2-4} \cline{5-7} \cline{8-10}
 $\left(a,b\right)$ & $\mathcal{D}\left(\widehat{\mathbf{R}},\mathbf{R}\right)$ & $\mathcal{D}\left(\widehat{\mathbf{C}},\mathbf{C}\right)$ & $\left\Vert \widehat{\mathbf{F}}_t-\mathbf{H}_R^{-1}\mathbf{F}_t\mathbf{H}_C^{-1\top}\right\Vert$& $\mathcal{D}\left(\widehat{\mathbf{R}},\mathbf{R}\right)$ & $\mathcal{D}\left(\widehat{\mathbf{C}},\mathbf{C}\right)$ & $\left\Vert \widehat{\mathbf{F}}_t-\mathbf{H}_R^{-1}\mathbf{F}_t\mathbf{H}_C^{-1\top}\right\Vert$& $\mathcal{D}\left(\widehat{\mathbf{R}},\mathbf{R}\right)$ & $\mathcal{D}\left(\widehat{\mathbf{C}},\mathbf{C}\right)$ & $\left\Vert \widehat{\mathbf{F}}_t-\mathbf{H}_R^{-1}\mathbf{F}_t\mathbf{H}_C^{-1\top}\right\Vert$ \\ \hline
  (50,50) &\sethlcolor{pink}\hl{0.08(0.042)}&\sethlcolor{pink}\hl{0.01(0.0011)}&0.13(0.088)&\sethlcolor{pink}\hl{0.048(0.018)}&0.0076(0.00077)&0.07(0.033)&\sethlcolor{pink}\hl{0.037(0.015)}&0.0048(0.0005)&\sethlcolor{pink}\hl{0.052(0.026)} \\
          &0.13(0.031)&0.01(0.0011)&0.17(0.1)&0.13(0.025)&0.0064(0.00072)&0.11(0.054)&0.12(0.021)&0.0059(0.00058)&0.057(0.03) \\
          &0.13(0.036)&0.011(0.0013)&\sethlcolor{pink}\hl{0.1(0.05)}&0.13(0.025)&0.0079(0.00085)&\sethlcolor{pink}\hl{0.069(0.033)}&0.11(0.015)&0.0044(0.0004)&0.058(0.029) \\
          &0.13(0.031)&0.01(0.0012)&0.098(0.047)&0.13(0.032)&0.0061(0.00061)&0.085(0.041)&0.12(0.03)&\sethlcolor{pink}\hl{0.0041(0.00039)}&0.061(0.032) \\
          &0.14(0.042)&0.01(0.0011)&0.097(0.049)&0.13(0.029)&\sethlcolor{pink}\hl{0.006(0.00061)}&0.09(0.043)&0.12(0.022)&0.0047(0.00046)&0.062(0.03) \\  \hline

(100,100)&\sethlcolor{pink}\hl{0.068(0.027)}&0.0065(0.0006)&0.14(0.086)&\sethlcolor{pink}\hl{0.047(0.019)}&\sethlcolor{pink}\hl{0.0047(0.00036)}&0.078(0.036)&\sethlcolor{pink}\hl{0.032(0.011)}&0.0034(0.00025)&\sethlcolor{pink}\hl{0.051(0.023)} \\
         &0.14(0.034)&\sethlcolor{pink}\hl{0.006(0.00054)}&0.18(0.1)&0.13(0.024)&0.005(0.00045)&0.1(0.053)&0.12(0.018)&0.0033(0.00024)&0.065(0.032) \\
         &0.14(0.036)&0.0076(0.00073)&0.099(0.044)&0.13(0.021)&0.005(0.00042)&\sethlcolor{pink}\hl{0.076(0.033)}&0.11(0.015)&0.0032(0.00022)&0.054(0.023) \\
         &0.14(0.034)&0.0067(0.00056)&0.11(0.051)&0.13(0.025)&0.0049(0.00042)&0.081(0.038)&0.12(0.019)&\sethlcolor{pink}\hl{0.0031(0.00022)}&0.059(0.028) \\
       &0.14(0.032)&0.0068(0.00061)&\sethlcolor{pink}\hl{0.1(0.044)}&0.13(0.021)&0.0047(0.00037)&0.074(0.029)&0.12(0.017)&0.0034(0.00024)&0.057(0.026) \\  \hline
(200,200) &\sethlcolor{pink}\hl{0.065(0.029)}&\sethlcolor{pink}\hl{0.0041(0.00033)}&0.13(0.078)&\sethlcolor{pink}\hl{0.05(0.02)}&0.0033(0.0002)&\sethlcolor{pink}\hl{0.072(0.032)}&\sethlcolor{pink}\hl{0.034(0.012)}&0.0023(0.00013)&\sethlcolor{pink}\hl{0.049(0.02)}\\
        &0.14(0.034)&0.0051(0.00038)&0.17(0.1)&0.13(0.022)&0.0034(0.00025)&0.1(0.052)&0.12(0.016)&0.0025(0.00013)&0.064(0.033) \\
         &0.14(0.035)&0.0048(0.00035)&0.1(0.048)&0.13(0.024)&0.0034(0.00021)&0.076(0.034)&0.12(0.014)&0.0022(0.00012)&0.053(0.022) \\
        &0.14(0.033)&0.0047(0.00034)&0.11(0.043)&0.13(0.022)&\sethlcolor{pink}\hl{0.0031(0.00019)}&0.082(0.033)&0.12(0.016)&0.0023(0.00012)&0.056(0.021) \\
         &0.15(0.038)&0.005(0.00043)&\sethlcolor{pink}\hl{0.1(0.041)}&0.13(0.022)&0.0034(0.00022)&0.077(0.031)&0.12(0.016)&\sethlcolor{pink}\hl{0.0022(0.00011)}&0.055(0.022) \\  \hline
\multicolumn{10}{c}{Setting=2($\psi=0.5$), Missing pattern=II}\\
   &  \multicolumn{3}{c}{T=50} & \multicolumn{3}{c}{T=100}& \multicolumn{3}{c}{T=200}\\
     \cline{2-4} \cline{5-7} \cline{8-10}
 $\left(a,b\right)$ & $\mathcal{D}\left(\widehat{\mathbf{R}},\mathbf{R}\right)$ & $\mathcal{D}\left(\widehat{\mathbf{C}},\mathbf{C}\right)$ & $\left\Vert \widehat{\mathbf{F}}_t-\mathbf{H}_R^{-1}\mathbf{F}_t\mathbf{H}_C^{-1\top}\right\Vert$& $\mathcal{D}\left(\widehat{\mathbf{R}},\mathbf{R}\right)$ & $\mathcal{D}\left(\widehat{\mathbf{C}},\mathbf{C}\right)$ & $\left\Vert \widehat{\mathbf{F}}_t-\mathbf{H}_R^{-1}\mathbf{F}_t\mathbf{H}_C^{-1\top}\right\Vert$& $\mathcal{D}\left(\widehat{\mathbf{R}},\mathbf{R}\right)$ & $\mathcal{D}\left(\widehat{\mathbf{C}},\mathbf{C}\right)$ & $\left\Vert \widehat{\mathbf{F}}_t-\mathbf{H}_R^{-1}\mathbf{F}_t\mathbf{H}_C^{-1\top}\right\Vert$ \\ \hline
 (50,50)
&\sethlcolor{pink}\hl{0.091(0.047)}&0.0088(0.0011)&0.39(0.29)&\sethlcolor{pink}\hl{0.063(0.029)}&0.0064(0.00074)&0.18(0.1)&\sethlcolor{pink}\hl{0.04(0.015)}&0.0042(0.00041)&\sethlcolor{pink}\hl{0.1(0.047)} \\
&0.15(0.049)&\sethlcolor{pink}\hl{0.0078(0.00087)}&0.53(0.35)&0.13(0.031)&0.0064(0.00065)&0.24(0.15)&0.12(0.025)&\sethlcolor{pink}\hl{0.0041(0.00041)}&0.15(0.082) \\
&0.14(0.041)&0.0083(0.00091)&0.19(0.098)&0.14(0.037)&\sethlcolor{pink}\hl{0.0058(0.0006)}&0.14(0.073)&0.12(0.02)&0.0041(0.00043)&0.1(0.051) \\
&0.16(0.052)&0.0086(0.00096)&0.19(0.094)&0.13(0.031)&0.0069(0.0007)&\sethlcolor{pink}\hl{0.12(0.061)}&0.12(0.025)&0.0042(0.00045)&0.1(0.053) \\
&0.17(0.062)&0.01(0.0012)&\sethlcolor{pink}\hl{0.17(0.083)}&0.14(0.037)&0.0062(0.0006)&0.14(0.066)&0.12(0.022)&0.0046(0.00043)&0.11(0.053) \\  \hline
 (100,100)
&\sethlcolor{pink}\hl{0.093(0.046)}&0.007(0.00076)&0.38(0.29)&\sethlcolor{pink}\hl{0.061(0.027)}&0.0044(0.00036)&0.17(0.12)&\sethlcolor{pink}\hl{0.041(0.015)}&0.0029(0.00024)&\sethlcolor{pink}\hl{0.095(0.042)} \\
&0.16(0.052)&\sethlcolor{pink}\hl{0.006(0.00059)}&0.51(0.35)&0.14(0.029)&0.0044(0.00035)&0.25(0.15)&0.12(0.019)&0.003(0.0002)&0.14(0.077) \\
&0.17(0.057)&0.0071(0.00072)&0.18(0.092)&0.14(0.029)&0.0047(0.0004)&\sethlcolor{pink}\hl{0.13(0.058)}&0.12(0.02)&0.0029(0.00021)&0.098(0.044) \\
&0.16(0.047)&0.0063(0.00062)&\sethlcolor{pink}\hl{0.18(0.077)}&0.14(0.027)&\sethlcolor{pink}\hl{0.0039(0.00032)}&0.15(0.06)&0.12(0.02)&0.003(0.00021)&0.1(0.049) \\
&0.17(0.061)&0.0065(0.00064)&0.18(0.089)&0.14(0.032)&0.0047(0.00041)&0.13(0.061)&0.13(0.022)&\sethlcolor{pink}\hl{0.0028(0.0002)}&0.099(0.043) \\ \hline
 (200,200)
 &\sethlcolor{pink}\hl{0.095(0.048)}&0.0049(0.00044)&0.4(0.33)&\sethlcolor{pink}\hl{0.065(0.028)}&0.0032(0.00021)&0.16(0.1)&\sethlcolor{pink}\hl{0.043(0.017)}&\sethlcolor{pink}\hl{0.0021(0.00011)}&\sethlcolor{pink}\hl{0.088(0.041)} \\  &0.15(0.045)&0.0049(0.00044)&0.43(0.31)&0.15(0.036)&0.0032(0.00024)&0.25(0.14)&0.13(0.02)&0.0022(0.00013)&0.13(0.068) \\
&0.16(0.044)&0.0045(0.00039)&\sethlcolor{pink}\hl{0.17(0.085)}&0.14(0.032)&0.0031(0.00022)&\sethlcolor{pink}\hl{0.13(0.059)}&0.13(0.02)&0.0023(0.00014)&0.096(0.04) \\  &0.16(0.049)&0.0044(0.00037)&0.18(0.079)&0.14(0.031)&\sethlcolor{pink}\hl{0.0029(0.00019)}&0.14(0.061)&0.13(0.019)&0.0021(0.00011)&0.093(0.037) \\
&0.17(0.053)&\sethlcolor{pink}\hl{0.0042(0.00036)}&0.2(0.086)&0.14(0.033)&0.0031(0.00021)&0.14(0.054)&0.13(0.021)&0.0021(0.00012)&0.097(0.039) \\  \hline
   \multicolumn{10}{c}{Setting=3, Missing pattern=II}\\
    &  \multicolumn{3}{c}{T=50} & \multicolumn{3}{c}{T=100}& \multicolumn{3}{c}{T=200}\\
     \cline{2-4} \cline{5-7} \cline{8-10}
 $\left(a,b\right)$ & $\mathcal{D}\left(\widehat{\mathbf{R}},\mathbf{R}\right)$ & $\mathcal{D}\left(\widehat{\mathbf{C}},\mathbf{C}\right)$ & $\left\Vert \widehat{\mathbf{F}}_t-\mathbf{H}_R^{-1}\mathbf{F}_t\mathbf{H}_C^{-1\top}\right\Vert$& $\mathcal{D}\left(\widehat{\mathbf{R}},\mathbf{R}\right)$ & $\mathcal{D}\left(\widehat{\mathbf{C}},\mathbf{C}\right)$ & $\left\Vert \widehat{\mathbf{F}}_t-\mathbf{H}_R^{-1}\mathbf{F}_t\mathbf{H}_C^{-1\top}\right\Vert$& $\mathcal{D}\left(\widehat{\mathbf{R}},\mathbf{R}\right)$ & $\mathcal{D}\left(\widehat{\mathbf{C}},\mathbf{C}\right)$ & $\left\Vert \widehat{\mathbf{F}}_t-\mathbf{H}_R^{-1}\mathbf{F}_t\mathbf{H}_C^{-1\top}\right\Vert$ \\ \hline
  (50,50) &\sethlcolor{pink}\hl{0.069(0.03)}&\sethlcolor{pink}\hl{0.011(0.0011)}&0.12(0.063)&\sethlcolor{pink}\hl{0.042(0.015)}&\sethlcolor{pink}\hl{0.0066(0.00057)}&\sethlcolor{pink}\hl{0.068(0.031)}&\sethlcolor{pink}\hl{0.037(0.014)}&0.0061(0.00058)&0.046(0.022) \\
          &0.13(0.033)&0.012(0.0014)&0.16(0.094)&0.12(0.024)&0.0097(0.00092)&0.082(0.047)&0.11(0.016)&0.0057(0.00053)&0.055(0.027) \\
          &0.14(0.039)&0.011(0.0011)&0.11(0.058)&0.13(0.026)&0.007(0.00068)&0.077(0.039)&0.12(0.022)&0.0059(0.00052)&0.051(0.026) \\
          &0.16(0.081)&0.012(0.0013)&0.12(0.23)&0.12(0.023)&0.0076(0.00079)&0.075(0.037)&0.11(0.016)&0.0057(0.00052)&\sethlcolor{pink}\hl{0.045(0.022)} \\
          &0.14(0.035)&0.013(0.0014)&\sethlcolor{pink}\hl{0.09(0.042)}&0.13(0.03)&0.01(0.0011)&0.069(0.037)&0.11(0.016)&\sethlcolor{pink}\hl{0.0056(0.00053)}&0.057(0.028) \\  \hline
(100,100)&\sethlcolor{pink}\hl{0.063(0.025)}&\sethlcolor{pink}\hl{0.0078(0.00065)}&0.1(0.056)&\sethlcolor{pink}\hl{0.047(0.019)}&\sethlcolor{pink}\hl{0.0051(0.00038)}&\sethlcolor{pink}\hl{0.066(0.032)}&\sethlcolor{pink}\hl{0.035(0.013)}&0.0038(0.00027)&\sethlcolor{pink}\hl{0.049(0.021)} \\
         &0.14(0.032)&0.0082(0.00069)&0.15(0.089)&0.13(0.022)&0.0052(0.00038)&0.096(0.05)&0.12(0.016)&\sethlcolor{pink}\hl{0.0036(0.00024)}&0.059(0.031) \\
         &0.14(0.038)&0.0086(0.00079)&\sethlcolor{pink}\hl{0.1(0.051)}&0.13(0.023)&0.0052(0.00037)&0.075(0.031)&0.13(0.024)&0.0036(0.00025)&0.052(0.023) \\
         &0.13(0.03)&0.0083(0.00072)&0.091(0.039)&0.13(0.023)&0.0055(0.00038)&0.067(0.03)&0.12(0.015)&0.0038(0.00026)&0.051(0.023) \\
         &0.14(0.033)&0.008(0.00076)&0.095(0.041)&0.13(0.023)&0.0057(0.00044)&0.07(0.032)&0.12(0.014)&0.0035(0.00024)&0.051(0.021) \\ \hline
(200,200)&\sethlcolor{pink}\hl{0.07(0.031)}&0.0057(0.00042)&0.1(0.063)&\sethlcolor{pink}\hl{0.048(0.019)}&\sethlcolor{pink}\hl{0.0037(0.00022)}&\sethlcolor{pink}\hl{0.069(0.029)}&\sethlcolor{pink}\hl{0.033(0.011)}&0.0029(0.00015)&\sethlcolor{pink}\hl{0.044(0.017)} \\
          &0.14(0.033)&0.0057(0.00042)&0.15(0.091)&0.13(0.022)&0.004(0.00025)&0.085(0.045)&0.12(0.014)&0.0027(0.00013)&0.054(0.023) \\
          &0.14(0.032)&0.0055(0.00043)&0.099(0.043)&0.13(0.023)&0.0041(0.00026)&0.069(0.032)&0.11(0.013)&\sethlcolor{pink}\hl{0.0025(0.00014)}&0.051(0.02) \\
          &0.14(0.032)&0.0056(0.00042)&\sethlcolor{pink}\hl{0.095(0.039)}&0.13(0.025)&0.0039(0.00023)&0.072(0.029)&0.12(0.016)&0.0028(0.00015)&0.051(0.021) \\
           &0.14(0.033)&\sethlcolor{pink}\hl{0.0055(0.00039)}&0.099(0.042)&0.13(0.023)&0.0039(0.00023)&0.075(0.031)&0.12(0.015)&0.0027(0.00015)&0.053(0.022) \\ \hline
   \hline
  \end{tabular}
  }
\end{table}

\begin{figure}
\includegraphics[scale=0.3]{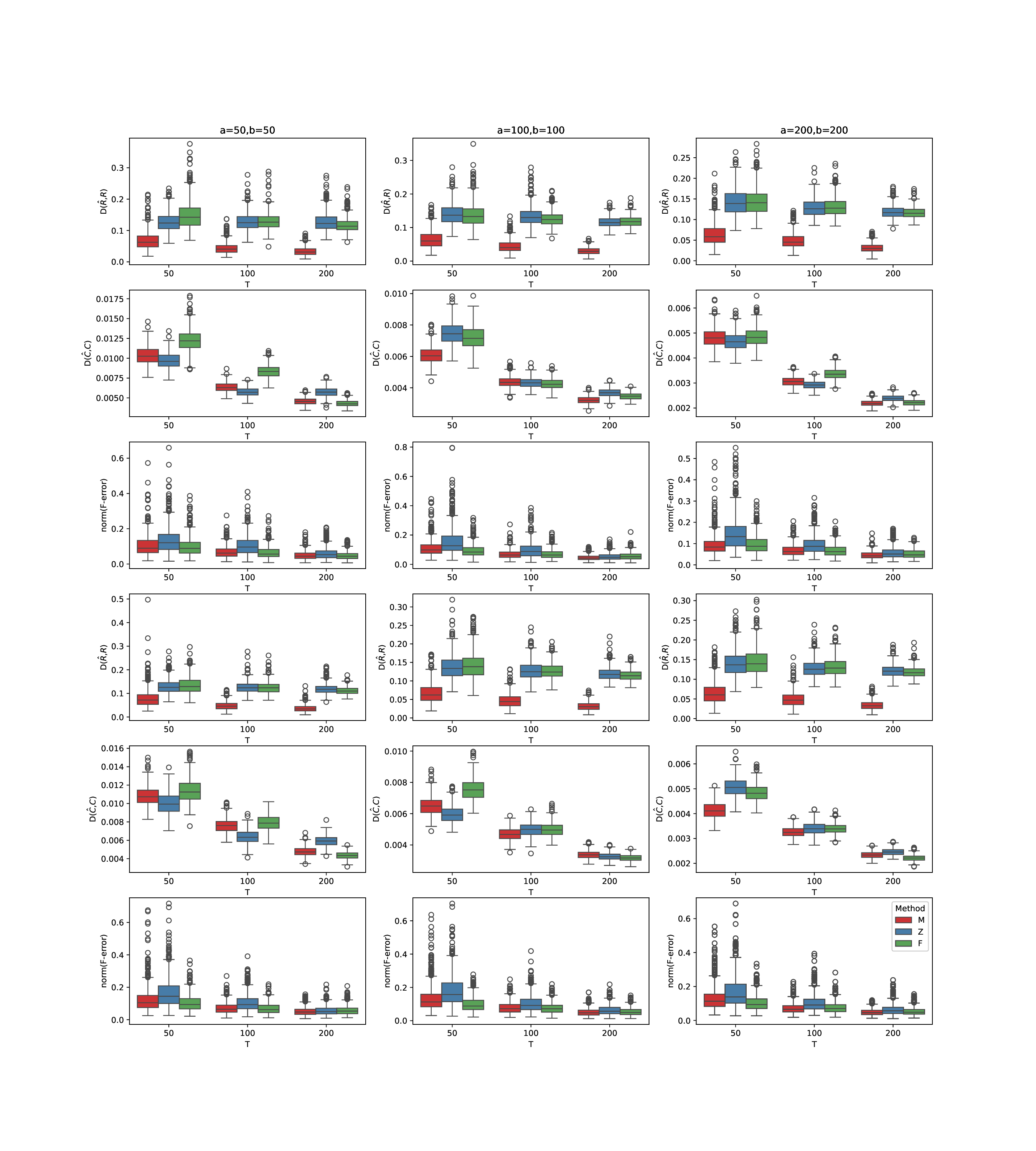}
\caption{Boxplots of errors of loading matrices estimation and factor matrices estimation for Setting 1 and Setting 2 with $\psi=0.1$ under missing pattern II.}
\label{fig:4-3}
\end{figure}

\begin{figure}
\includegraphics[scale=0.3]{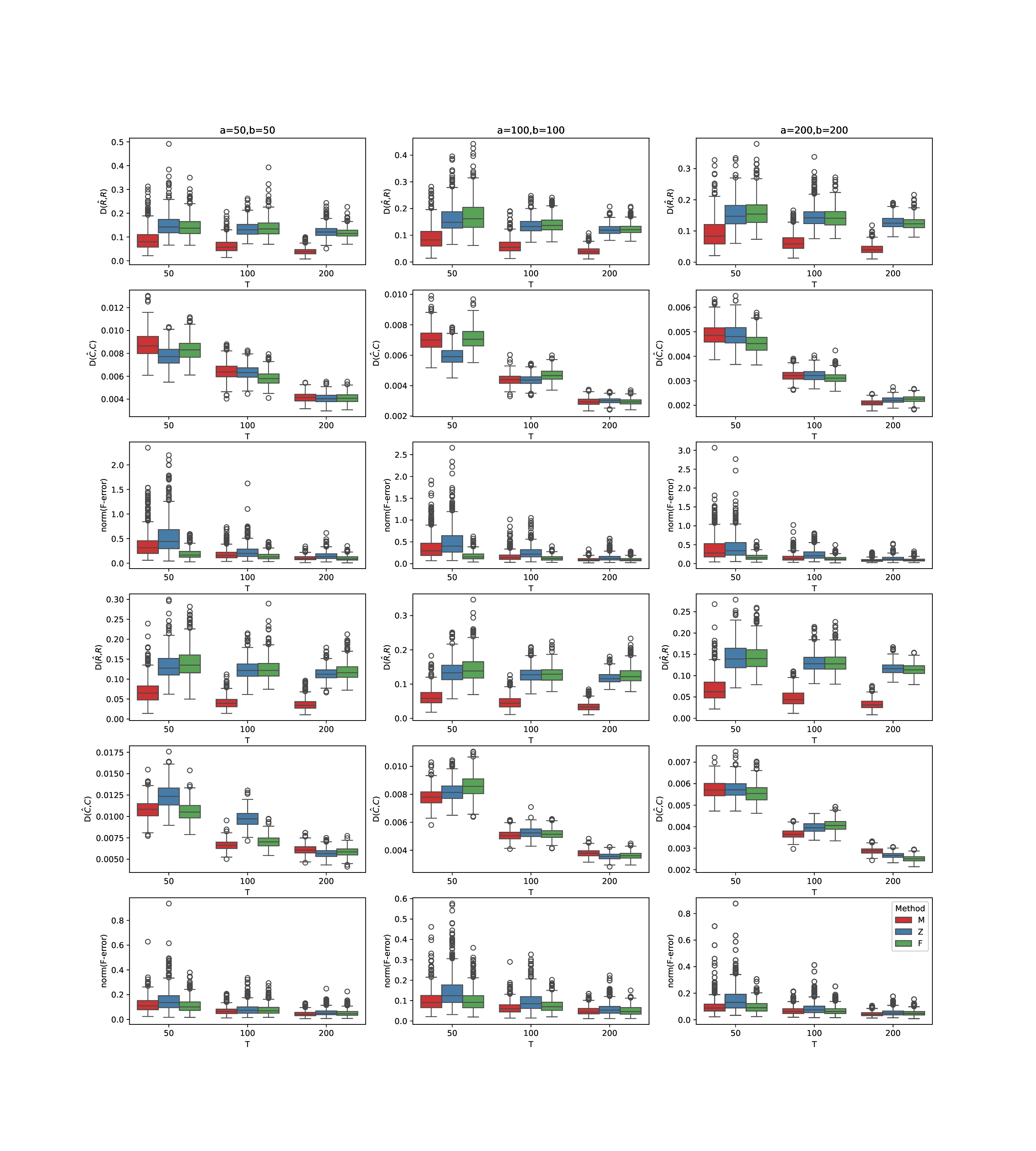}
\caption{Boxplots of errors of loading matrices estimation
and factor matrices estimation for Setting 2 with $\psi=0.5$ and Setting 3 under missing pattern II.}
\label{fig:4-4}
\end{figure}

\subsection{The QQ Plots and Histograms for Different Settings under Missing Pattern II}
This section presents all the QQ plots and histograms for the various Settings under missing pattern II of three elements of the first row of $\widehat{\mathbf{R}}-\mathbf{H}_R^{\top}\mathbf{R}$.

\begin{figure}
\includegraphics[scale=0.3]{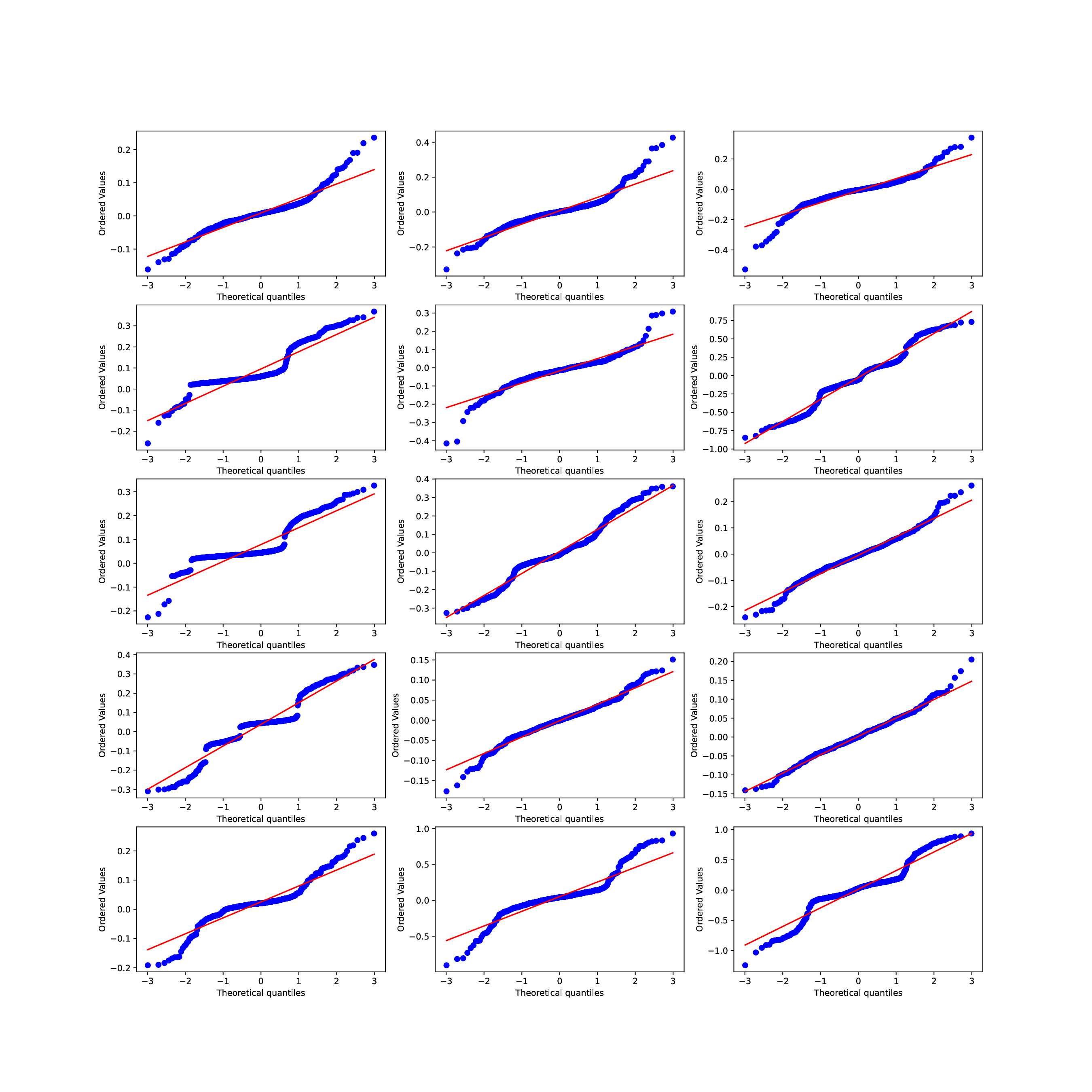}
\caption{The QQ plots for Setting 1, missing pattern II. }
\label{fig:6-1}
\end{figure}
\begin{figure}
\includegraphics[scale=0.3]{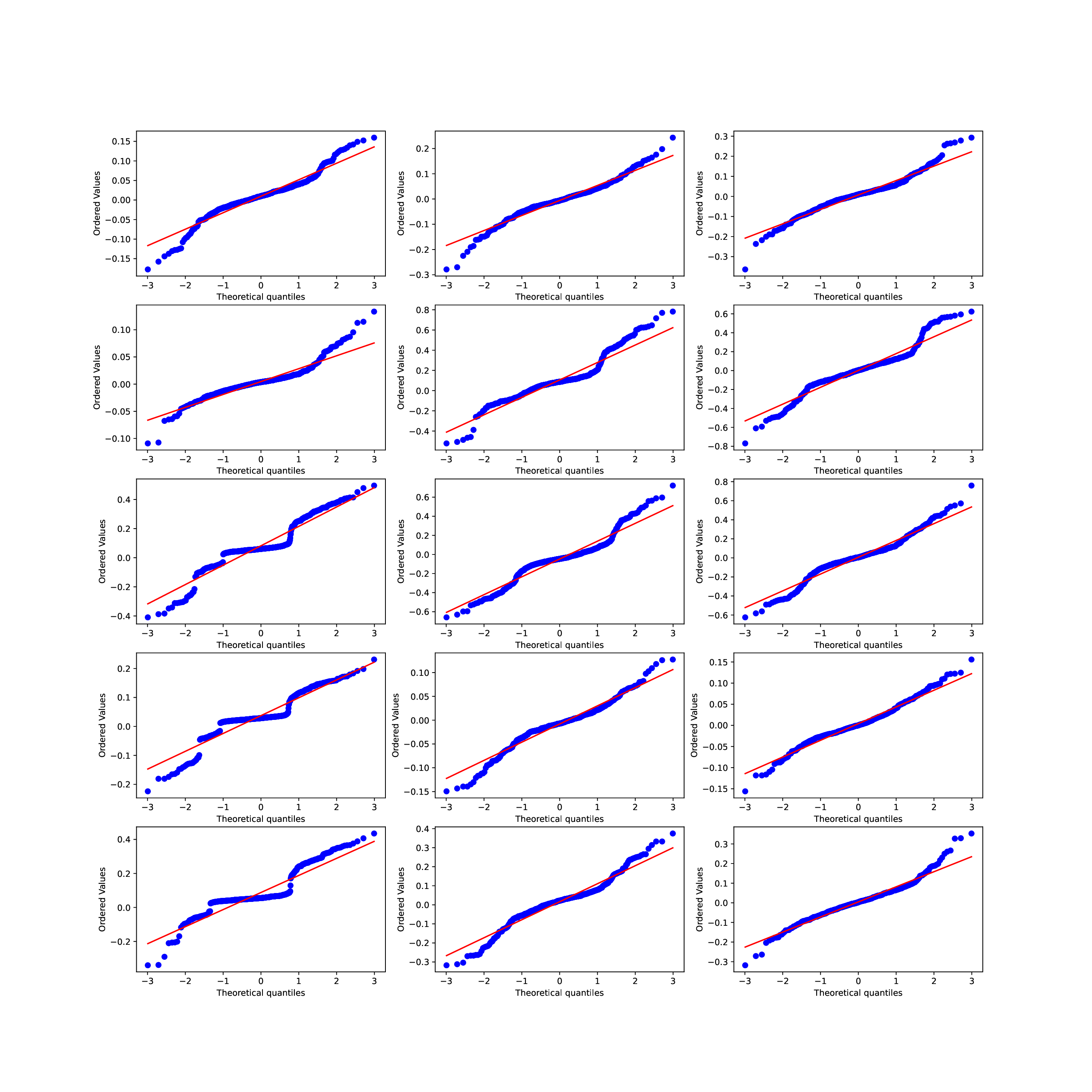}
\caption{The QQ plots for Setting 2 with $\psi=0.1$, missing pattern II.}
\label{fig:6-2}
\end{figure}
\begin{figure}
\includegraphics[scale=0.3]{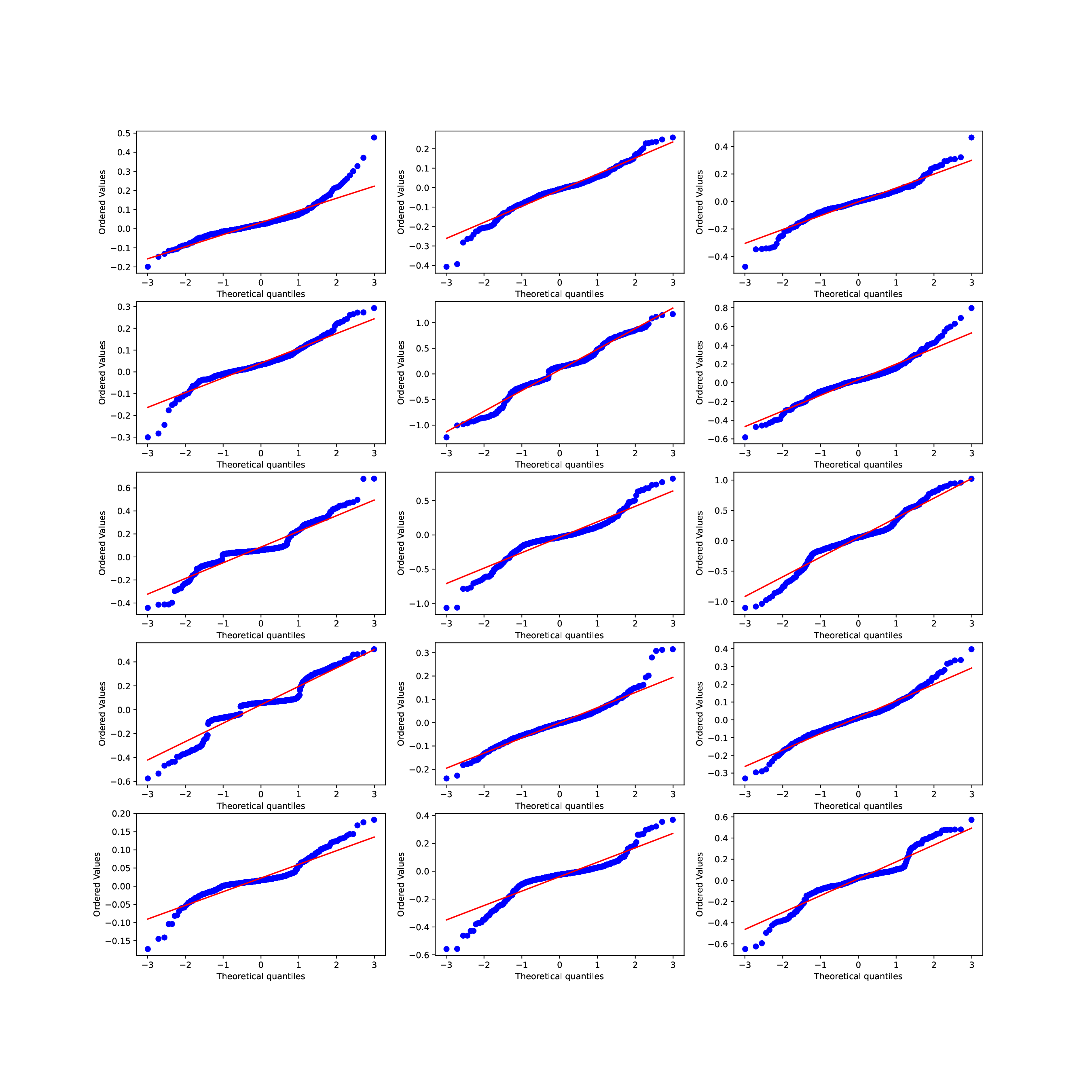}
\caption{The QQ plots for Setting 2 with $\psi=0.5$, missing pattern II.}
\label{fig:6-3}
\end{figure}
\begin{figure}
\includegraphics[scale=0.3]{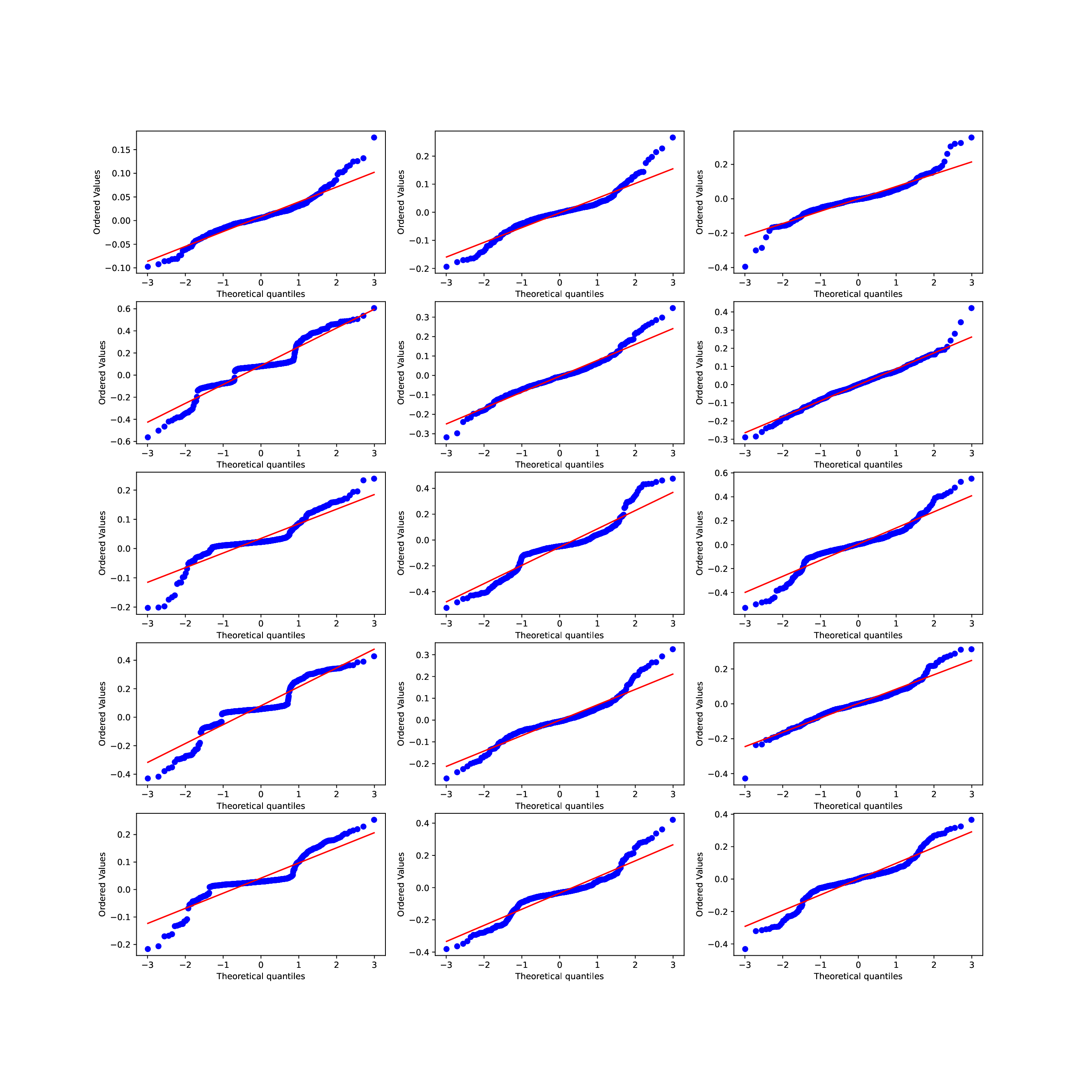}
\caption{The QQ plots for Setting 3, missing pattern II.}
\label{fig:6-4}
\end{figure}

\begin{figure}
\includegraphics[scale=0.4]{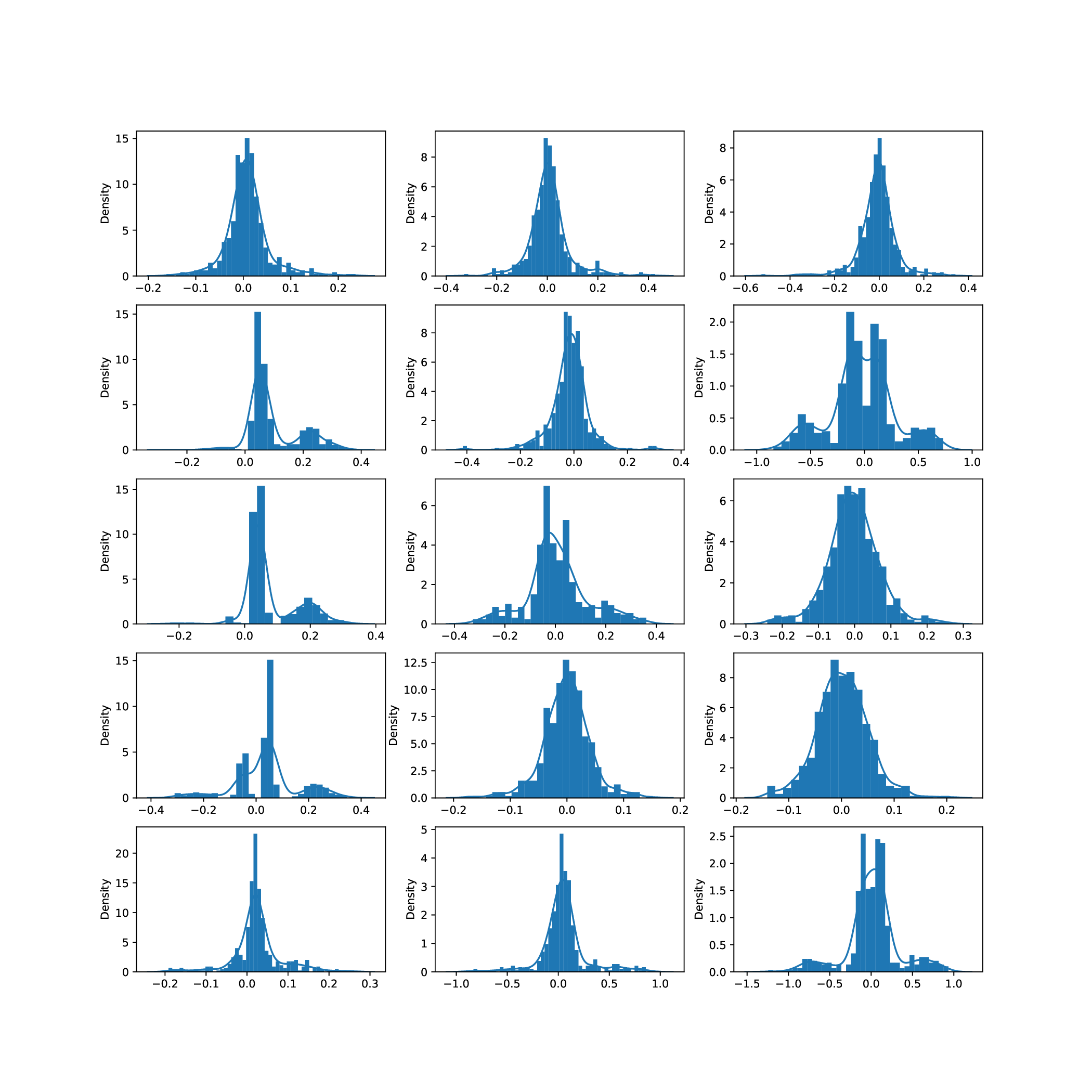}
\caption{The histograms for Setting 1, missing pattern II.}
\label{fig:6-5}
\end{figure}
\begin{figure}
\includegraphics[scale=0.4]{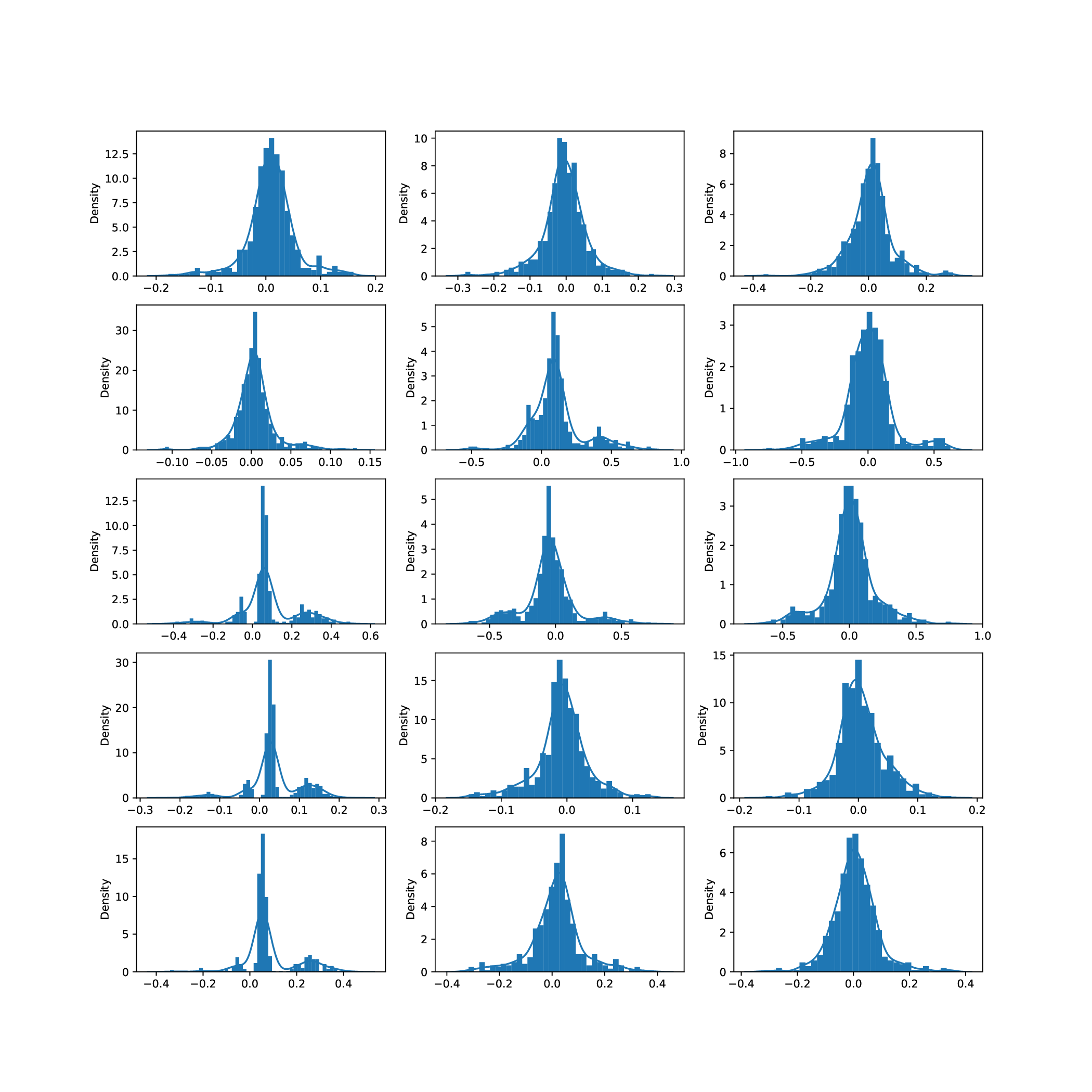}
\caption{The histograms for Setting 2 with $\psi=0.1$, missing pattern II.}
\label{fig:6-6}
\end{figure}
\begin{figure}
\includegraphics[scale=0.4]{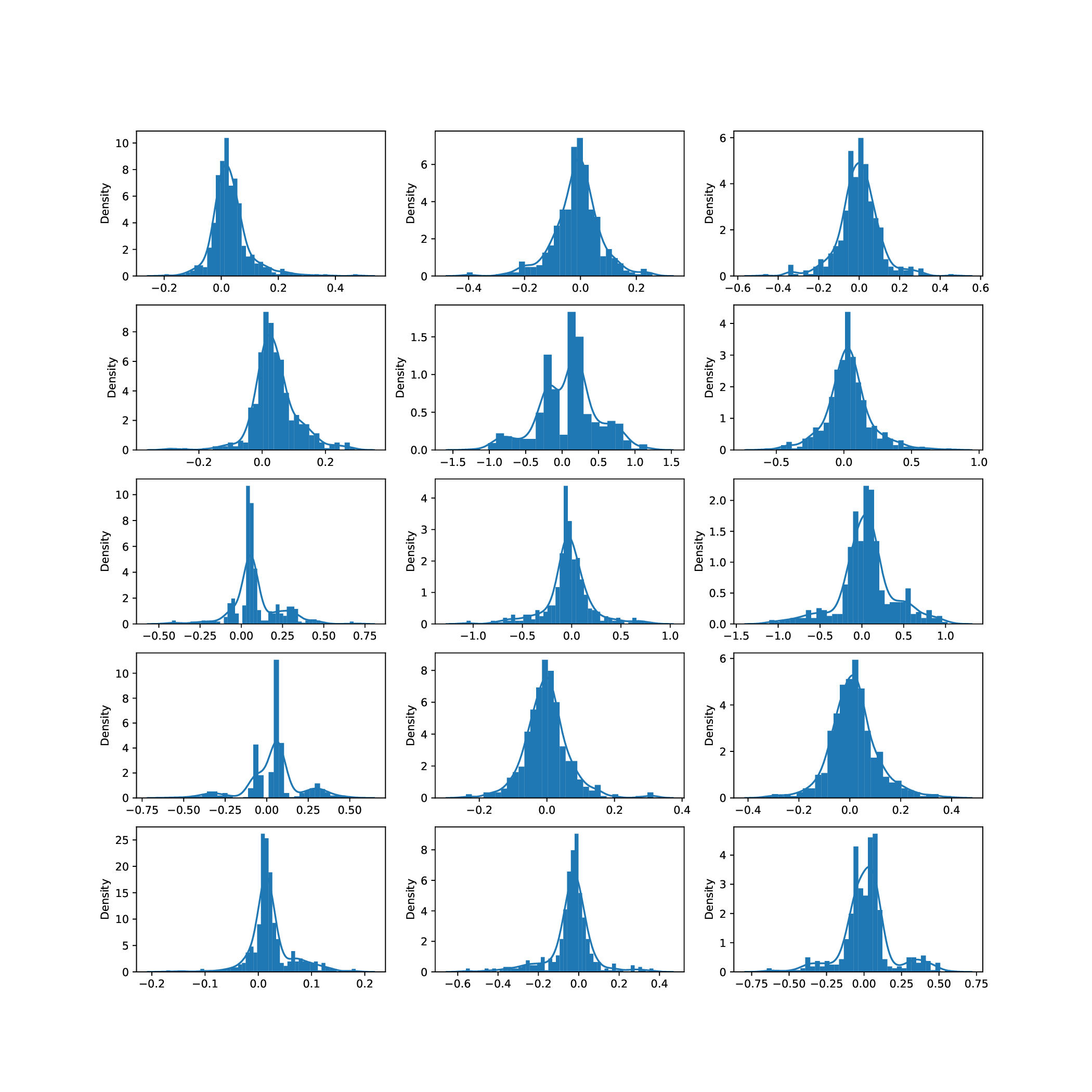}
\caption{The histograms for Setting 2 with $\psi=0.5$, missing pattern II.}
\label{fig:6-7}
\end{figure}
\begin{figure}
\includegraphics[scale=0.4]{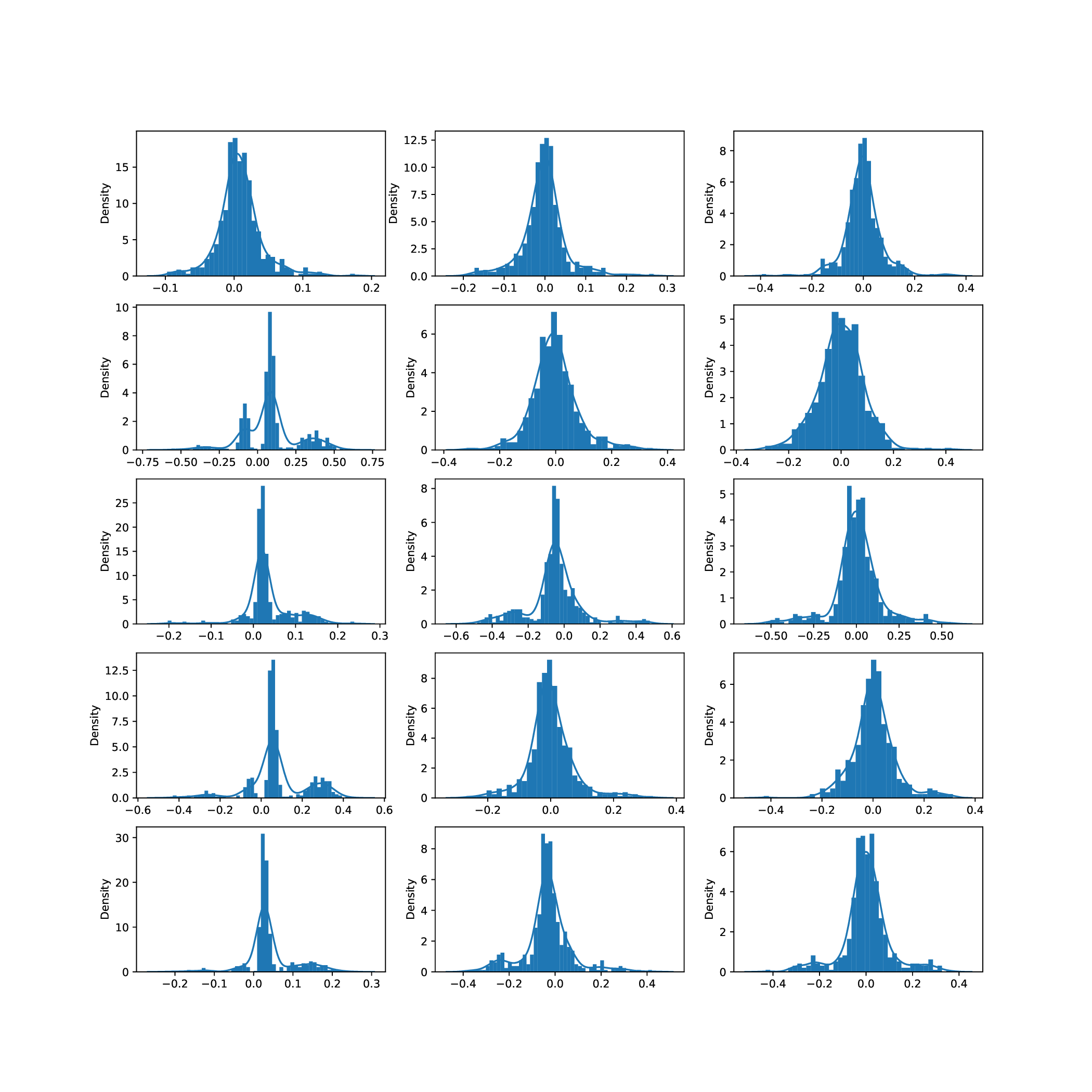}
\caption{The histograms for Setting 3, missing pattern II.}
\label{fig:6-8}
\end{figure}

\section{Multinational Macroeconomic Indexes Dataset}\label{A:3}%
Table \ref{tabA3-1} lists the transformation of all the macroeconomic indices, and a brief data description. In the transformation column, $\bigtriangleup$ denotes the first difference.
\begin{table}
\caption{Data transformations, and variable definitions\label{tabA3-1}}%
\resizebox{.9\linewidth}{!}{
\begin{tabular}{lll}
 \hline
Name & Transformation  &  Description\\
 \hline
CPI: Food      & $\bigtriangleup^2\ln$    & Consumer Price Index: Food,seasonally adjusted\\
CPI: Energy    & $\bigtriangleup^2\ln$    &  Consumer Price Index: Energy,seasonally adjusted\\
CPI: Total     & $\bigtriangleup^2\ln$    &Consumer Price Index: Total,seasonally adjusted\\
IR: Long trates & $\bigtriangleup$        & Interest Rates: Long-term gov bond yields\\
IR: Short time trates & $\bigtriangleup$  & Interest Rates: 3-month Inter bank rates and yields\\
P: Industrial production & $\bigtriangleup\ln$ & Production:Total industry excl construction\\
P: Total manufacturing  & $\bigtriangleup\ln$ & Production:Total manufacturing\\
GDP  & $\bigtriangleup\ln$ &  GDP: Index\\
IT: Trade-export  & $\bigtriangleup\ln$ & International Trade: Total Exports Value(goods)\\
IT: Trade-import  & $\bigtriangleup\ln$ &International Trade: Total Imports Value(goods)\\
 \hline
\end{tabular}
}
\end{table}

\begin{figure}
\includegraphics[scale=0.5]{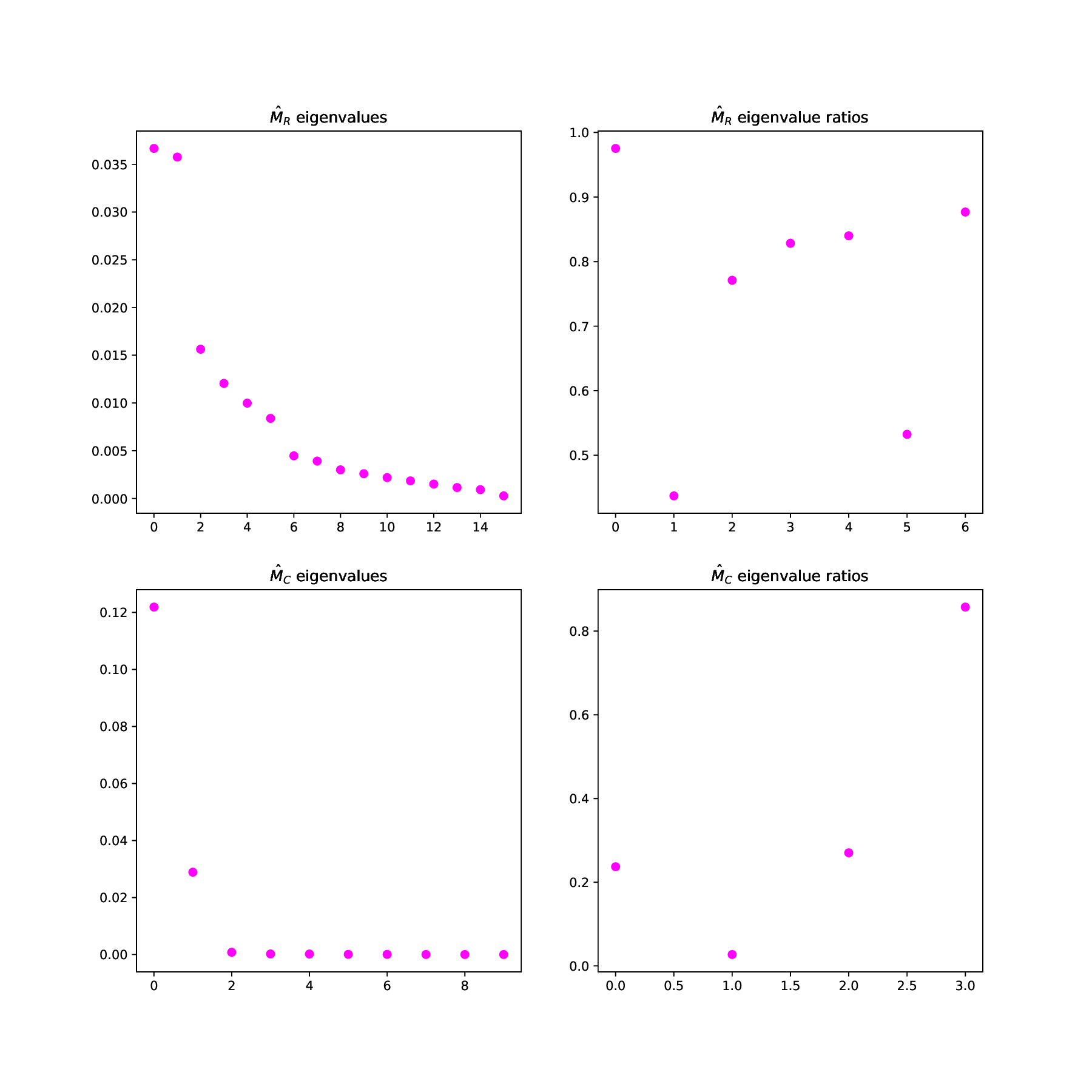}
\caption{Eigenvalues}
\label{fig:5-2}
\end{figure}

%
%
%
%
%

\end{document}